\documentclass[%
 reprint,
 superscriptaddress,
 amsmath,amssymb,
 aps,
]{revtex4-2}

\usepackage{graphicx}
\usepackage{dcolumn}
\usepackage{bm, xcolor}

\usepackage[version=4]{mhchem}

\usepackage{subfig} 
\usepackage{verbatim}
\usepackage{graphics,setspace,float,enumerate,enumitem}
\usepackage[export]{adjustbox}
\usepackage[framemethod=TikZ]{mdframed}
\usepackage{tcolorbox}
\usepackage{commath,mathrsfs} 
\usepackage[noabbrev,capitalize]{cleveref}
\usepackage{multirow}

\creflabelformat{equation}{#2\textup{#1}#3}

\newcommand{\refSI}[2]{\ref{#1}}

\newcommand{\angstrom}{\mbox{\normalfont\AA}}
\newcommand{\rom}[1]{\uppercase\expandafter{\romannumeral #1\relax}} 
\newcommand{\muB}{$\mu_\text{B}$}

\newcommand{\Fsf}[1]{#1\textsubscript{SF}}
\newcommand{\FpUJ}[1]{#1+$U$+$J$}
\newcommand{\FsfUJ}[1]{\FpUJ{\Fsf{#1}}}

\newcommand{\pderiv}[2]{\frac{\partial #1}{\partial #2}}

\usepackage{xr}
\makeatletter

\newcommand*{\addFileDependency}[1]{
\typeout{(#1)}
\@addtofilelist{#1}
\IfFileExists{#1}{}{\typeout{No file #1.}}
}\makeatother

\newcommand*{\myexternaldocument}[2]{%
\externaldocument[#2]{#1}%
\addFileDependency{#1.tex}%
\addFileDependency{#1.aux}%
}

\myexternaldocument{suppinfo}{si-}

\makeatother

\begin{document}

\preprint{APS/123-QED}

\title{
Realistic non-collinear ground states of solids \\ with source-free exchange correlation functional
%
%
%
}

\author{Guy C. Moore}
\affiliation{Department of Materials Science and Engineering, University of California Berkeley, Berkeley, CA 94720, USA}
\affiliation{Materials Science Division, Lawrence Berkeley National Laboratory, Berkeley, CA 94720, USA}

\author{Matthew K. Horton}
\affiliation{Department of Materials Science and Engineering, University of California Berkeley, Berkeley, CA 94720, USA}
\affiliation{Materials Science Division, Lawrence Berkeley National Laboratory, Berkeley, CA 94720, USA}

\author{Aaron D. Kaplan}
\affiliation{Materials Science Division, Lawrence Berkeley National Laboratory, Berkeley, CA 94720, USA}

\author{Sin\'ead M. Griffin}
\affiliation{Department of Physics, University of California Berkeley, Berkeley, CA 94720, USA}
\affiliation{Molecular Foundry, Lawrence Berkeley National Laboratory, Berkeley, CA 94720, USA}

\author{Kristin A. Persson}
\affiliation{Department of Materials Science and Engineering, University of California Berkeley, Berkeley, CA 94720, USA}
\affiliation{Molecular Foundry, Lawrence Berkeley National Laboratory, Berkeley, CA 94720, USA}


\date{\today}

\begin{abstract}

In this work, we expand upon the source-free (SF) exchange correlation (XC) functional developed by Sangeeta Sharma and co-workers to plane-wave density functional theory (DFT) based on the projector augmented wave (PAW) method. This constraint is implemented by the current authors within the VASP source code, using a fast Poisson solver that capitalizes on the parallel three-dimensional fast Fourier transforms (FFTs) implemented in VASP. Using this modified XC functional, we explore the improved convergence behavior that results from applying this constraint to the GGA-PBE+$U$+$J$ functional. In the process, we compare the non-collinear magnetic ground state computed by each functional and their SF counterpart for a select number of magnetic materials in order to provide a metric for comparing with experimentally determined magnetic orderings. We observe significantly improved agreement with experimentally measured magnetic ground state structures after applying the source-free constraint. Furthermore, we explore the importance of considering probability current densities in spin polarized systems, even under no applied field. We analyze the XC torque as well, in order to provide theoretical and computational analyses of the net XC magnetic torque induced by the source-free constraint. Along these lines, we highlight the importance of properly considering the real-space integral of the source-free local magnetic XC field. Our analysis on probability currents, net torque, and constant terms draws additional links to the rich body of previous research on spin-current density functional theory (SCDFT), and paves the way for future extensions and corrections to the SF corrected XC functional.

\end{abstract}


\maketitle

\section{Introduction}

It is a well-known physical fact that Maxwell's equations preclude the existence of unphysical magnetic monopoles. It is less conventional to apply this divergence-free constraint to density functional theory functionals for \textit{ab initio} calculations. It has recently been a topic of exploration to apply this constraint to the exchange correlation component of the effective internal magnetic field, $\bm B_{xc}$ \cite{sharmaSourceFreeExchangeCorrelationMagnetic2018a, krishnaCompleteDescriptionMagnetic2019}. Previous studies have shown that applying this physically inspired constraint to $\bm B_{xc}$ results in improved agreement with the majority of a small test set of over twenty experimentally measured magnetic structures, both in terms of magnetic moment magnitudes \cite{sharmaSourceFreeExchangeCorrelationMagnetic2018a} and non-collinear ground states \cite{krishnaCompleteDescriptionMagnetic2019}, which can be challenging to accurately match experiment with conventional DFT approaches.

There are two primary approaches for incorporating magnetism in density functional theory (DFT). The first approach is spin density functional theory (SDFT), in which a functional is defined with respect to spin-up and spin-down electron densities, $\rho_{\uparrow} (\bm r)$ and $\rho_{\downarrow} (\bm r)$, respectively. The extension of SDFT to non-collinear magnetism motivates a spinor representation of the functional, inspired by the Pauli matrix spin-1/2 formalism. Under this reformulation, functionals of $\rho_{\uparrow} (\bm r)$ and $\rho_{\downarrow} (\bm r)$ can be expressed in terms of total electron density \mbox{$\rho(\bm r) = \rho_\uparrow (\bm r) + \rho_\downarrow (\bm r)$}, and magnetization, which in both collinear and non-collinear formulations obeys the relationship \mbox{$|\bm m (\bm r)| = |\rho_\uparrow (\bm r) - \rho_\downarrow (\bm r)|$}, under a local diagonalization of the spinor $2 \times 2$ representation \cite{peraltaNoncollinearMagnetismDensity2007}. Details on the connection between the spinor and density/magnetization formulation of non-collinear DFT is touched on in Equation~\refSI{eq:rho_spinor}{S2}, Equation~\refSI{eq:s_mag}{S4}, and Equation~\refSI{eq:rho_up_down}{S5} in Section~\refSI{sec:KinEprojSpinCurrents}{S2 B}, contained in the Supplementary Information.

The second DFT formulation that incorporates non-collinear magnetism is current density functional theory (CDFT). This methodology is commonly used to incorporate electrodynamic effects in time-dependent density functional theory (TDDFT) \cite{furnessCurrentDensityFunctional2015a, tchenkoueForceBalanceApproach2019a}. The ``current" in CDFT conveys the reformulation of the functional to be minimized with respect to the spin current density $\bm j_s (\bm r) \propto \nabla \times \bm m (\bm r)$, rather than $\bm m (\bm r)$, the magnetization field itself \cite{sharmaSourceFreeExchangeCorrelationMagnetic2018a}.

Several works have explored the theoretical justification for reformulating SDFT functionals within the CDFT setting \cite{sharmaSourceFreeExchangeCorrelationMagnetic2018a, capelleSpinDensityFunctionalsCurrentDensity1997a, eschrigCurrentDensityFunctional1985}. In particular, Sharma and coauthors show through variational calculus that applying a divergence-free constraint to $\bm B_{xc}$ is equivalent to redefining the exchange correlation energy functional $E_{xc}[\rho, \bm m]$ in terms of $\nabla \times \bm m (\bm r)$, $E_{xc}[\rho, \nabla \times \bm m]$, which is consistent with CDFT methodology \cite{capelleSpinDensityFunctionalsCurrentDensity1997a}.

In conventional SDFT, the exchange correlation local potential fields can be expressed as follows
\begin{align}
    v_{xc} (\bm r) &= \frac{1}{2} \left( v^\uparrow_{xc} (\bm r) + v^\downarrow_{xc} (\bm r) \right) \nonumber \\
    \bm B_{xc} (\bm r) &= \frac{1}{2} \left( v^\uparrow_{xc} (\bm r) - v^\downarrow_{xc} (\bm r) \right) \hat{\bm m} (\bm r),
    \label{eq:sdft_xc}
\end{align}
which are the complementary local potential fields for total electron density, $\rho$, and magnetization density, $\bm m$, respectively, and $\hat{\bm m} (\bm r)$ is the unit vector in the direction of the magnetization. 

The source-free constraint is applied by projecting the original exchange-correlation magnetic field onto a divergence-free field. Consistent with the work of Sharma and others, we use the fundamental theorem of vector calculus, the Helmholtz identity, to reformulate our equations in terms of vector and scalar fields. This theorem states that any once differentiable, $\mathcal C^1$, vector field can be decomposed into a divergence free and a curl free component, 
\begin{align}
    \label{eq:helmholtz}
    \bm B (\bm r)  = - \nabla \phi (\bm r) + \nabla \times \bm A (\bm r) + \overline{\bm B}
\end{align}
where $\phi$ is a scalar field, and $\bm A$ is a vector potential field. The identity is conveniently written in this way due to two important mathematical properties: the curl of a gradient is zero everywhere in the domain $\nabla \times  \nabla \phi = 0$ and likewise for the divergence of the curl of a vector field $\nabla \cdot (\nabla \times \bm A) = 0$. For a magnetic field $\bm B$, such that $\nabla \cdot \bm B = 0$, we can define $\bm B = \nabla \times \bm A$, where $\bm A$ is the well-known magnetic vector potential. 

One may ask why we include a constant $\overline{\bm B}$, which is not included in most statements of the Helmholtz identity. This comes down to the fact that under periodic boundary conditions, 
\begin{align}
    \int_\Omega d \bm r \ \nabla \phi (\bm r) &= \bm 0 \\
    \int_\Omega d \bm r \ \nabla \times \bm A (\bm r) &= \bm 0
\end{align}
by Equations~\refSI{eq:grad_int_periodic}{S12}~\&~\refSI{eq:stokes_periodic}{S13} in Section~\refSI{sec:VanishIntPBC}{S2 C}. Therefore, we must include a constant term, such that \mbox{$\int_\Omega d \bm r \ \bm B (\bm r)$} need not be the zero vector.

By taking the divergence of both sides of \cref{eq:helmholtz}, we see that, for the exchange-correlation magnetic field,
\begin{align}
    \nabla \cdot \bm B_{xc} &= - \nabla \cdot (\nabla \phi) = -\nabla^2 \phi. \label{eq:poisson}
\end{align}
Therefore, solving for $\phi$ requires the solution to the Poisson equation above, which is the method suggested by Sharma et al. for applying the source-free constraint \cite{sharmaSourceFreeExchangeCorrelationMagnetic2018a}. It is interesting to note that $\bm B_{xc}' (\bm r)$ must be non-collinear, as we show in Section~\refSI{sec:NclBxc}{S2 A}.

\cref{eq:helmholtz} can be employed rigorously, because while the Helmholtz decomposition is not unique, it can be under the correct constraints. The uniqueness of the gradient of the solution to the Poisson equation (e.g. $\nabla \phi$) can be used to prove this. In summary, each field in \cref{eq:helmholtz} can be uniquely solved for under periodic boundary conditions (PBCs), as follows
\begin{enumerate}
    \item The curl-free term:
    \begin{align}
        \nabla^2 \phi &= - \nabla \cdot \bm B_{xc}.
        \label{eq:phiPoisson}
    \end{align}
    Therefore, $\nabla \phi$ is unique under PBCs.
    \item The divergence-free term:
    \begin{align}
        \nabla^2 \bm A_{xc} &= - \nabla \times \bm B_{xc},
        \label{eq:AxcPoisson}
    \end{align}
    subject to the Coulomb gauge constraint, \mbox{$\nabla \cdot \bm A_{xc} = 0$} (see Appendix~\ref{sec:GaugeAxc}). 
    Under this constraint, $\nabla A_{xc,i}$ are unique under PBCs.
    \item The constant term:
    \begin{align}
        \overline{\bm B}_{xc} &= \frac{1}{V_\Omega} \int_\Omega d \bm r \ \bm B_{xc} (\bm r).
        \label{eq:BxcConstIntegral}
    \end{align}
\end{enumerate}
Therefore, in order to compute $\bm B_{xc}'$, the source-free projection of $\bm B_{xc}$, one simply needs to compute $\nabla \phi$
\begin{align}
    \bm B_{xc}' (\bm r) &= \bm B_{xc} (\bm r) + \nabla \phi (\bm r) \nonumber \\
    &= \bm B_{xc} (\bm r) + \bm H_{xc} (\bm r) \nonumber \\
    &= \nabla \times \bm A_{xc} (\bm r) + \overline{\bm B}_{xc}
    \label{eq:sf_bxc_defn}
\end{align}
In other words, we leave $\overline{\bm B}_{xc}$ untouched from SDFT, to be consistent with Ref.~\onlinecite{sharmaSourceFreeExchangeCorrelationMagnetic2018a}. In Sections~\ref{sec:q0-discussion}~\&~\ref{sec:SFimplement},
we explore the justifications and implications of this, and lay some potential groundwork for future studies to employ in a more careful treatment of the constant term, $\overline{\bm B'}_{xc}$. In a concrete sense, we explore the role of $\overline{\bm B'}_{xc}$ on the converged magnetic ground states in Section~\ref{sec:q0-discussion}.

In the following sections, we examine some of the details of CDFT, and SCDFT by extension, which are important to the study of the source-free constraint on SDFT, and possible future directions of this work. The topics of the following subsections of the introduction are outlined below:
\begin{itemize}
  \item In Section~\ref{sec:DegenSdftCdft}, we explore the possible theoretical justifications for the degeneracies of non-collinear SDFT, based on the statements put forth by earlier studies.
  \item Section~\ref{sec:ConsiderCDFT} provides an introduction to the probability current, and its contributions to $E_{xc}$, which are not explicitly accounted for in SDFT. Next, we explore extensions to metaGGA functionals. 
  This extension will require the explicit treatment of the probability current, which enters into the kinetic energy density. Therefore, we do not include source-free metaGGA in our implementation. 
  \item In Section~\ref{sec:ZTTintro}, we explore the zero torque theorem, and extensions thereof. We demonstrate that local magnetic torques can arise without violating conservation laws. 
  \item Finally, Section~\ref{sec:OrbitalMagmom} provides an overview of the relationship between the orbital magnetic moment and probability current, to further illustrate where these additional currents could play a more crucial role. 
\end{itemize}
Next, following the Methods section, we highlight the results of this study, which include the following subsections
\begin{itemize}
    \item Section~\ref{sec:monopoles} provides a quantitative analysis of the monopole density that arises in an SDFT description of \ce{Mn3ZnN}, and the effect of the source-free correction.
    \item In Section~\ref{sec:convergence-testset}, we examine the improved convergence properties of the source-free functional for a representative test set of non-collinear magnetic structures.
    \item  We also interrogate the improved convergence speed using the source-free functional in Section~\ref{sec:convergence-YMnO3}. The well-known hexagonal manganite \ce{YMnO3} was chosen as a material test case \cite{YMnO3-exp}.
    \item In Section~\ref{sec:UO2-exploration}, we explore the computed orbital moments and probability current that arises in \ce{UO2} to provide material grounds for future studies to examine the coupling between the probability current and XC vector potential.
    \item Since the net XC torque induced by the source-free correction hasn't been examined by other studies, to the best of our knowledge, we provide a analysis of the net XC torque in Section~\ref{sec:SF-torque}, based on the theory presented in Section~\ref{sec:ZTTintro}.
    \item In addition to global XC torque, in Section~\ref{sec:q0-discussion} we provide computational evidence for the importance of properly considering $\overline{\bm B'}_{xc}$, which will be the subject of future studies.
\end{itemize}

\subsection{The comparative degeneracies of $E_{xc}$ for SDFT versus CDFT with respect to $\bm m$}
\label{sec:DegenSdftCdft}

Capelle and Gross (CG) show that within SDFT, the exchange-correlation energy, $E_{xc}$ only depends on the magnetization via the ``spin vorticity" \cite{capelleSpinDensityFunctionalsCurrentDensity1997a}
\begin{align}
    \bm \nu_s (\bm r)
    &= \nabla \times \left( \frac{\bm j_s (\bm r)}{\rho} \right)
    = \frac{c}{q} \nabla \times \left( \frac{\nabla \times \bm m (\bm r)}{\rho (\bm r)} \right),
    \label{eq:spin_vort}
\end{align}
where $c$ and $q$ are the speed of light and elementary charge, respectively. Therefore, transformations of the magnetization density, $\bm m (\bm r) \mapsto \bm m' (\bm r)$, of the form
\begin{align}
    \bm m' (\bm r) &= \bm m (\bm r) + \nabla \alpha (\bm r) + \bm \Gamma (\bm r) \label{eq:m_dof_sdft} \\
    \text{where } & \nabla \times \bm \Gamma (\bm r) = \rho (\bm r) \nabla \gamma (\bm r) \nonumber
\end{align}
will have no effect on XC contributions to the SDFT functional, $E^S_{xc}$, i.e. \mbox{$E^S_{xc}[\rho,\bm m'] = E^S_{xc}[\rho,\bm m]$}. $\alpha (\bm r)$ and $\bm \Gamma (\bm r)$ are arbitrary functions, which are only subject to the relevant aforementioned constraints. 

By comparison, in CDFT, $E^C_{xc}$ depends on $\bm m$ via the spin current, $\bm j_s = \frac{c}{q} \nabla \times \bm m$ \cite{capelleSpinDensityFunctionalsCurrentDensity1997a, sharmaSourceFreeExchangeCorrelationMagnetic2018a}. Therefore, $E^C_{xc}$ is invariant to transformations of the form, $\bm m (\bm r) \mapsto \bm m' (\bm r)$,
\begin{align}
    \bm m' (\bm r) &= \bm m (\bm r) + \nabla \alpha (\bm r), \label{eq:m_dof_cdft}
\end{align}
for arbitrary $\nabla \alpha (\bm r)$. Therefore, $\bm \Gamma (\bm r)$ of the form in \cref{eq:m_dof_sdft} provides the additional degree of freedom that introduces ambiguity into the non-collinear magnetic ground state obtained using SDFT compared with CDFT.

Again, we emphasize that the functions $\alpha$, $\bm \Gamma$, and $\gamma$ are all completely arbitrary, and have been introduced to illustrate the gauge invariance of non-collinear SDFT \cite{capelleSpinDensityFunctionalsCurrentDensity1997a}. The high number of degrees of freedom associated with $\bm \Gamma (\bm r)$ provides an explanation for the highly degenerate energy landscape in non-collinear SDFT. The additional gauge invariance explains why the site-projected magnetic moments -- related to $\bm m$ -- rotate very little during convergence, compared with the source-free functional, which is in fact a current density functional, to reiterate \cite{sharmaSourceFreeExchangeCorrelationMagnetic2018a, capelleSpinDensityFunctionalsCurrentDensity1997a}. This is the core computational exploration of the present study.

\subsection{Consideration of additional currents arising in CDFT}
\label{sec:ConsiderCDFT}

Within the framework of current density functional theory (CDFT), we can introduce the probability (denoted with subscript $p$) current density  \cite{vignaleMagneticFieldsDensity1990, capelleSpinDensityFunctionalsCurrentDensity1997a}. The classical field of the probability current, (also known as the ``paramagnetic current") can be expressed as the expectation of the quantum mechanical operator
\begin{align}
    {\bm j}_p ({\bm r}) &= \left< \hat{\bm j}_p ({\bm r}) \right> \nonumber \\
    &= \frac{\hbar}{2mi} \left< \hat \Psi ({\bm r})^\dagger \nabla \hat \Psi ({\bm r}) - \nabla \hat \Psi ({\bm r})^\dagger \hat \Psi ({\bm r}) \right> \nonumber \\
    &= \frac{\hbar}{2m} \left< \Im \left( \hat \Psi ({\bm r})^\dagger \nabla \hat \Psi ({\bm r}) \right) \right>. \label{eq:jp}
\end{align}
The probability current can be motivated by expressing the Schr\"odinger equation as a conservation law, \mbox{$\partial \rho / \partial t = \nabla \cdot \bm j_p$} \cite{SakuraiNapolitano2017}. Therefore, $\bm j_p$ is directly related to the probability flux of an electron in real space. It is also known as the ``paramagnetic current," because $\bm j_p$ couples in a Zeeman-like manner to the external vector potential in a similar fashion to \cref{eq:Ep}. Therefore, the magnetization induced by $\bm j_p$ will preferentially align with the external magnetic field. 

Capelle and Gross explore the form of $\bm j_p$ within Kohn-Sham (KS) SDFT \cite{capelleSpinDensityFunctionalsCurrentDensity1997a},
\begin{align}
    \bm j_p^{KS} &= \frac{i\hbar}{2m} \sum_{k=1}^N \left( \phi_k \nabla \phi_k^* - \phi_k^* \nabla \phi_k \right),
\end{align}
where $\phi_k$ are Kohn-Sham orbitals, and the sum is over the $N$ lowest-energy bands. $\hbar$ is the familiar reduced Planck constant, and $m$ is the electron mass.
Any complex function, and therefore $\phi_k$, can be expressed as $\phi_k (\bm r) = |\phi_k (\bm r)| e^{i S_k'(\bm r)}$. Letting $S_k(\bm r) = \hbar S_k'(\bm r)$, it is possible to show that \cite{SakuraiNapolitano2017}
\begin{align}
    {\bm j}_p^{KS}(\bm r) = \sum_k^N \frac{1}{m} \left| \phi_k (\bm r) \right|^2 \nabla S_k(\bm r). \label{eq:jp_rho_grad}
\end{align}
Next, we consider the case in which all $\nabla S_k$ are equal to the same function, $\nabla S$. In this case, the following simplification can be made
\begin{align}
    \widetilde {\bm j}_p^{KS}(\bm r) = \frac{1}{m} \rho (\bm r) \nabla S(\bm r). \label{eq:jp_rho_grad_special}
\end{align}
We have drawn on the fact that $ \rho (\bm r) = \sum_k^N \left| \phi_k (\bm r) \right|^2 $, and are reminded that $\nabla S$ is linked to classical momentum \cite{SakuraiNapolitano2017}. Therefore, ${\bm j}_p^{KS}(\bm r) = \widetilde {\bm j}_p^{KS}(\bm r)$ only in the case when these $\nabla S_k$ momenta are equal across all ground state KS orbitals, $\phi_k (\bm r)$.

In their original seminal work, Vignale and Rasolt (VR) consider the physical constraint that the current density functional should be invariant to gauge transformations of the external magnetic vector potential \mbox{$\bm A' \mapsto \bm A + \nabla \Lambda$}, where $\Lambda (\bm r)$ is an arbitrary function \cite{vignaleDensityfunctionalTheoryStrong1987}. Under this gauge invariance constraint, VR demonstrate that $E_{xc}$ depends on $\bm j_p$ through the probability/paramagnetic current vorticity alone, which is defined as such \cite{vignaleDensityfunctionalTheoryStrong1987}
\begin{align}
\bm \nu_p (\bm r) = \nabla \times \left[ \frac{\bm j_p (\bm r)}{\rho (\bm r)} \right]. \label{eq:vorticity}
\end{align} 
This result is a foundational pillar of CDFT, and provides a basis for useful results in other key works, such as Ref.~\onlinecite{capelleSpinDensityFunctionalsCurrentDensity1997a}. 

Interestingly, in the hypothetical case in which ${\bm j}_p^{KS}(\bm r) = \widetilde {\bm j}_p^{KS}(\bm r)$ (\cref{eq:jp_rho_grad_special}) we see that $\bm \nu_p (\bm r) = \bm 0$ everywhere, and therefore we may neglect any ${\bm j}_p$ contributions to $E_{xc}$ entirely. However, this is not the case for differing $\nabla S_k$. We presume that, among other effects, the inhomogeneity across ground state $\nabla S_k$ is accentuated by spin-orbit coupling (SOC) \cite{socVASP}, which will introduce an angular-momentum dependence on $\phi_k$. This hypothesis can most likely be further explored by the machinery proposed in Ref.~\onlinecite{bencheikhSpinOrbitCoupling2003}, which provides a framework for rigorously incorporating SOC within SCDFT. However, on a more elementary level, even a non-interacting homogeneous electron gas will obey Fermi statistics. Therefore the electron gas will possess a distribution of momenta, which can be directly related to single-electron $\nabla S_k$ \cite{AshcroftMermin1976, SakuraiNapolitano2017}.

Under no external magnetic field, i.e. a curl-free $\bm A$, $\bm j_p$ enters into the exchange-correlation component of the CDFT functional in the following form \cite{vignaleMagneticFieldsDensity1990, vignaleDensityfunctionalTheoryStrong1987, capelleSpinDensityFunctionalsCurrentDensity1997a},
\begin{align}
    E_p &\equiv - \int_\Omega \bm{j}_p (\bm r) \cdot \bm{A}_{xc} (\bm r) \ d \bm r. \label{eq:Ep}
\end{align}
In VR's extension of SCDFT to non-collinear magnetism, a set of spin-projected currents, $\bm j_{p,\lambda}$, and their associated $\bm A_{xc,\lambda}$ are introduced as additional quantities. As we will see later, $\bm j_{p,\lambda}$ play a crucial role in the extension of the zero torque theorem, \cref{eq:ztt_general}. Furthermore, in Ref.~\onlinecite{bencheikhSpinOrbitCoupling2003}, Bencheikh demonstrates that $\bm A_{xc,\lambda}$ become essential in the inclusion of SOC effects in SCDFT. However, because SOC is treated differently in VASP \cite{socVASP}, we will not explore this further here.


The probability current, $\bm j_p$, has been introduced as the gauge-invariant kinetic energy density in the metaGGA extension of SCDFT \cite{taoExplicitInclusionParamagnetic2005, furnessCurrentDensityFunctional2015a}. Following from Ref.~\cite{taoExplicitInclusionParamagnetic2005},
\begin{align}
  \tilde \tau_\sigma &= \tau_\sigma - m \frac{\left| \bm j_{p,\sigma} \right|^2}{\rho_\sigma} \label{eq:kin_density_spin}
\end{align}
where $\sigma$ denotes the different spin channels, $\sigma \in \left\{ \uparrow, \downarrow \right\}$. Starting from the definitions of VR \cite{vignaleCurrentSpindensityfunctionalTheory1988}, it is straightforward to show that the squared magnitude of the spin probability currents $\left| \bm j_{p,\sigma} \right|^2$ in \cref{eq:kin_density_spin} can be related to $\left| \bm j_{p,\lambda} \right|^2$, which we explore in Section~\refSI{sec:KinEprojSpinCurrents}{S2 B}. Therefore, we hope it to be the subject of future studies to explore the extensions of this work to metaGGA functionals.


Furthermore, in Appendix~\ref{sec:DecompProbCurrent}, we abstract from the definition in \cref{eq:jp}, and consider a separation of $\bm j_{p,\lambda}$ into divergence and curl-free contributions. We provide these small derivations to the reader in the hope that may be useful for the reformulation of SDFT metaGGA functionals to their SCDFT counterparts.



\subsection{Zero torque theorem and $\bm \tau_{xc}$}
\label{sec:ZTTintro}

Within full magnetostatic spin-current DFT (SCDFT), additional spin currents arise subject to the following continuity equation
\begin{align}
    \nabla \cdot \left[ \bm{j}_{p,\lambda} + m_\lambda \bm{A}_{xc} \right] &= \left( \bm {m} \times \bm{B}_{xc} \right)_\lambda \label{eq:cons_torque}
\end{align}
where $\lambda = 1,2,3$ and corresponds to the three components of the non-collinear magnetic fields. Some SCDFT formulations include additional $\bm{A}_{xc,\lambda}$ fields as well \cite{vignaleCurrentSpindensityfunctionalTheory1988}. \cref{eq:cons_torque} is an extension of the zero torque theorem (ZTT), in which local torques may arise without violating conservation laws. 

The magnetic torque due to the XC component of the functional can be expressed as $\bm \tau_{xc} = \bm m \times \bm B_{xc}$ \cite{pluharExchangecorrelationMagneticFields2019}. We call on the definition of $\bm B_{xc}$ in \cref{eq:sdft_xc}, as defined in conventional SDFT, to show that $\bm \tau_{xc} = \bm m \times \bm B_{xc} = \bm 0$. This is simply due to the fact that $\bm B_{xc} \parallel \bm m$ everywhere in $\Omega$ (the periodic domain) at every self-consistency step. The zero magnetic torque theorem \cite{pluharExchangecorrelationMagneticFields2019},
\begin{align}
    \int_{\Omega} \bm \tau_{xc} \ d\bm r &= \int_{\Omega} \bm m \times \bm B_{xc} \ d\bm r = \bm 0 
    \label{eq:zero_torque}
\end{align}
states that ``a system cannot exert a net torque on itself." Since $\bm \tau_{xc} = \bm m \times \bm B_{xc} = \bm 0$ then \cref{eq:zero_torque} is trivially satisfied at every step of the DFT minimization algorithm. 

By comparison, within the source-free implementation, local torques may arise \cite{sharmaSourceFreeExchangeCorrelationMagnetic2018a}, in which case, from \cref{eq:sf_bxc_defn} we see that
\begin{align}
    \bm \tau_{xc}' &= \bm m \times \bm B_{xc}' \nonumber \\
    &= \bm m \times \bm B_{xc} + \bm m \times \bm H_{xc} \nonumber \\
    &= \bm m \times \nabla \phi \nonumber \\
    &= \phi (\nabla \times \bm m) - \nabla \times (\phi \bm m).
    \label{eq:sf_torque}
\end{align}
Under periodic boundary conditions on $\Omega$, we may show that
\begin{align}
    \int_\Omega \bm \tau_{xc}' \ d \bm r 
    &= \int_\Omega \bm \tau_{xc} \ d \bm r 
    +\int_\Omega \bm m \times \bm \nabla \phi \ d \bm r \nonumber \\
    &= \int_\Omega \phi \left( \nabla \times \bm m \right) \ d \bm r.
    \label{eq:sf_torque_effective}
\end{align}
Again, we have used the property that the surface integral vanishes under periodic boundary conditions, \mbox{$\int_\Omega d \bm r \ \nabla \times \left( \phi \bm m \right) = \bm 0$}, via Equation~\refSI{eq:stokes_periodic}{S13} provided in Section~\refSI{sec:VanishIntPBC}{S2 C}. It is not apparent why \cref{eq:sf_torque_effective} should be zero, and therefore why ZTT should be obeyed for the source-free functional. From \cref{eq:sf_torque_effective}, it is clear that the net XC torque possesses the same gauge invariance of $E^C_{xc}$ (\cref{eq:m_dof_cdft}), and therefore it cannot be eliminated via the addition of a gradient field to $\bm m$. In Section~\ref{sec:SF-torque}, we explore the adherence to the ZTT computationally, and possible ways to maintain the zero torque condition in Appendix \ref{sec:ZTTfix}. 

Several previous studies have sought to address adherence to the ZTT with various proposed XC non-collinear functionals. For example, in Ref. ~\onlinecite{capelleSpinCurrentsSpin2001}, the authors rigorously explore the dependence of $\bm B_{xc}$ on $\bm m$ that would result in satisfying the ZTT for all $\bm m$. Additionally, Ref. ~\onlinecite{pluharExchangecorrelationMagneticFields2019} proposed a way to maintain the zero global torque constraint in Section~III of their Supplemental Material. While their approach is promising, the method that we propose in Appendix \ref{sec:ZTTfix} leverages the convenient property that volume integrals of derivatives disappear under periodic boundary conditions PBCs. Therefore, in comparison to Ref. ~\onlinecite{pluharExchangecorrelationMagneticFields2019}, we are able to ensure that 
\begin{enumerate}
    \item The resultant ${\bm B'}_{xc}$ is indeed source-free, i.e. \mbox{$\nabla \cdot {\bm B'}_{xc} = 0$}, by solving for the auxiliary vector potential, ${\bm A'}$, rather than the magnetic field itself.
    \item Our least-squares solution only requires solving $3$ linear equations, whereas Ref. ~\onlinecite{pluharExchangecorrelationMagneticFields2019} requires solving $3P-3$ linear equations, where $P$ is the size of the real-space grid \cite{pluharExchangecorrelationMagneticFields2019}.
    \item We treat all points on the real-space grid on the same footing. In other words, we do not solve for the field to maintain ZTT on ``boundary" grid points separately from the bulk, as proposed in Ref. ~\onlinecite{pluharExchangecorrelationMagneticFields2019}. In our setting of PBCs, there is no notion of ``edge" versus ``bulk" to begin with.
\end{enumerate}
We emphasize again that our proposal to maintain the ZTT in Appendix~\ref{sec:ZTTfix} is not currently implemented in our VASP code patch. We hope that this will be the subject of subsequent studies. 

All this being said, ZTT may not be a strict constraint on the spin-current density functional. We see this by starting from Equation 33 of Ref.~\onlinecite{bencheikhSpinOrbitCoupling2003}, which is also Equation 6.10b of Ref.~\onlinecite{vignaleCurrentSpindensityfunctionalTheory1988},
\begin{align}
    & \nabla \cdot \bm{j}_{p, \lambda}+\frac{2}{\hbar} \nabla \cdot\left(m_\lambda \bm{A}_{x c}\right)+\frac{q}{m c} \nabla \cdot\left(\rho \bm{A}_{x c, \lambda}\right) \nonumber \\
    &\quad =-\frac{2}{\hbar}\left[\bm{m} \times \bm{B}_{x c}\right]_\lambda+\frac{2 q}{\hbar c} \sum_{\mu, \nu} \epsilon_{\lambda \mu \nu} \bm{j}_{p, \nu} \cdot \bm{A}_{x c, \mu}.
\end{align}
where $\bm{j}_{p, \lambda}$ are defined in Section~\refSI{sec:KinEprojSpinCurrents}{S2 B}, and \mbox{$\frac{q}{c} \bm A_{xc,\lambda} = \delta E_{xc} / \delta \bm{j}_{p, \lambda}$}. Additionally, $\epsilon_{\lambda \mu \nu}$ is the Levi-Civita symbol. 
In the spirit of Ref.~\onlinecite{capelleSpinCurrentsSpin2001}, we see that by taking the integral of both sides, and imposing periodic boundary conditions on the region of integration $\Omega$, we arrive at
\begin{align}
    \int_\Omega d \bm{r} \  \left[ \bm{m} \times \bm{B}_{x c} \right]_\lambda =\frac{q}{c} \int_\Omega d \bm{r} \  \sum_{\mu, \nu} \epsilon_{\lambda \mu \nu} \ \bm{j}_{p, \nu} \cdot \bm{A}_{x c, \mu}.
    \label{eq:ztt_general}
\end{align}
There is no reason for the right hand side of this equation to be zero. Therefore, the above continuity equation provides an extension of the ``zero torque theorem." In other words, in order to satisfy the steady state conservation law, it is no longer necessary for the net torque to be zero, i.e. $\int_\Omega d \bm{r} \  \bm{m} \times \bm{B}_{x c} = \bm 0$ need not be obeyed.

\subsection{Orbital magnetic moments}
\label{sec:OrbitalMagmom}


Having explored the connections to CDFT, we will now turn our attention to the orbital magnetic moments, which are a measurable link to the orbital current density, $\bm j_{\text{orb}}$. The orbital character of the magnetic moment can be teased out using x-ray magnetic circular dichroism (XMCD) \cite{PhysRevLett.69.2307, PhysRevLett.123.207201}, and sometimes in combination with other x-ray spectroscopy techniques, such as Resonant Inelastic X-ray Scattering (RIXS) \cite{PhysRevLett.123.207201}.

It is possible to express the orbital magnetic moment in terms of the orbital current \cite{sharmaComparisonExactexchangeCalculations2007}
\begin{align}
    \label{eq:orbmom_jorb}
    \bm m_i^{\text{orb}} &= \int_{\Omega_i} \bm r_i \times \bm j_{\text{orb}} (\bm r_i) \ d \bm r_i,
\end{align}
where $\Omega_i$ is a region surrounding the magnetic site $i$, such as a PAW sphere. These orbital moments are known to arise due to spin-orbit coupling. We will explore the effect of the source-free constraint on the ground state orbital moments in the \ce{UO2} test case explored in Section~\ref{sec:UO2-exploration}.

\section{Methods}

\subsection{Hubbard $U$, Hund $J$ and neglect of the exchange splitting parameter}

In our implementation of the source-free correction \cite{sf-patch-github}, we leave out the exchange-splitting scaling parameter $s$ included in the study of Sharma and coworkers. In other words, $s=1$ in our implementation. There are two primary reasons that we neglected this scaling parameters. The first reason for not including this parameter, is that while $s$ is not material-specific in theory, it requires fitting to the magnetic structure of a finite set of material systems. In addition to this ambiguity, it has been shown that Hubbard $U$ and Hund $J$ values have a significant effect on the magnitude of magnetic moments, as well as non-collinear magnetic structure \cite{krishnaCompleteDescriptionMagnetic2019}. For this reason, we utilize pre-computed, and/or custom-computed, $U$ and $J$ values for the magnetic systems that we explore in this study.

That is not to say that the inclusion of $s$ is unjustified -- its inclusion maintains the variational nature of the XC functional with respect to the magnetization field, $\bm m$ \cite{sharmaSourceFreeExchangeCorrelationMagnetic2018a}. However, we choose to not include this feature, due to its ambiguous nature, and the fact that it has nothing to do with the source-free constraint itself \footnote{After all, Sharma et al. describe the $s$ scaling as an ``additional modification'' to the XC functional \cite{sharmaSourceFreeExchangeCorrelationMagnetic2018a}. This was further confirmed in conversation with S. Sharma and J. K. Dewhurst.}. In our implementation and test cases, we did not find a need for $s \ne 1$.

\subsection{Source-free implementation}
\label{sec:SFimplement}

In Appendix~\ref{sec:NumericalDetails}, we provide the details of the source-free $\bm B_{xc}$ implementation in VASP \cite{hafner_vienna_1997}, which, generally speaking, should be consistent with both References~\onlinecite{sharmaSourceFreeExchangeCorrelationMagnetic2018a}~\&~\onlinecite{hawkhead2023firstprinciples}. As stated in \cref{eq:sf_correct}, we do not modify the \mbox{$\bm q = \bm 0$} component of ${\bm B'}_{xc}$. In other words, we set \mbox{$\widehat{\bm B'}_{xc} (\bm q = \bm 0) = \widehat{\bm B}_{xc} (\bm q = \bm 0)$}, which still satisfies the divergence-free constraint. Our choice appears to be consistent with the implementation of Ref.~\onlinecite{sharmaSourceFreeExchangeCorrelationMagnetic2018a} in the \texttt{Elk} source code \cite{elk-code, dewhurstDevelopmentElkLAPW}.

Our justification for leaving \mbox{$\overline{\bm B'}_{xc} = \overline{\bm B}_{xc}$} is the arbitrary choice of $\widehat{\bm B'}_{xc} (\bm q)$, which is apparent from the singularity with respect to $|\bm q|^2$ in \cref{eq:sf_correct}. However, one can entertain possible physically motivated constraints on $\overline{\bm B'}_{xc}$. For example, $\overline{\bm B'}_{xc}$ will enter into the net torque expression, \cref{eq:zero_torque}. Therefore, one may consider a least-squares constraint on $\overline{\bm B'}_{xc}$, based on the ZTT, as well as a possible term that preserves the net XC energy expression from SDFT. An in-depth and rigorous treatment of $\overline{\bm B'}_{xc}$, the real-space integral of ${\bm B'}_{xc}$, will be left to future studies.



\section{Results}

These magnetic structures were obtained from the Bilbao MAGNDATA database \cite{BilbaoServer}. All structures in this study contained less than sixteen atoms in their unit cell. The materials contained in this data set include metallic systems, i.e. \ce{Mn3Pt} \cite{Mn3Pt-exp}, as well as insulators, i.e. \ce{MnF2} \cite{MnF2-exp}. We take an in-depth focus of \ce{Mn3ZnN} \cite{Mn3ZnN-exp}, because its non-collinear ground state is accomodated by a relatively small unit cell comprised of only three symmetrically disctinct magnetic Mn ions.

Furthermore, for \ce{Mn3ZnN} (\cref{fig:Mn3ZnN_comparison}) it is clear that the curl of the magnetization, $\nabla \times \bm m$, varies by a much larger relative magnitude than that of \ce{Mn3Pt} (\cref{fig:Mn3Pt_comparison}). This can be ascertained by the circulation of spins around the [1~1~1] direction in \ce{Mn3ZnN}, which can be seen in \cref{fig:Mn3ZnN_fieldlines} and \cref{fig:Mn3ZnN_comparison} \cite{Mn3ZnN-exp}. We care about a large variance in the spin current for multiple reasons. The first reason has to do with the gauge symmetries of SDFT and CDFT, as explored in Ref.~\onlinecite{capelleSpinDensityFunctionalsCurrentDensity1997a}. After all, our goal in this study is to computationally explore these degeneracies. The second reason is that the curl of the magnetization enters into the expression for the net XC torque, as stated in \cref{eq:sf_torque_effective}. Hence, we surmise that the magnetic structure of \ce{Mn3ZnN} will test the limits of the ZTT, and to what degree it is, or isn't upheld at the point of self-consistency.

\subsection{Quantifying the effects of monopoles}
\label{sec:monopoles}

We found that for the \ce{Mn3ZnN} calculations, the PBE+U+J calculation yielded {$|\nabla \cdot \bm B_{xc} (\bm r)|_{\infty} \approx$ 104 eV/(\muB{} \angstrom$^4$)}, and for the source-free PBE+U+J counterpart, {$|\nabla \cdot \bm B_{xc} (\bm r)|_{\infty} $ $<$ $10^{-13}$ eV/(\muB{} \angstrom$^4$)}, where {$|\nabla \cdot \bm B_{xc} (\bm r)|_{\infty} = \max_{\bm r \in \mathbb R^3} \left| \nabla \cdot \bm B_{xc} (\bm r) \right|$}. This numerical comparison confirms that the source-free constraint is working as expected, and draws attention to the large density of magnetic monopoles that form in conventional non-collinear PBE+U+J. A visualization of the monopole density in the \ce{Mn3ZnN} test case is provided in \cref{fig:Mn3ZnN_monopoles}. We hone in on this material for reasons of computational cost and clear visualization. Namely, \ce{Mn3ZnN} exhibits a highly non-collinear spin texture, describable with a commensurate unit cell of only three magnetic atoms. Furthermore, this magnetic antiperovskite is known to exhibit exceptionally large and exotic magnetostriction effects \cite{hamadaPhaseInstabilityMagnetic2012b}, which is relevant to magnetostructural phase transitions and therefore magnetocalorics \cite{lawQuantitativeCriterionDetermining2018}.

\subsection{Local convergence test}
\label{sec:convergence-testset}

In order to compare the convergence of \Fsf{GGA-PBE} to conventional non-collinear GGA-PBE, we performed tests on the set of commensurate magnetic structures containing transition metal elements. To test the convergence of all structures, we apply a random perturbation from the experimental structure to all magnetic moments. The rotations of local moments are performed by an implementation of the Rodrigues' rotation formula within a cone angle of 45\textdegree.


To compare the performance of the source-free functional versus its SDFT counterpart, we provide a magnetic moment comparison (in \cref{fig:moment_comparison}) as well as a symmetry comparison (in \cref{fig:symmetry_comparison}). For the magnetic moment comparison, we plot the mean difference across magnetic moments computed using \FsfUJ{GGA-PBE} versus \FpUJ{GGA-PBE}. We observe that for all of these magnetic systems, the source-free functional predicts a moment that is slightly smaller than its source-free counterpart, bringing it in closer agreement with experiment, for most systems. 

In order to probe the comparison between non-collinear ground states themselves, we show a symmetry metric comparison between converged structures from \FsfUJ{GGA-PBE} and \FpUJ{GGA-PBE} in \cref{fig:symmetry_comparison}. For this study, we used the \texttt{findsym} program from the ISOTROPY software suite developed by Stokes et al. \cite{stokesISOTROPYSoftwareSuite}. We define this symmetry metric to be the minimum tolerance (in \muB{}), normalized by the absolute maximum magnitude of the individual magnetic moments within the structures, and scaled as a percentage value. Therefore, 0\% implies perfect agreement with experimentally resolved magnetic space-group, whereas 100\% implies poor agreement. We emphasize that for this study, we use mean $U$ and $J$ values taken from Ref.~\onlinecite{mooreHighthroughputDeterminationHubbard2022a}. In principle, one should calculate these $U$ and $J$ values for each structure, as $U$ and $J$ are very sensitive to local chemical environment, and therefore oxidation and spin states \cite{mooreHighthroughputDeterminationHubbard2022a}. However, to reduce immediate computational cost, we save this exploration for future studies. 

Figures \ref{fig:Mn3ZnN_comparison} and \ref{fig:MnF2_comparison} show improved agreement with experimentally measured magnetic ground states using \FsfUJ{PBE} compared to \FpUJ{PBE}, as well as better consistency of the output computed spin configuration. Improved performance of the SF functional  is also achieved in Figures \ref{fig:MnF2_comparison} and \ref{fig:Mn3As_comparison}. In the case of \ce{MnF2} and \ce{Mn3As}, we performed structural relaxations, with the spins initialized in the symmetric and/or experimental configuration. We explored the effect of structural optimization in these two materials, because without allowing for spin-lattice relaxation, a ground state with a strong ferromagnetic component was stabilized, as was the case for \ce{MnF2}, which is shown in Figure~\ref{fig:MnF2_comparison}.

\subsection{Convergence case study: \ce{YMnO3}}
\label{sec:convergence-YMnO3}

To examine whether a tighter energy convergence threshold improves the converged structure for GGA, we imposed a $10^{-8}$ eV energy cut off to \ce{YMnO3}, comparing the convergence behavior between GGA and it source-free counterpart. The convergence behavior of GGA compared to \Fsf{GGA} is shown in \cref{fig:convergence_sf_YMnO3}. In this plot, the absolute relative energy between subsequent self-consistency steps is plotted on a logarithmic scale. For the source-free functional, it is interesting to note that there appears to be a slight energy barrier that the algorithm climbs, only to descend to the symmetric experimentally reported ground state \cite{YMnO3-exp}, just before 200 self-consistency steps. 

To the reader, this may seem to be a long convergence time, however, we direct the attention to the conventional GGA counterpart. While convergence to 1~$\mu$eV is achieved rather rapidly, we see that the magnetic spins move very little in the convergence process. Furthermore, with the tighter energy cut off, the DFT calculation does not converge within the 600 electronic self-consistency step limit. At one point, the energy does dip below the tolerance energy threshold, but this is not simultaneously true for the band-structure convergence metric, which stays above the 0.01 $\mu$eV energy cut off. 

It is worth noting that to improve convergence for this particular calculation, we used a ``sigma" value smearing of 0.2 near to the Fermi level. This is standard practice, and a different smearing can be used by continuing the DFT calculation with a smaller ``sigma" value, and a different smearing method. Additionally, we used Gaussian smearing, which is known to be more robust across different material chemistries, such as between insulators and conductors, according to VASP documentation. 

\subsection{Augmented orbital moments in \ce{UO2}}
\label{sec:UO2-exploration}

Thus far, we have focused on the spin component of the magnetic moment. However, in reality, there is an orbital component to the magnetic moment that supplements the spin contribution \cite{restaElectricalPolarizationOrbital2010, changBerryCurvatureOrbital2008}, especially in materials with strong spin-orbit coupling (SOC). In many $3d$ transition metal oxides, it can be theoretically and experimentally shown that the orbital moment is ``quenched'' \cite{AshcroftMermin1976, streltsovOrbitalPhysicsTransition2017}, in which case it is fair to neglect the orbital contributions. However, in $f$-block species, SOC can become much more prevalent. Additionally, we have explored the connections between the orbital moment and the paramagnetic current density, $\bm j_p$ in \ref{sec:OrbitalMagmom}. Therefore, it would behoove us to explore how the source-free functional affects the orbital magnetic moments.

\ce{UO2} has become the archetype of correlated oxides with strong spin-orbit coupling, which gives rise to the strongly non-collinear ground state of \ce{UO2} \cite{UO2-exp}. Therefore, we apply the source-free functional towards \ce{UO2}, which we found to exhibit exceptionally large orbital magnetic moments. Specifically, we report the computed spin and orbital magnetic moments in \cref{tab:magmomUO2} using \FpUJ{GGA} and \FsfUJ{GGA}.
Due to the connection between orbital moments and $\bm j_p$ in \cref{eq:orbmom_jorb}, we plot $\bm j_p (\bm r)$ in \cref{fig:UO2_jparamag}. We see that in this visualization, there is strong circulation of $\bm j_p$ around the uranium atoms, shown in grey. It is interesting to note that we achieve this improved agreement with experiment, even without the direct coupling to $\bm A_{xc}$ in the form of \cref{eq:Ep}.





\subsection{Local and global torques arising from the source-free constraint}
\label{sec:SF-torque}

In \cref{fig:tauxc_sf} we observe that it is indeed not the case that $\bm \tau_{xc}' = \bm 0 \ \forall \ \bm r \in \Omega$. However, for the systems we tested, we found the net (integrated) torque, as stated in \cref{eq:sf_torque_effective}, to be orders of magnitude smaller than the largest local torque, i.e. \mbox{$\max_{\bm r \in \Omega} \left\{ \left| \bm \tau_{xc}' (\bm r) \right| \right\}$}.  We found that for our \ce{Mn3ZnN} test case, the self-consistent $\bm \tau_{xc}$ obeyed the following inequality $\int_\Omega \bm \tau_{xc} \ d \bm r <$ $5 \times 10^{-5}$ eV, even though $\max_\Omega \left\{ \bm \tau_{xc} \right\} \approx$ 5 eV. 
The energy convergence tolerance for this calculation was $1 \times 10^{-6}$ eV. Therefore, the net torque could simply be an artifact of numerical convergence.

Despite the ``small" net torque relative to the energy convergence tolerance, it is nontrivial to determine whether the right-hand side of \cref{eq:sf_torque_effective} will be ``small enough" in general. Additionally, further investigation should examine the effects of enforcing the ZTT at every self-consistency step. In Appendix \ref{sec:ZTTfix}, 
we propose possible approaches to ensure that the ZTT is upheld at every self-consistency step. The method in Appendix~\ref{sec:ZTTfix} should be the most general and robust, with added computational cost. We have not implemented these ZTT corrections at this time. However, we imagine that a careful adherence to the ZTT will be important for the calculation of magnetocrystalline anisotropy energy (MAE), which is on the order of $\mu$eV, and therefore this additional physical constraint should be considered.

\subsection{Importance of the constant, $\bm q = \bm 0$, component of ${\bm B'}_{xc}$}
\label{sec:q0-discussion}

In \cref{fig:q0_comparison}, it is clear that when applying the \mbox{$\overline{\bm B'}_{xc} = \bm 0$} constraint for \ce{MnF2}, the SF functional converges to the correct collinear AFM ground state \cite{MnF2-exp}. For \mbox{$\overline{\bm B'}_{xc} = \overline{\bm B}_{xc}$}, the canted FM configuration is erroneously stabilized, which is remedied by structural relaxations, as shown in \cref{fig:MnF2_comparison}. On the other hand, for \ce{MnPtGa}, if one applies the additional \mbox{$\overline{\bm B'}_{xc} = \bm 0$} constraint, we see that the ground state is a structure with much stronger AFM character. This differs significantly from the canted ferromagnetic configuration obtained using \mbox{$\overline{\bm B'}_{xc} = \overline{\bm B}_{xc}$}. This canted FM spin configuration is much closer to the magnetic ordering resolved using neutron diffraction \cite{MnPtGa-exp}.

To generalize this behavior, \mbox{$\overline{\bm B'}_{xc} = \overline{\bm B}_{xc}$} erroneously stabilizes ferromagnetism in antiferromagnetics, and \mbox{$\overline{\bm B'}_{xc} = \bm 0$} stabilizes antiferromagnetism in ferromagnets, which is equally problematic. Due to the obvious importance of carefully considering $\overline{\bm B'}_{xc}$, we plan to address this in the future. For all other test cases in this study, we simply set \mbox{$\overline{\bm B'}_{xc} = \overline{\bm B}_{xc}$}, in order to maintain consistency with Ref.'s~\onlinecite{sharmaSourceFreeExchangeCorrelationMagnetic2018a}~\&~\onlinecite{hawkhead2023firstprinciples}. As we mention in the introduction, it would be possible to solve for $\overline{\bm B'}_{xc}$ that mitigates the net XC torque, while preserving the XC spin-splitting energy from SDFT, in accordance with \cref{eq:sdft_xc}. We plan to explore this rigorously in a follow-up study.

\section{Conclusions}

Significantly improved convergence to the non-collinear magnetic structure has been achieved with the application of the source-free constraint to $\bm B_{xc}$ \cite{sharmaSourceFreeExchangeCorrelationMagnetic2018a} to the PAW DFT formulation implemented in VASP. With the use of parallel three-dimensional FFTs as the basis for the fast Poisson solver, it is possible to apply this constraint with little additional computational cost, and no reduction of the parallel scalability of the DFT code. While we have focused on \FsfUJ{GGA-PBE} in this study, this constraint is generalizable to other SDFT functionals in non-collinear implementations, such as meta-GGA. 

Subsequent studies will combine the improved local convergence of \Fsf{DFT} with global optimization algorithms in order to achieve a unified and robust determination of non-collinear ground states without any prior experimental knowledge.

We hope that the augmented magnetoelectric coupling predicted using the source-free functional lays the groundwork for future studies to perform an in-depth investigation of the magnetoelectric figures of merit calculated using this modified functional, especially considering that the Berry phase provides a convenient and rigorous theoretical link between the modern theory of polarization and the modern theory of magnetization \cite{changBerryCurvatureOrbital2008, restaElectricalPolarizationOrbital2010}, which can be explicitly expressed in terms of the spin current \cite{PhysRevLett.95.057205}. Additionally, the unified theory provides a robust and solid theoretical description of the orbital magnetic moment \cite{restaElectricalPolarizationOrbital2010}, which extends the semiclassical theory touched on in Section~\ref{sec:OrbitalMagmom}. Future studies are encouraged to further examine the role of spin and orbital currents as they pertain to magnetoelectric coupling.

\section{Acknowledgements}

We would like to thank Dr. S. Sharma and Dr. J. K. Dewhurst for taking the time to discuss the source-free implementation over video conference call. We appreciate their support and encouragement. G.M. acknowledges support from the Department of Energy Computational Science Graduate Fellowship (DOE CSGF) under grant DE-SC0020347. Computations in this paper were performed using resources of the National Energy Research Scientific Computing Center (NERSC), a U.S. Department of Energy Office of Science User Facility operated under contract no. DE-AC02-05CH11231. Expertise in high-throughput calculations, data and software infrastructure was supported by the U.S. Department of Energy, Office of Science, Office of Basic Energy Sciences, Materials Sciences and Engineering Division under Contract DE-AC02-05CH11231: Materials Project program KC23MP.


\appendix

\subsection{Numerical details of source-free constraint}
\label{sec:NumericalDetails}

Because plane-wave DFT is defined on periodic boundary conditions, we can start with the definition of the inverse discrete Fourier transform, because the density fields lie in regular three-dimensional grids,
\begin{align}
    \phi_{\bm n} &= \frac{1}{M} \sum_{\bm k} \widehat{\phi}_{\bm k} \exp \left( i 2\pi \bm k^T D^{-1} \bm n \right) \nonumber \\
    & = \mathcal{F}^{-1} \widehat{\phi} \\
    \text{where } D &= 
    \begin{pmatrix}
        N_x & 0 & 0 \\
        0 & N_y & 0 \\
        0 & 0 & N_z \\
    \end{pmatrix},
    \text{ and } M = N_x N_y N_z \nonumber
\end{align}
where $n_i, k_i \in \left[0, 1, ..., N_i \right]$, and $N_x, N_y, N_z$ are the dimensions of the 3D grid. If we define the real-space position vector as $\bm r = L D^{-1} \bm n$, where $L$ has columns as lattice vectors $L = [\bm a \ \bm b \ \bm c]$. Therefore, $\bm k^T D^{-1} \bm n = \bm k^T D^{-1} D L^{-1} \bm r = \bm k^T L^{-1} \bm r$. From here, we can apply the convenient ``scaling" property describing how differential operators commute with the inverse discrete Fourier transform, where $l=x,y,z$ is the dimension of the partial derivative, 
\begin{align}
    \frac{\partial}{\partial r_l} \phi (\bm r) &\approx i 2\pi \cdot \mathcal{F}^{-1}  \left( q_l \widehat{\phi} \right) \\ 
    \text{where } q_l &= \sum_j L^{-1}_{jl} k_j, \quad \bm q = L^{-T} \bm k \nonumber
\end{align}
To obtain the (discrete) spectral approximation of the divergence of the $\bm{B}$ field:
\begin{align}
    \nabla \cdot \bm B_{xc} &= \frac{\partial B_x}{\partial x} + \frac{\partial B_y}{\partial y} + \frac{\partial B_z}{\partial z} \nonumber \\ 
    &\approx i 2\pi \cdot \mathcal{F}^{-1}  \left(
    q_x \widehat{B}_x +
    q_y \widehat{B}_y +
    q_z \widehat{B}_z
    \right) \nonumber \\
    &\approx i 2\pi \cdot \mathcal{F}^{-1}  \left( \bm q^T \widehat{\bm B}_{xc}\right) \label{eq:dft_div}
\end{align}
In order to solve the Poisson equation, as posed in \cref{eq:poisson}, we apply a similar reasoning as before, and see that 
\begin{align}
    \nabla^2 \phi &= \frac{\partial^2 \phi}{\partial x^2} + \frac{\partial^2 \phi}{\partial y^2} + \frac{\partial^2 \phi}{\partial z^2} \nonumber \\
    &\approx - (2\pi)^2 \cdot \mathcal{F}^{-1}  \left(
    (q_x^2 + q_y^2 + q_z^2) \widehat{\phi} \right) \nonumber \\
    &\approx - (2\pi)^2 \cdot \mathcal{F}^{-1}  \left(
    |\bm q|^2 \widehat{\phi} \right) \label{eq:dft_laplacian}
\end{align}
Therefore, combining \cref{eq:dft_div,eq:dft_laplacian,eq:poisson}, we find that the discrete Fourier transform of $\phi$ can be expressed as
\begin{align}
    \widehat{\phi} (\bm q) &=
    \begin{cases}
    \frac{i}{2\pi} \frac{\bm q \cdot \widehat{\bm B}_{xc} (\bm q)}{|\bm q|^2}, & | \bm q|  \ne 0 \\
    0, & | \bm q| = 0
    \end{cases}
\end{align}
We are interested in $\nabla \phi$. Which we can approximate as
\begin{align}
    - \nabla \phi &\approx \mathcal{F}^{-1} \left( \frac{\bm q^T \widehat{\bm B}_{xc} (\bm q)}{|\bm q|^2} \bm q \right)
\end{align}
Therefore, the source-free correction is achieved by performing the following
\begin{align}
    \widehat{\bm B'}_{xc} (\bm q) = \widehat{\bm B}_{xc} (\bm q) - 
    \begin{cases}
    \frac{\bm q \cdot \widehat{\bm B}_{xc} (\bm q)}{|\bm q|^2} \bm q, & | \bm q|  \ne 0 \\
    0, & | \bm q| = 0
    \end{cases}
    \label{eq:sf_correct}
\end{align}
and as a result, $\nabla \cdot {\bm B_{xc}}' = 0$ is obtained in the discrete numerical sense.

\section{A general approach to satisfy the zero torque theorem}
\label{sec:ZTTfix}

Our goal is to solve for an auxillary field $\bm A'$ such that the zero torque condition is obeyed
\begin{align}
\int_{\Omega} d \bm{r} \ \bm{m} \times \bm{B} = \int_{\Omega} d \bm{r} \ \bm{m} \times(\nabla \phi+\nabla \times {\bm A'}) &= \bm0.
\end{align}
The trivial solution, $\nabla \times {\bm A'} = - \nabla \phi$, can only hold if  $\nabla \phi = \bm 0$. Therefore, we recast the problem as such
\begin{align}
\int_{\Omega} d \bm{r} \ \bm{m} \times(\nabla \times {\bm A'}) &=-\int_{\Omega} 
d \bm{r} \ \bm{m}\times \nabla \phi = - \bm{\tau}'.
\label{eq:ZTT_problem}
\end{align}
By expanding the triple product, we see that
\begin{align}
\bm{m} \times(\nabla \times {\bm A'}) 
&= \left[m_j \partial_i {A'}_j -m_j \partial_j {A'}_i \right] \hat{\bm e}_i  \nonumber \\
&= \nabla_{{\bm A'}}(\bm{m} \cdot {\bm A'})-(\bm{m} \cdot \nabla) {\bm A'}.
\end{align}
At this step, we can apply the product rule, $u \partial_x v=\partial_x(u v)-v \partial_x u$, to show that integrals of the form \mbox{$\int d \bm{r} \ \partial_x(u v)$} vanish under periodic boundary conditions (PBCs) by Equation~\refSI{eq:grad_int_periodic}{S12}.
Having leveraged this convenient property of the PBCs, it is possible to transfer the derivatives from $\bm A'$ to $\bm m$ such that we can restate \cref{eq:ZTT_problem} as
\begin{align}
&\int d \bm{r} \ \bm{m} \times(\nabla \times {\bm A'}) \nonumber \\
&\quad = \int d \bm{r} \ \left[{A'}_i \partial_j m_j-{A'}_j \partial_i m_j\right] \hat{\bm e}_i 
= -\bm{\tau}'
\label{eq:ZTT_problem_pbcs}
\end{align}
Now, we will introduce the following discretized approximation for the $L^2$ inner product on regular grids
\begin{align}
\langle u,v \rangle &= \int_\Omega d \bm{r} \ u(\bm{r}) v(\bm{r})  \approx \frac{1}{\Delta V} \bm{u}^T\bm{v},
\end{align}
where $\Delta V = \Delta x \Delta y \Delta z$. We introduce this shorthand for the purposes of this study, however, it is just as applicable to continuous functions. With this definition, we can recast the problem in \cref{eq:ZTT_problem_pbcs} as a matrix-vector system
\begin{align}
M\bm{a} = -\bm{\tau}',
\end{align}
where $\bm a = \left[ {\bm A'}_x^T \ {\bm A'}_y^T \ {\bm A'}_z^T\right]^T$, and the matrix $M$ is defined as 
\begin{align}
M &= \frac{1}{\Delta V}
\begin{bmatrix}
{\bm m}_{y, y}^T+{\bm m}_{z, z}^T & -{\bm m}_{y, x}^T & -{\bm m}_{z, x}^T \\
-{\bm m}_{x, y}^T & {\bm m}_{z,z}^T+{\bm m}_{x, x}^T & -{\bm m}_{z, y}^T \\
-{\bm m}_{x, z}^T & -{\bm m}_{y, z}^T & {\bm m}_{x, x}^T+{\bm m}_{y, y}^T
\end{bmatrix},
\end{align}
in which case $\bm m_{i,j} = \pderiv{\bm m_i}{x_j}$. Because the system of equations is underdetermined (i.e. $M$ is ``short and fat"), we can can solve for $\bm a$ using a least-squares approach, 
\begin{align}
\bm{a} &= - M^{+} \bm{\tau}',
\end{align}
which solves for the solution $\bm a$ with minimal $L^2$ norm, 
\begin{align}
||\bm a||_2 = \left\{ \sum_{i=x,y,z} \int_\Omega d \bm r \ |\bm A'_i (\bm r)|^2 \right\}^{1/2}.
\end{align}
$M^{+}$ is the correspanding Moore-Penrose right-hand pseudoinverse $M^{+}=M^T\left(M M^T\right)^{-1}$ such that $M M^{+}=I$, as long as the rows of $M$ are linearly independent. 
We note that $M M^{T}$ is diagonally dominant, because it contains $L^2$ norms (which are guaranteed to be positive) of the spatial partial derivatives of $\bm m$ along the diagonal. Therefore, $M M^{T}$ should be invertible, so long as the magnetization varies in all spatial directions over the domain, which it should for spin-polarized systems.

In conclusion, by applying both source-free and ZTT corrections to the exchange-correlation magnetic field, $\bm B_{xc}$,
\begin{align}
    \bm B_{xc} &\mapsto {\bm B'}_{xc} \nonumber \\
    {\bm B'}_{xc} &= \bm B_{xc} + \nabla \phi + \nabla \times {\bm A'}
    \label{eq:Bxc_sf_ztt}
\end{align}
we can simultaneously satisfy
\begin{enumerate}[label=\Roman*.]
    \item The source free constraint: 
    \begin{align*}
        \nabla \cdot {\bm B'}_{xc} = 0
    \end{align*}
    \item The zero torque theorem: 
    \begin{align*}
        \int_\Omega d \bm r \ \bm m \times {\bm B'}_{xc} = \bm 0.
    \end{align*}
\end{enumerate}
However, we should stress that the second ZTT constraint is not implemented in the code at present. In other words, we set $\bm A' (\bm r) = \bm 0$ in the context of this study. 

\section{Choice of $\bm{A}_{xc}$ gauge}
\label{sec:GaugeAxc}

It is worth noting that two gauge choices of $\bm A_{xc}$ have been presented in the literature. Within the original works of Vignale and Rasolt \cite{vignaleMagneticFieldsDensity1990, vignaleDensityfunctionalTheoryStrong1987, vignaleCurrentSpindensityfunctionalTheory1988}, the $\nabla \cdot \left( \rho \bm{A}_{xc} \right) = 0$ arises naturally. However, by comparison, in \cite{capelleSpinDensityFunctionalsCurrentDensity1997a}, the Coulomb gauge $\nabla \cdot \bm{A}_{xc} = 0$ is implied by the Helmholtz decomposition.

Under the $\nabla \cdot \bm{A}_{xc} = 0$ gauge, it is possible to solve for $\bm{A}_{xc}$ using the following Poisson equation, 
\begin{align}
    \nabla \times \left(\nabla \times \bm A_{xc}\right) &= \nabla \times \bm B_{xc} \nonumber \\
    -\nabla^2 \bm A_{xc} + \nabla \left(\nabla \cdot \bm A_{xc} \right) &= \nabla \times \bm B_{xc} \nonumber \\
    \nabla^2 \bm A_{xc} &= - \nabla \times \bm B_{xc} \label{eq:poisson_Axc}
\end{align}
In order to solve for $\bm{A}_{xc}$ subject to $\nabla \cdot \left( \rho \bm{A}_{xc} \right) = 0$, we can employ another Helmholtz decomposition
\begin{align}
    \bm A_{xc} &= \bm A_{xc}' + \nabla \xi \nonumber \\
\end{align}
where $\nabla \cdot \bm A_{xc}' = 0$ and is solved using \cref{eq:poisson_Axc}, and $\xi$ is determined from the following elliptical equation
\begin{align}
    \nabla \cdot \left( \rho \nabla \xi \right) &= - \nabla \cdot \left( \rho \bm A_{xc}' \right). \label{eq:ell_rhoAxc}
\end{align}
However, the spatial dependence of $\rho$ in \cref{eq:ell_rhoAxc} makes the equation more difficult to solve using a single spectral solve step. Instead, an iterative spectral solver could be used to solve this elliptical equation, using the subtractive solver presented in Ref.~\onlinecite{bravermanFastSpectralSubtractional2004}, for example.





\section{Considerations for $\bm j_p \cdot \bm A_{xc}$ energy contributions under periodic boundary conditions}
\label{sec:PeriodicBCjpAxc}

We will use the following result, \cref{eq:int_grad_vec}, to draw a few conclusions.
\begin{align}
    &\int_{\Omega} \nabla \alpha (\bm r) \cdot \bm L(\bm r) \ d\bm r \nonumber \\
    &= \int_{\Omega} \left\{ \nabla \cdot \left[ \alpha(\bm r) \bm L(\bm r) \right] - \alpha(\bm r) \left[ \nabla \cdot \bm L(\bm r) \right] \right\} \ d\bm r \nonumber \\
    &= \oint_{\partial \Omega} \left[ \alpha(\bm r) \bm L(\bm r) \right] \cdot d \bm S - \int_{\Omega} \alpha(\bm r) \left[ \nabla \cdot \bm L(\bm r) \right] \ d\bm r \label{eq:int_grad_vec}
\end{align}
If $\Omega$ obeys periodic boundary conditions, then $\oint_{\partial \Omega} \left[ \alpha(\bm r) \bm L(\bm r) \right] \cdot d \bm S = 0$.


We can start by considering the Coulomb gauge \mbox{$\nabla \cdot \bm{A}_{xc} = 0$}. If this gauge is chosen, charge conservation should still be obeyed through the following
\begin{align}
    \label{eq:jp_cons}
    \nabla \cdot \left( \bm{j}_p + \rho \bm{A}_{xc}\right) &= 0
\end{align}
By the Helmholtz identity, we can decompose $\bm{j}_p$ into the following
\begin{align}
    \bm{j}_p = \nabla \eta + \nabla \times \bm \Gamma. \label{eq:jp_helm}
\end{align}
Therefore, we may rewrite \cref{eq:jp_cons} as
\begin{align}
    \nabla^2 \eta = - \nabla \cdot \left( \rho \bm{A}_{xc} \right)
    \label{eq:poisson_jp}
\end{align}
Here, it is interesting to note that while $\eta$ can satisfy charge conservation, by \cref{eq:int_grad_vec}, under periodic boundary conditions and the $\nabla \cdot \bm{A}_{xc} = 0$ gauge, the following is true
\begin{align}
    \int_\Omega \nabla \eta \cdot \bm{A}_{xc} \ d \bm r = 0.
\end{align}
Therefore, $\eta$ will not enter into the energy term, \cref{eq:Ep}, allowing us to conclude that
\begin{align}
    \left. E_p \right|_{\nabla \cdot \bm{A}_{xc} = 0} &= \int_\Omega \left(\nabla \times \bm \Gamma\right) \cdot \bm{A}_{xc} \ d \bm r \nonumber \\
    & \qquad \forall \ 
    \left\{ 
    \bm{j}_p, \ \left. \bm{A}_{xc} \right|_{\nabla \cdot \bm{A}_{xc} = 0}
    \right\}
    \ . \label{eq:Ep_Axc_0}
\end{align}
\cref{eq:Ep_Axc_0} states that while the curl-free projection of $\bm j_p$, $\nabla \eta$, maintains local charge conservation, only the divergence-free projection, $\nabla \times \bm \Gamma$, enters into the expression for $E_p$.

\section{Considerations for $\bm j_{p,\lambda}$}
\label{sec:DecompProbCurrent}

Starting from a Helmholtz decomposition of the probability current density
\begin{align}
\bm{j}_{p,\lambda} &= \nabla \eta_\lambda + \nabla \times \bm \Gamma_\lambda
\end{align}
We see that the following conservation law
\begin{align}
&\nabla \cdot\left[\bm{j}_{p, \lambda}+m_\lambda \bm{A}_{xc}\right]=\left(\bm{m} \times \bm{B}_{x c}\right)_\lambda 
\end{align}
only places a constraint on the curl-free contribution to $\bm{j}_{p, \lambda}$, similarly to \cref{eq:poisson_jp},
\begin{align}
\nabla^2 \eta_\lambda &= -\nabla \cdot \left[m_\lambda \bm{A}_{xc}\right] + \left(\bm{m} \times \bm{B}_{x c}\right)_\lambda. \label{eq:poisson_grad_jplam}
\end{align}
Therefore, only the divergence-free component, $\nabla \times \bm \Gamma_\lambda$, is independent from this conservation law, and will contribute to the energy in an analogous form of \cref{eq:Ep_Axc_0}.





\begin{widetext}

\begin{figure}[h]
    \centering
    \subfloat[\centering \Fsf{PBE}]{{
    \adjincludegraphics[width=0.48\linewidth,trim={{.05\width} 0 {.05\width} 0},clip]{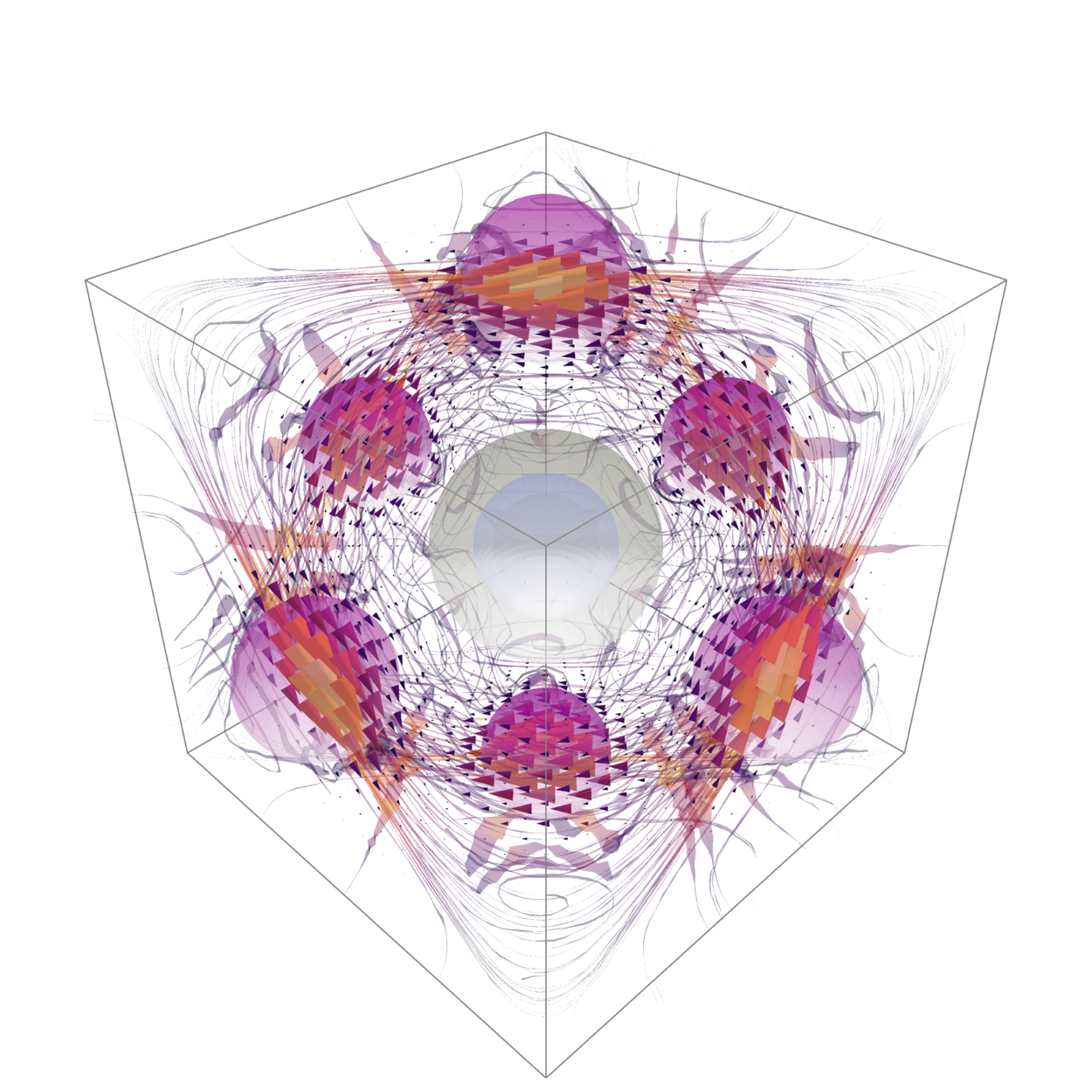}
    }}
    \subfloat[\centering PBE]{{
    \adjincludegraphics[width=0.48\linewidth,trim={{.05\width} 0 {.05\width} 0},clip]{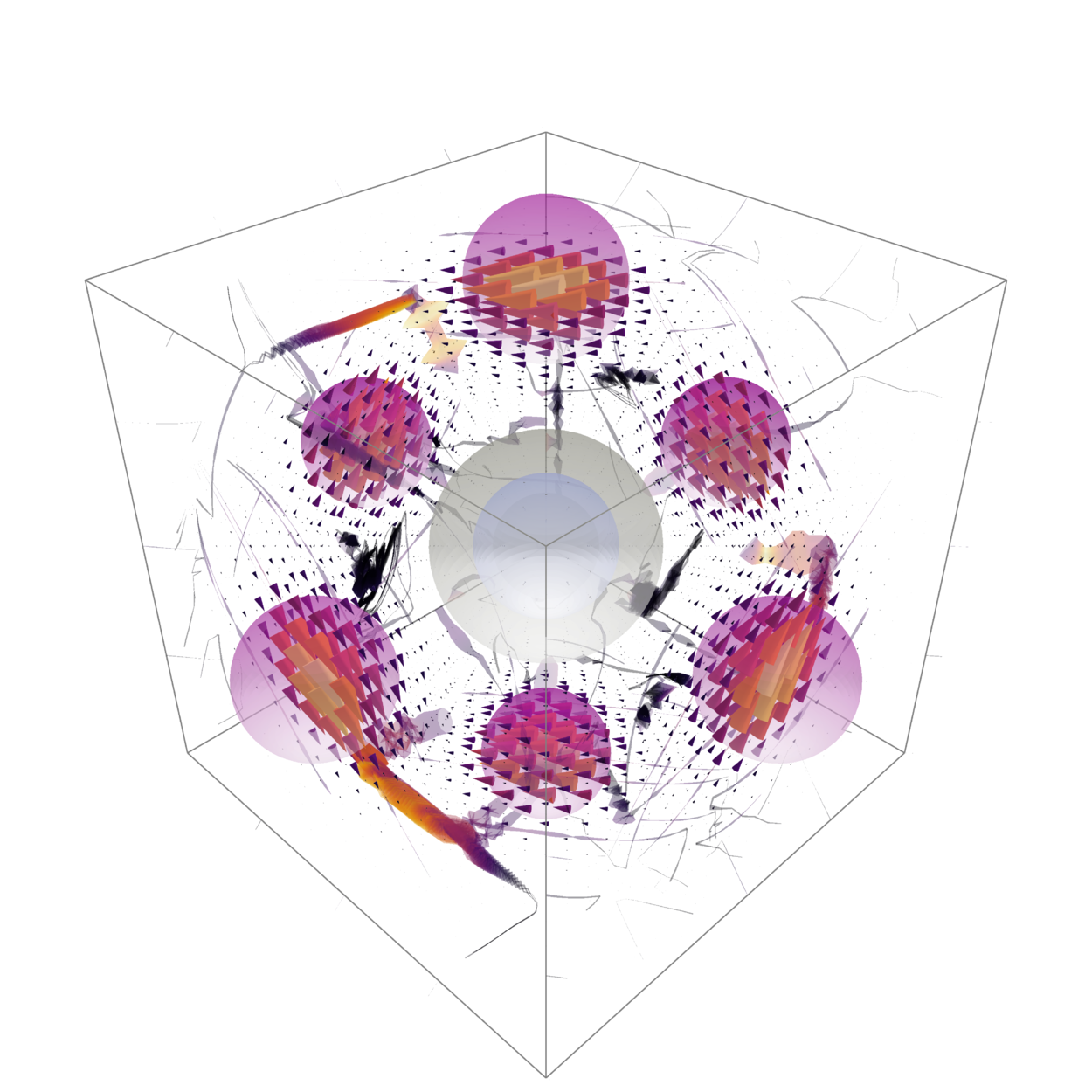}
    }}
    \caption{\Fsf{PBE} versus PBE comparison plot of ground state magnetization density $\bm m (\bm r)$ (vector field) and $\bm B_{xc} (\bm r)$ (streamlines) viewed along [111] direction for \ce{Mn3ZnN}}
     \label{fig:Mn3ZnN_fieldlines}
\end{figure}

\begin{figure}[h]
    \centering
    \includegraphics[width=0.58\linewidth]{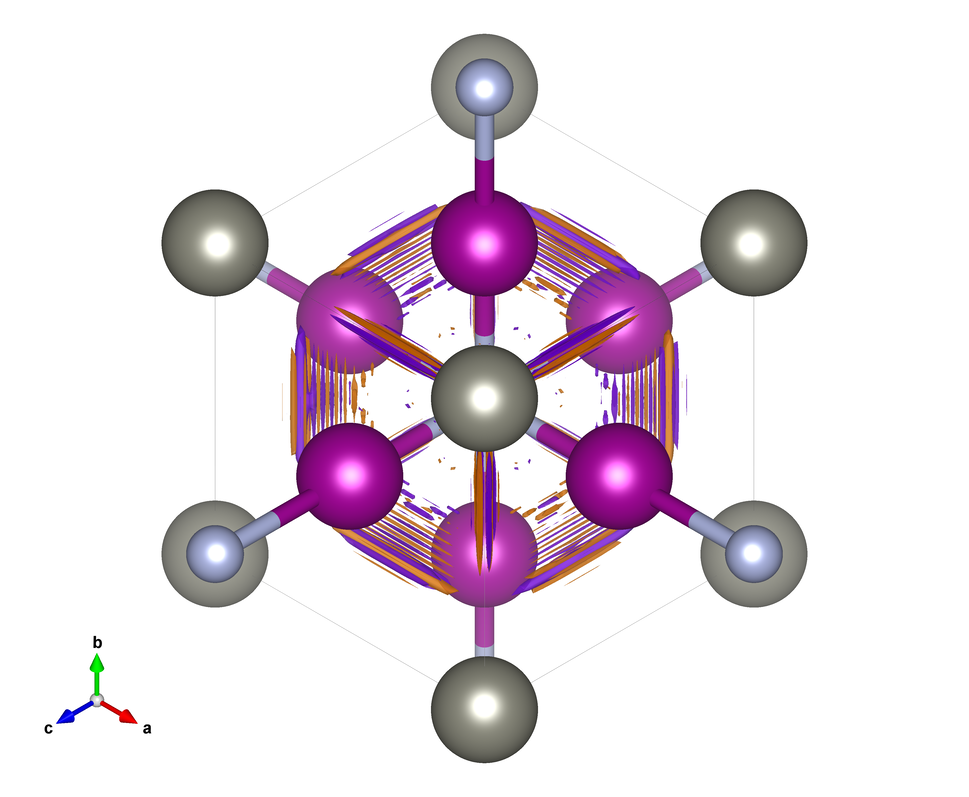} 
    \caption{Isosurface VESTA plot of PBE ground state $\nabla \cdot \bm B_{xc} (\bm r)$ viewed along [111] direction for \ce{Mn3ZnN}. Positive isosurface is indicated in violet, and negative counterpart in orange, at a fixed isosurface level.
    Manganense, zinc, and nitrogen atoms are color-coded in magenta, grey, and light blue, respectively.}
    \label{fig:Mn3ZnN_monopoles}
\end{figure}

\begin{figure}[h]
    \centering
    \subfloat[\centering \Fsf{PBE}]{{
    \label{fig:tauxc_sf}
    \adjincludegraphics[width=0.48\linewidth,trim={{.0\width} 0 {.0\width} 0},clip]{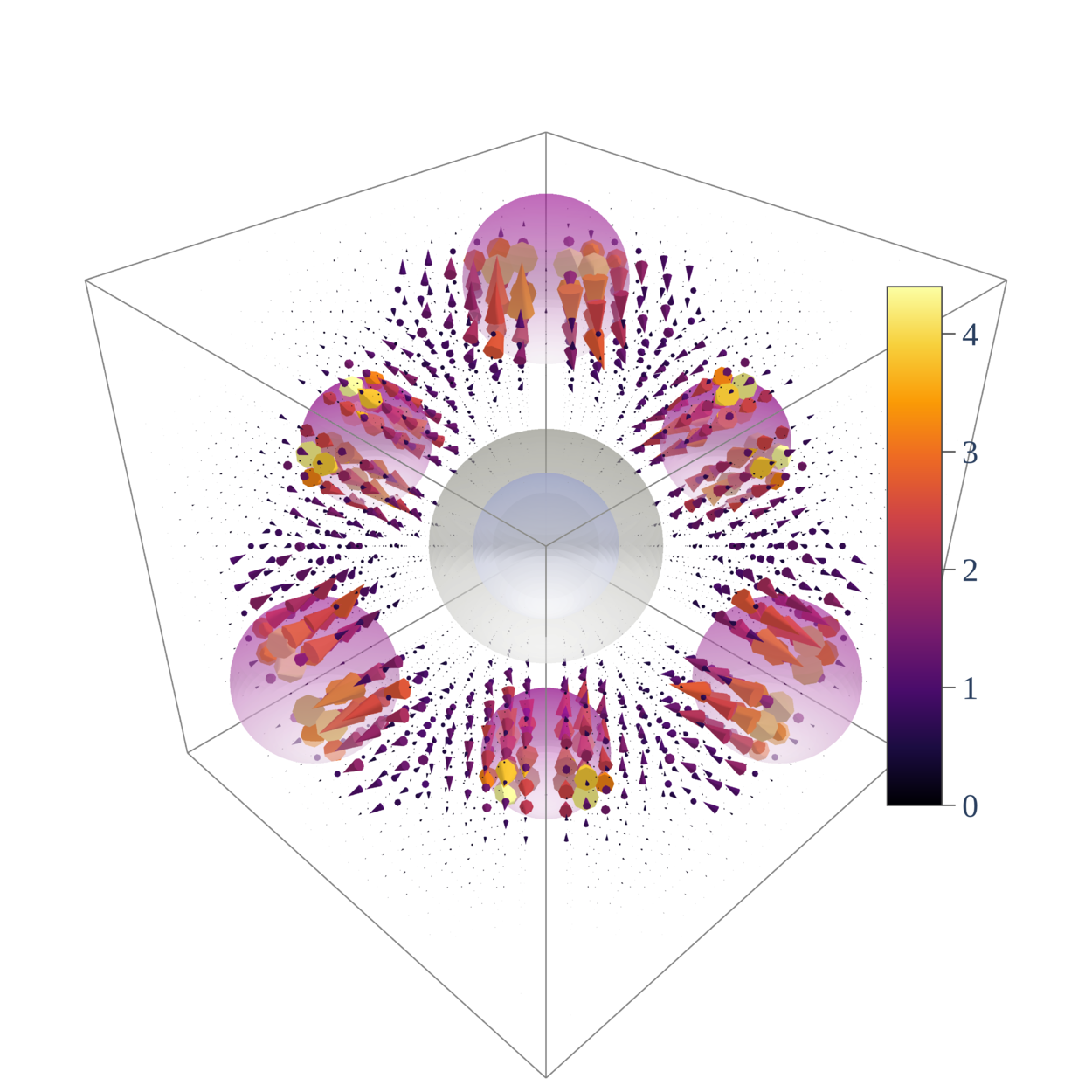}
    }}
    \subfloat[\centering PBE]{{
    \label{fig:tauxc_notsf}
    \adjincludegraphics[width=0.48\linewidth,trim={{.0\width} 0 {.0\width} 0},clip]{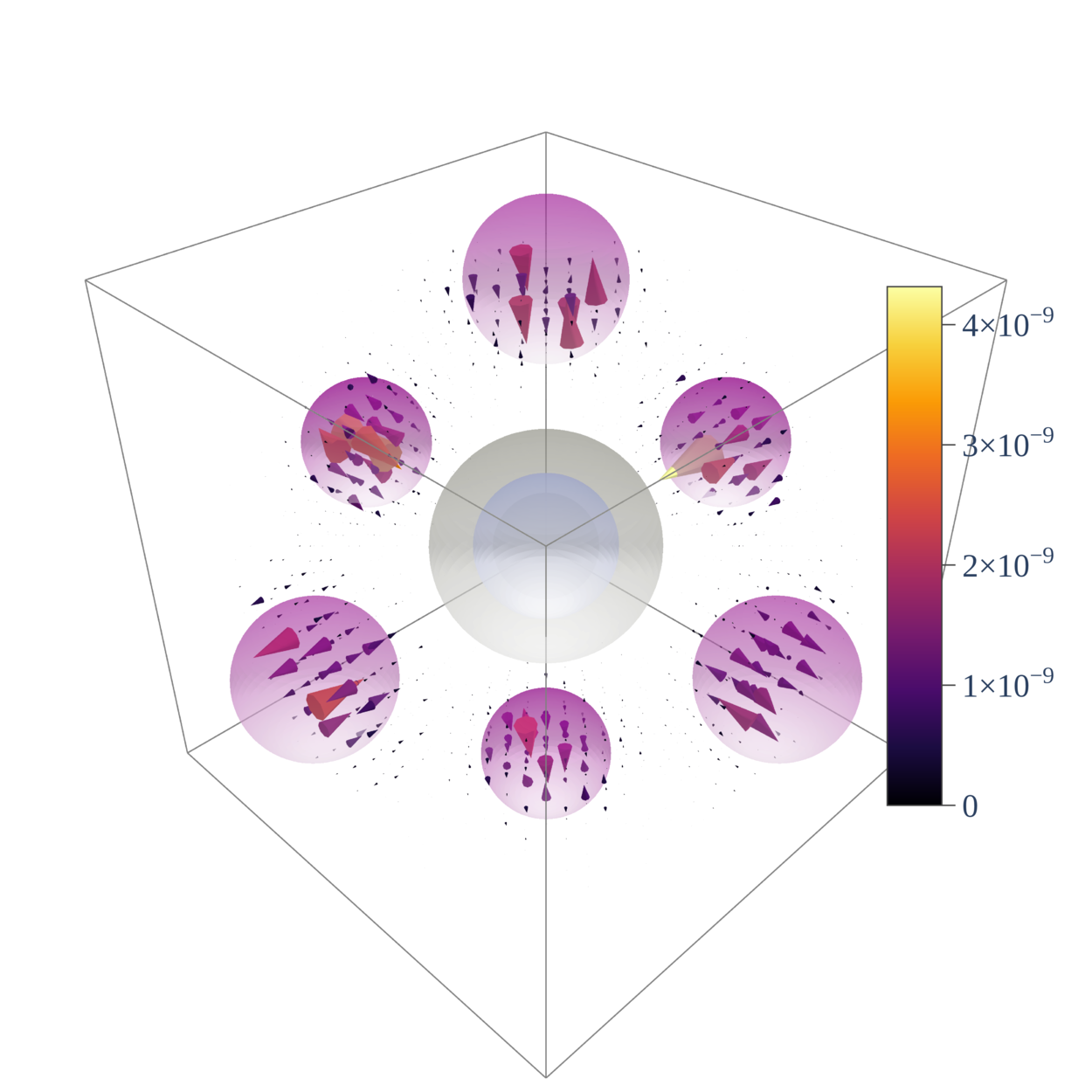}
    }}
    \caption{\Fsf{PBE} versus PBE comparison plot of ground state magnetic torque vector field $\bm \tau_{xc} = \bm m (\bm r) \times \bm B_{xc} (\bm r)$ (vector field) viewed along [111] direction for \ce{Mn3ZnN}. The units are in (\muB{} eV)/(\muB{} \angstrom$^3$).}
\end{figure}

\begin{figure}[h]
    \centering
    \includegraphics[width=0.98\linewidth]{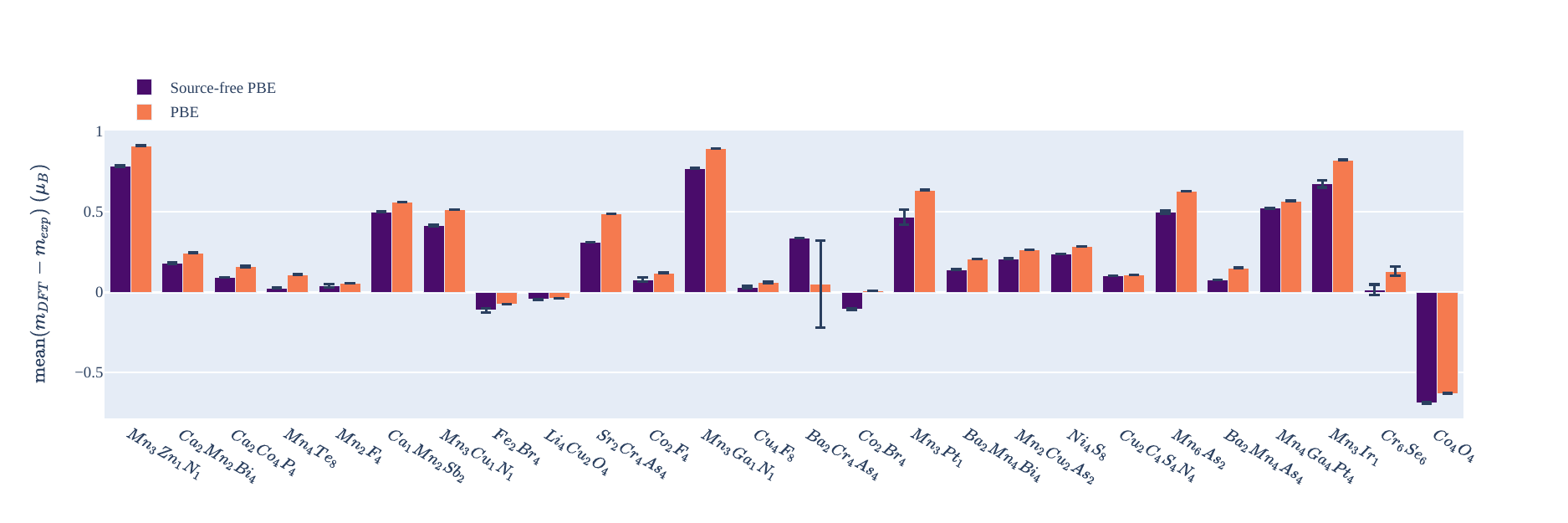}
    \caption{Comparison plot of magnetic moment differences with experiment for \FpUJ{PBE} and \FsfUJ{PBE} for a selected number of experimental structures from the MAGNDATA Bilbao Crystallographic Server \cite{BilbaoServer}. The $y$-axis is the difference metric $(1/N) \sum_i (m^i_{\text{DFT}} - m^i_{\text{exp}} )$, where $m^i_{\text{DFT}}$ and $m^i_{\text{exp}}$ are the magnetic moment magnitudes for site ``$i$" from the DFT-computed and experimental magnetic structures, respectfully.}
    \label{fig:moment_comparison}
\end{figure}

\begin{figure}[h]
    \centering
    \includegraphics[width=0.98\linewidth]{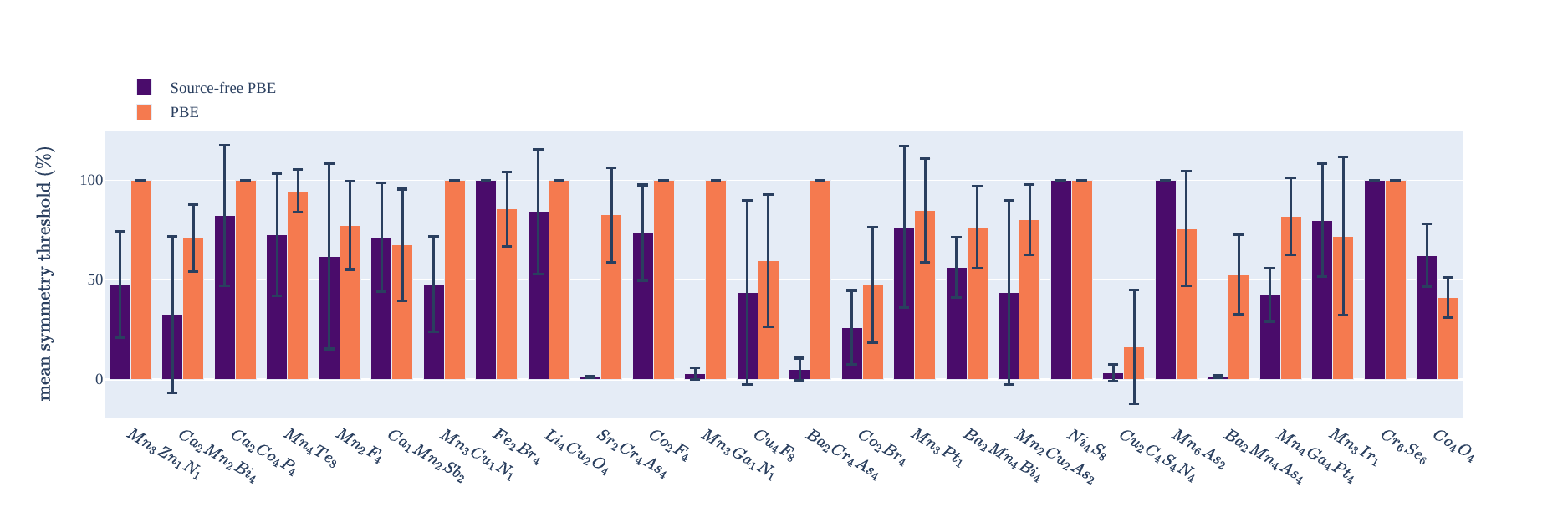}
    \caption{Comparison plot of minimum symmetry tolerance agreement with experimentally resolved magnetic space-group \FpUJ{PBE} and \FsfUJ{PBE} for a selected number of experimental structures from the MAGNDATA Bilbao Crystallographic Server \cite{BilbaoServer}. The $y$-axis is the minimum symmetry tolerance (in \muB{}) at which the experimental magnetic space-group is identified, normalized by $\max_i \{m^i_{\text{exp}}\}$. We use ISOTROPY's \texttt{findsym} to identify the magnetic space-group \cite{stokesISOTROPYSoftwareSuite}. }
    \label{fig:symmetry_comparison}
\end{figure}

\begin{figure}[h]
    \centering
    \centering
    {\large Trial I} \\
    \vspace{2ex}
    \begin{minipage}{0.39\linewidth}
    \begin{mdframed}[roundcorner=10pt, linewidth=1.5pt]
        \centering
        \textit{Converged structure:} \\
        PBE \\
        \subfloat[\centering PBE, trial I]{{
        \includegraphics[width=0.96\linewidth]{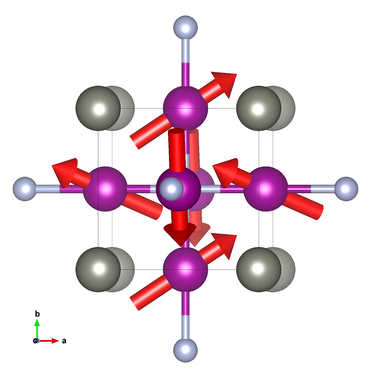}}}
    \end{mdframed}
    \end{minipage}
    \begin{minipage}{0.39\linewidth}
    \begin{mdframed}[roundcorner=10pt, linewidth=1.5pt]
        \centering
        \textit{Converged structure:} \\
        Source-free PBE \\
        \subfloat[\centering \Fsf{PBE}, trial I]{{
        \includegraphics[width=0.96\linewidth]{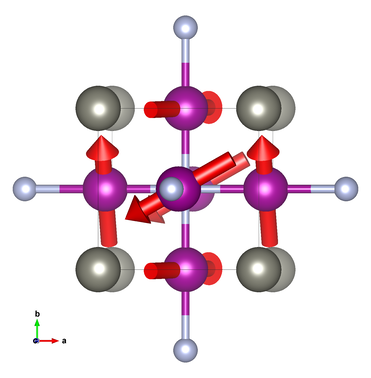}}}
    \end{mdframed}
    \end{minipage}
    \centering
    \vspace{2ex} \\
    {\large Trial II} \\
    \vspace{2ex}
    \begin{minipage}{0.39\linewidth}
    \begin{mdframed}[roundcorner=10pt, linewidth=1.5pt]
        \centering
        \textit{Converged structure:} \\
        PBE \\
        \subfloat[\centering PBE, trial II]{{
        \includegraphics[width=0.96\linewidth]{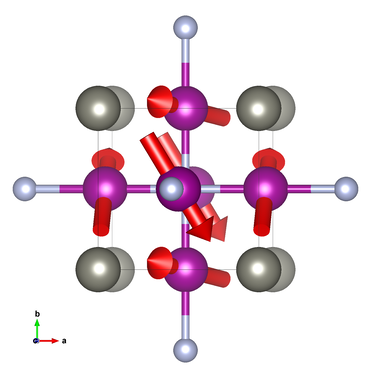}}}
    \end{mdframed}
    \end{minipage}
    \begin{minipage}{0.39\linewidth}
    \begin{mdframed}[roundcorner=10pt, linewidth=1.5pt]
        \centering
        \textit{Converged structure:} \\
        Source-free PBE \\
        \subfloat[\centering \Fsf{PBE}, trial II]{{
        \includegraphics[width=0.96\linewidth]{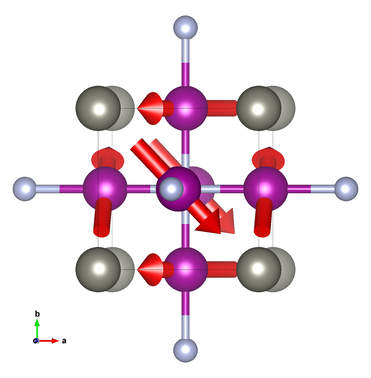}}}
    \end{mdframed}
    \end{minipage}
    \caption{Computed ground state magnetic configurations for two input structures (trials A and B) randomly perturbed from the experimentally measured magnetic structure for \ce{Mn3ZnN}; Comparisons are provided for the computed structures for \Fsf{PBE} versus PBE}
    \label{fig:Mn3ZnN_comparison}
\end{figure}

\begin{figure}[h]
    \centering
    \centering
    {\large Trial I} \\
    \vspace{2ex}
    \begin{minipage}{0.39\linewidth}
    \begin{mdframed}[roundcorner=10pt, linewidth=1.5pt]
        \centering
        \textit{Converged structure:} \\
        PBE \\
        \subfloat[\centering PBE, trial I]{{
        \includegraphics[width=0.96\linewidth]{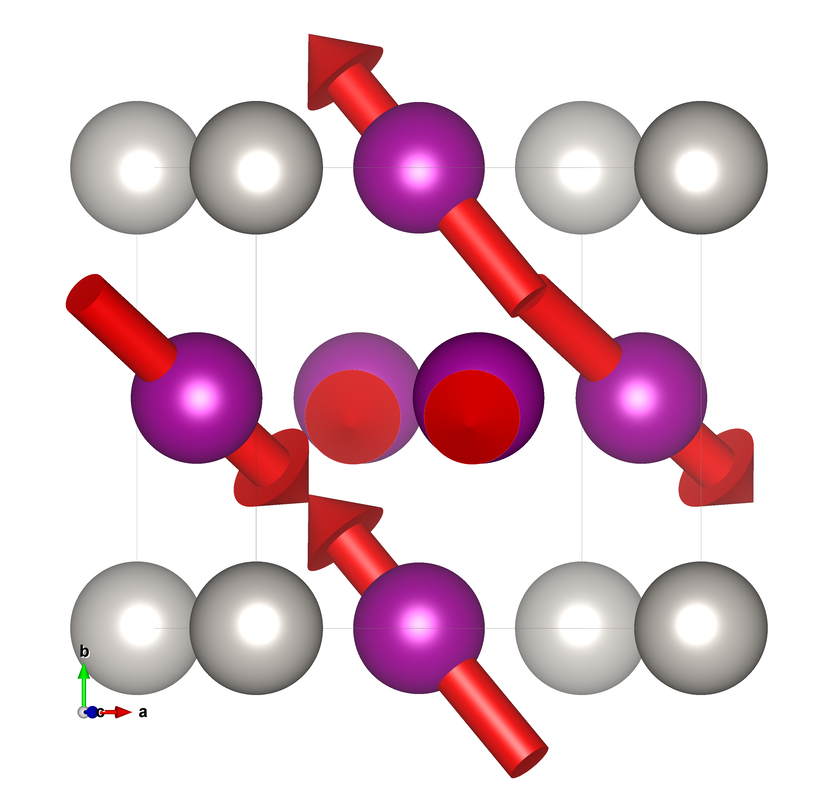}}}
    \end{mdframed}
    \end{minipage}
    \begin{minipage}{0.39\linewidth}
    \begin{mdframed}[roundcorner=10pt, linewidth=1.5pt]
        \centering
        \textit{Converged structure:} \\
        Source-free PBE \\
        \subfloat[\centering \Fsf{PBE}, trial I]{{
        \includegraphics[width=0.96\linewidth]{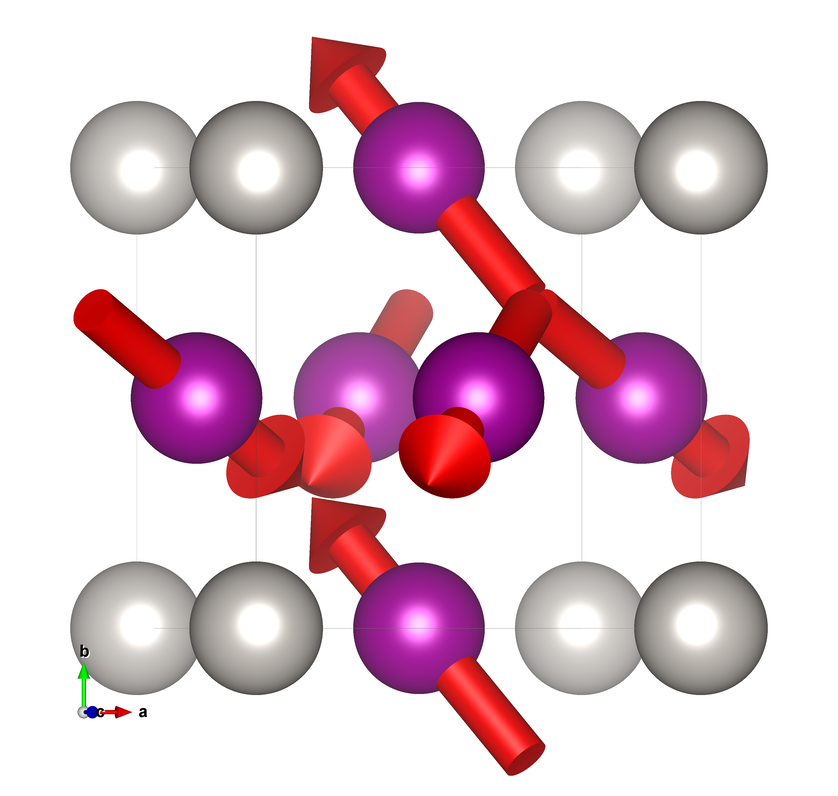}}}
    \end{mdframed}
    \end{minipage}
    \centering
    \vspace{2ex} \\
    {\large Trial II} \\
    \vspace{2ex}
    \begin{minipage}{0.39\linewidth}
    \begin{mdframed}[roundcorner=10pt, linewidth=1.5pt]
        \centering
        \textit{Converged structure:} \\
        PBE \\
        \subfloat[\centering PBE, trial II]{{
        \includegraphics[width=0.96\linewidth]{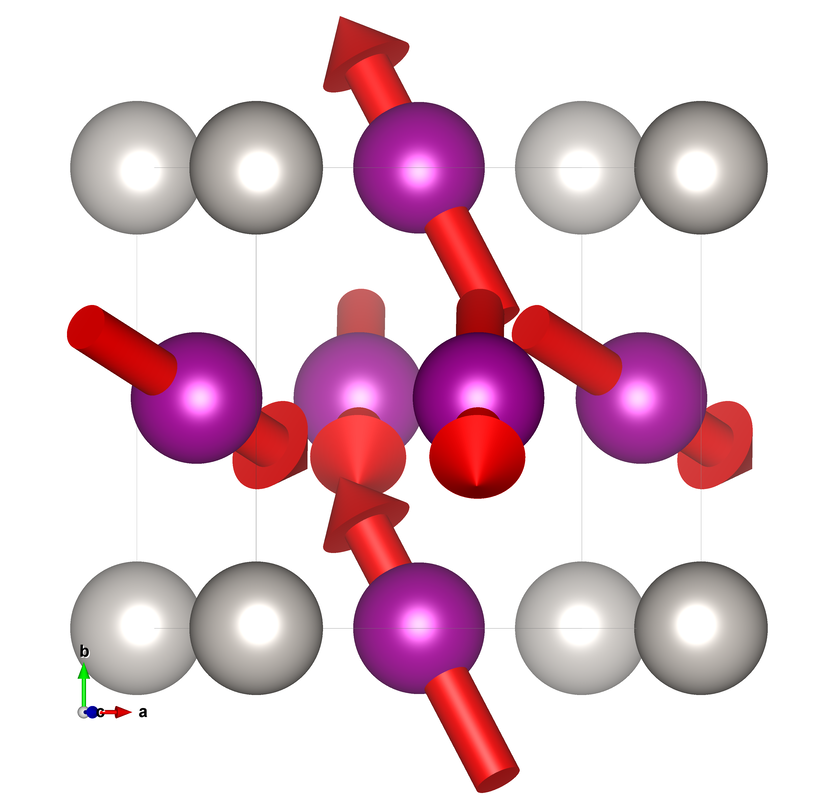}}}
    \end{mdframed}
    \end{minipage}
    \begin{minipage}{0.39\linewidth}
    \begin{mdframed}[roundcorner=10pt, linewidth=1.5pt]
        \centering
        \textit{Converged structure:} \\
        Source-free PBE \\
        \subfloat[\centering \Fsf{PBE}, trial II]{{
        \includegraphics[width=0.96\linewidth]{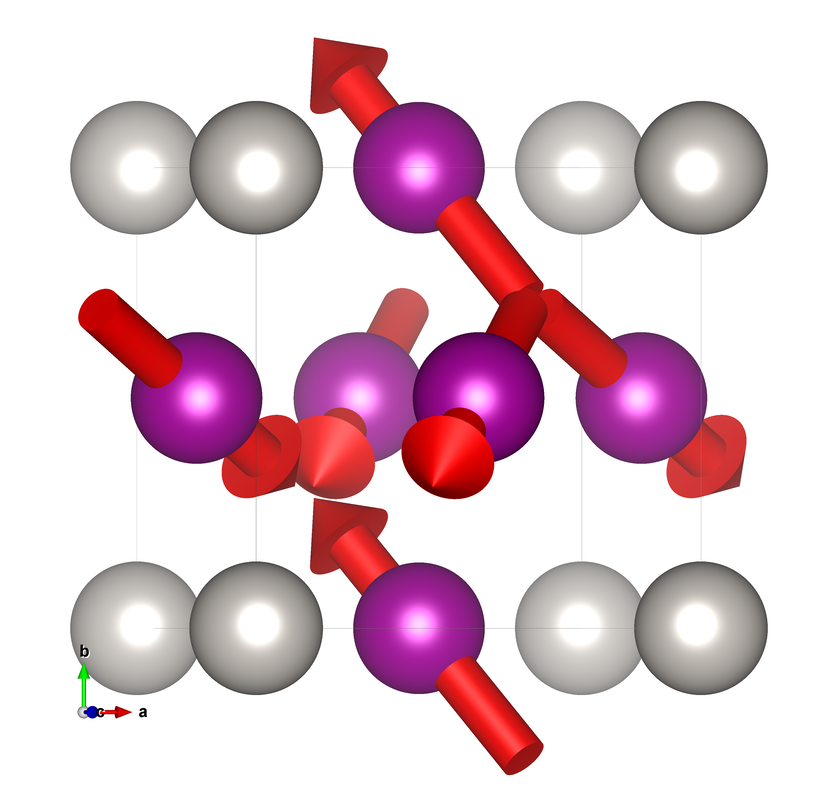}}}
    \end{mdframed}
    \end{minipage}
    \caption{Computed ground state magnetic configurations for two input structures (trials A and B) randomly perturbed from the experimentally measured magnetic structure for \ce{Mn3Pt}; Comparisons are provided for the computed structures for \Fsf{PBE} versus PBE}
    \label{fig:Mn3Pt_comparison}
\end{figure}

\begin{figure}[h]
    \centering
    \includegraphics[width=0.78\linewidth]{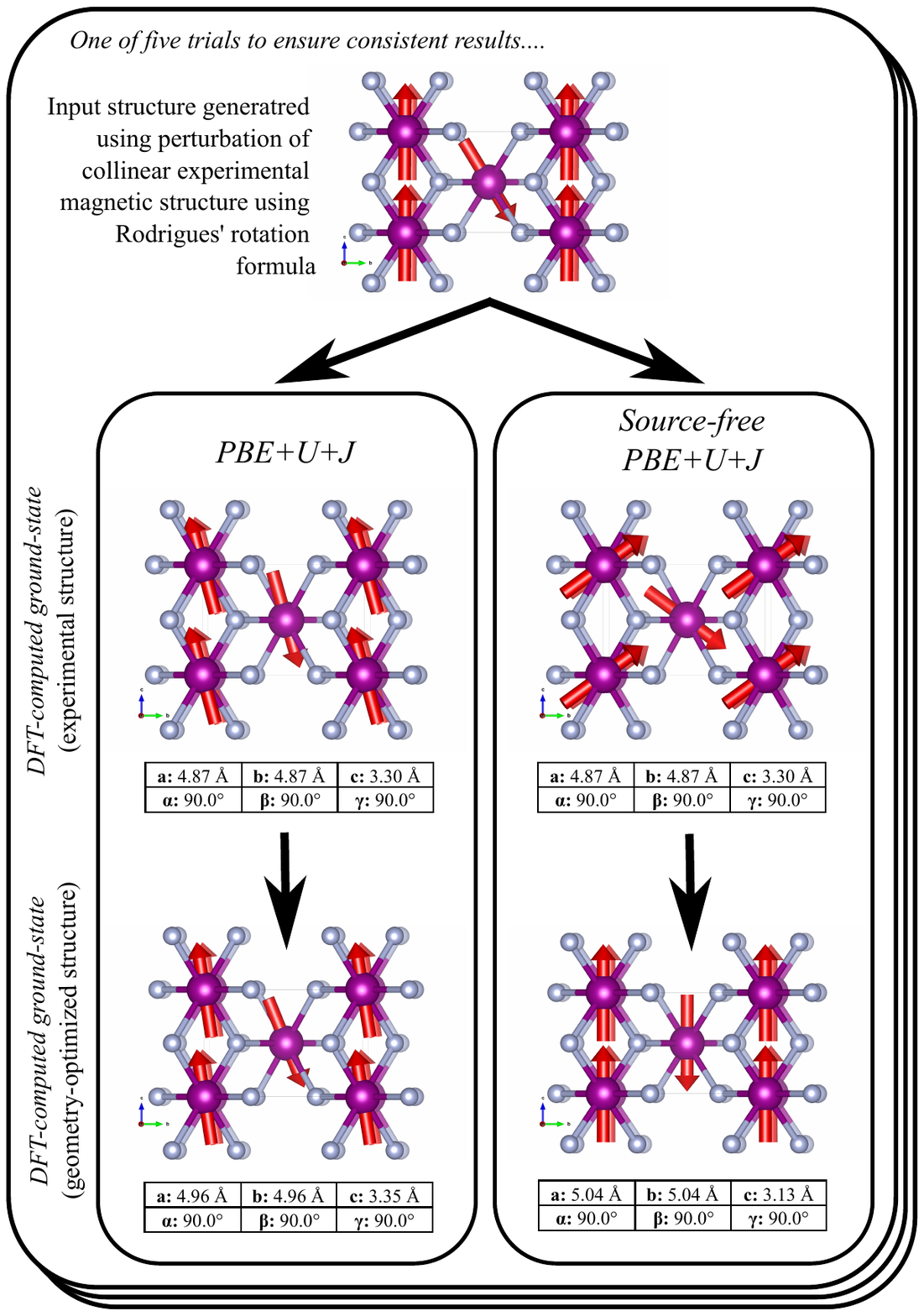}
    \caption{Computed ground state magnetic configurations for randomly perturbed input structure from the experimentally measured magnetic structure for \ce{MnF2}; Comparisons are provided for the computed structures for \Fsf{PBE} versus PBE}
    \label{fig:MnF2_comparison}
\end{figure}

\begin{figure}[h]
    \centering
    \includegraphics[width=0.78\linewidth]{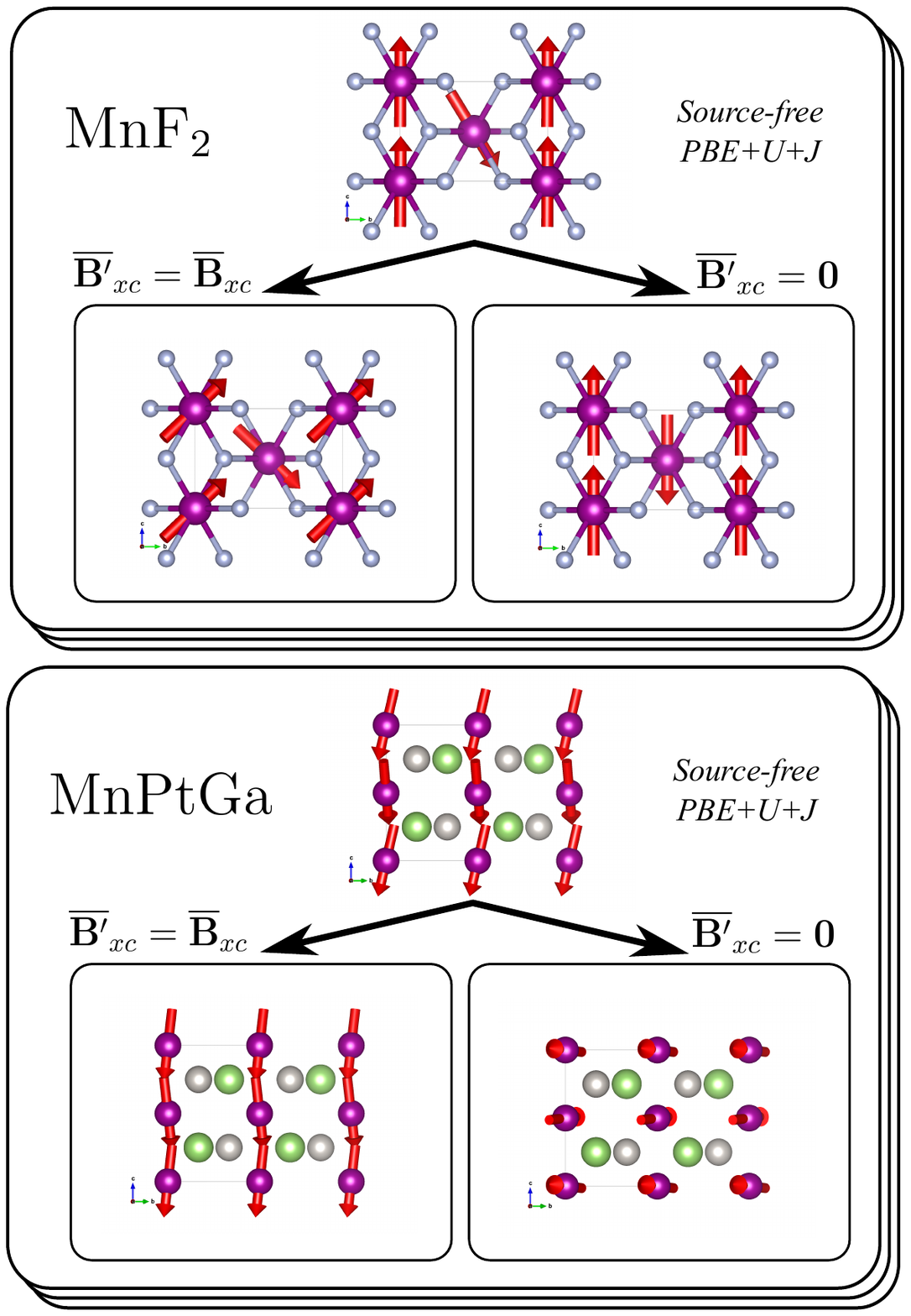}
    \caption{Comparison between the different choices of $\overline{\bm B'}_{xc}$ for antiferromagnetic \ce{MnF2} (top) and canted ferromagnetic \ce{MnPtGa} (bottom). Treatment of $\overline{\bm B'}_{xc}$ alone can dictate the FM or AFM character of spin-polarized systems. 
    As the inclusion of a $q=0$ component changes the Kohn-Sham effective potential, the absolute energies computed with or without $\overline{\bm B}_{xc}$ are not directly comparable.
    }
    \label{fig:q0_comparison}
\end{figure}

\begin{figure}[h]
    \centering
    \centering
    {\large Trial I} \\
    \vspace{2ex}
    \begin{minipage}{0.49\linewidth}
    \begin{mdframed}[roundcorner=10pt, linewidth=1.5pt]
        \centering
        \textit{Converged structure:} \\
        PBE \\
        \subfloat[\centering PBE, trial I]{{
        \includegraphics[width=0.96\linewidth]{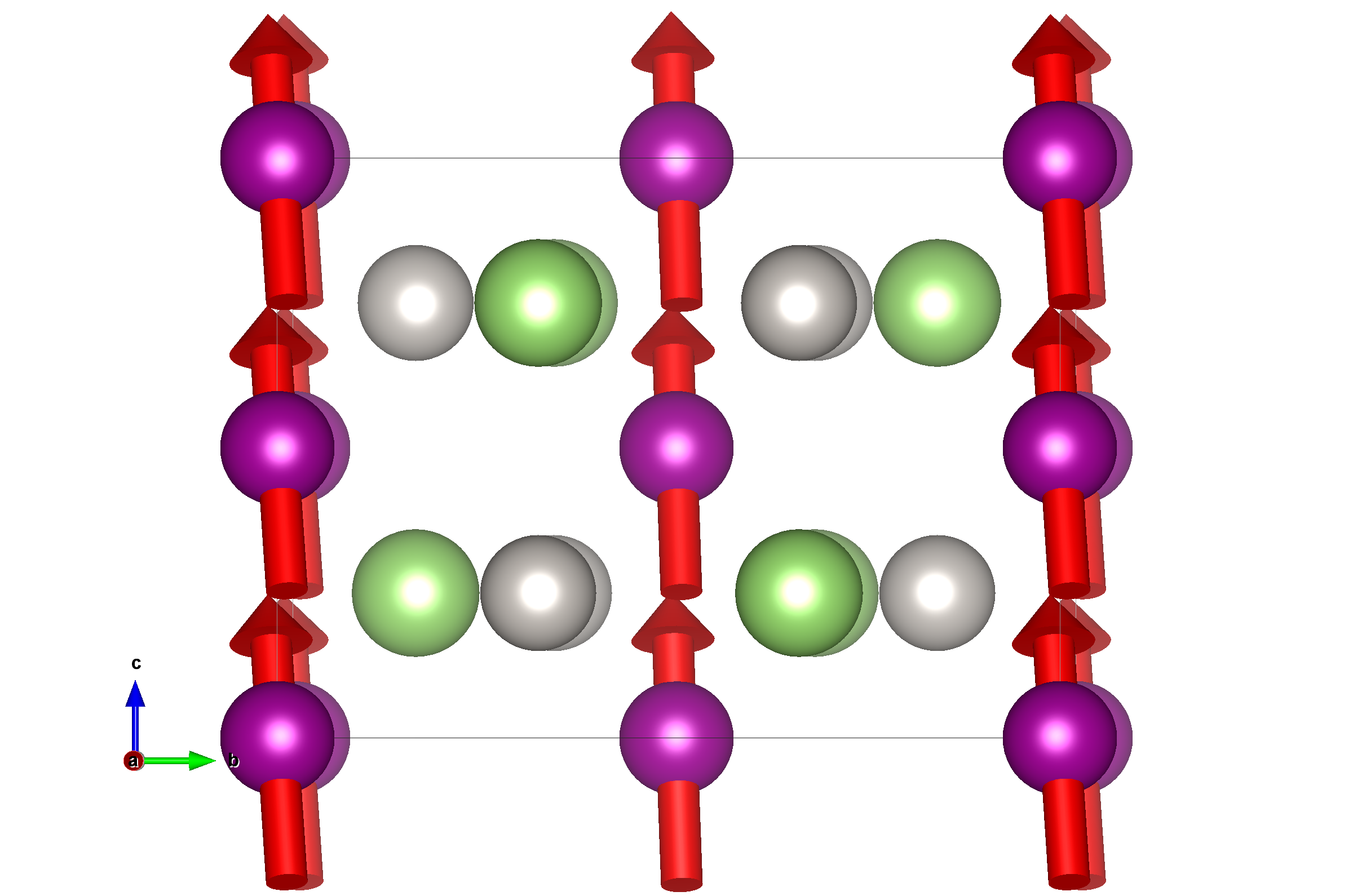}}}
    \end{mdframed}
    \end{minipage}
    \begin{minipage}{0.49\linewidth}
    \begin{mdframed}[roundcorner=10pt, linewidth=1.5pt]
        \centering
        \textit{Converged structure:} \\
        Source-free PBE \\
        \subfloat[\centering \Fsf{PBE}, trial I]{{
        \includegraphics[width=0.96\linewidth]{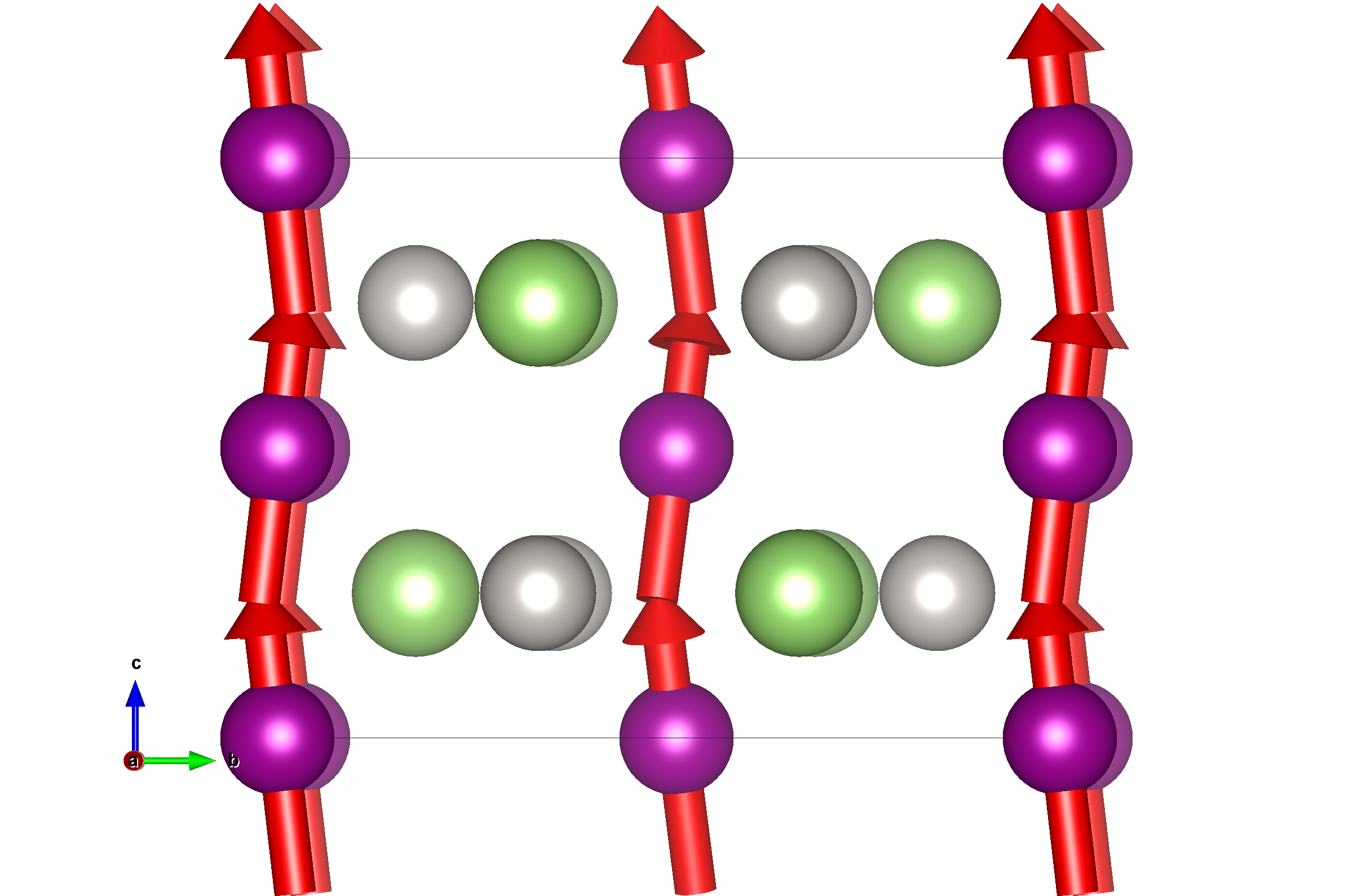}}}
    \end{mdframed}
    \end{minipage}
    \centering
    \vspace{2ex} \\
    {\large Trial II} \\
    \vspace{2ex}
    \begin{minipage}{0.49\linewidth}
    \begin{mdframed}[roundcorner=10pt, linewidth=1.5pt]
        \centering
        \textit{Converged structure:} \\
        PBE \\
        \subfloat[\centering PBE, trial II]{{
        \includegraphics[width=0.96\linewidth]{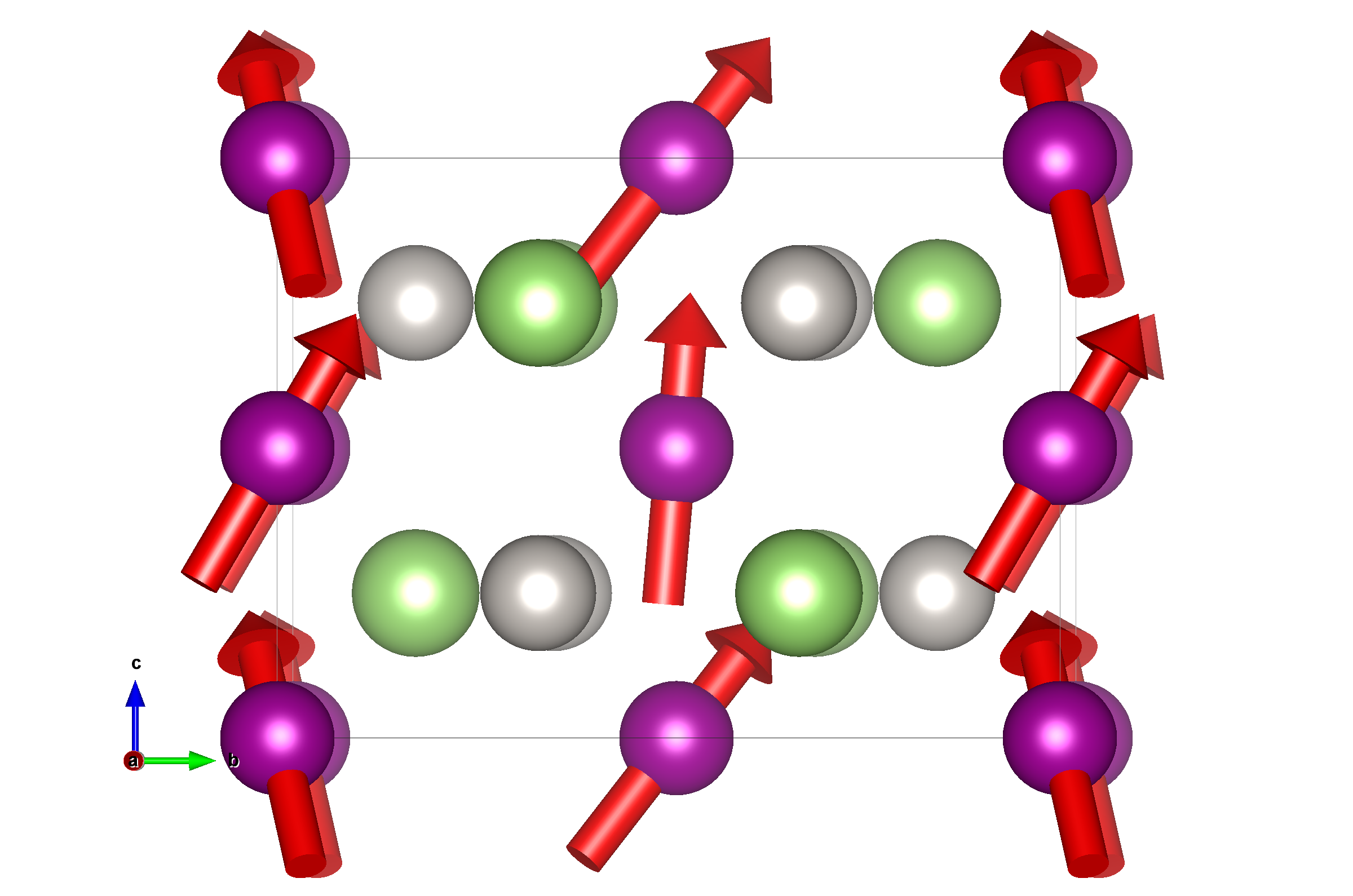}}}
    \end{mdframed}
    \end{minipage}
    \begin{minipage}{0.49\linewidth}
    \begin{mdframed}[roundcorner=10pt, linewidth=1.5pt]
        \centering
        \textit{Converged structure:} \\
        Source-free PBE \\
        \subfloat[\centering \Fsf{PBE}, trial II]{{
        \includegraphics[width=0.96\linewidth]{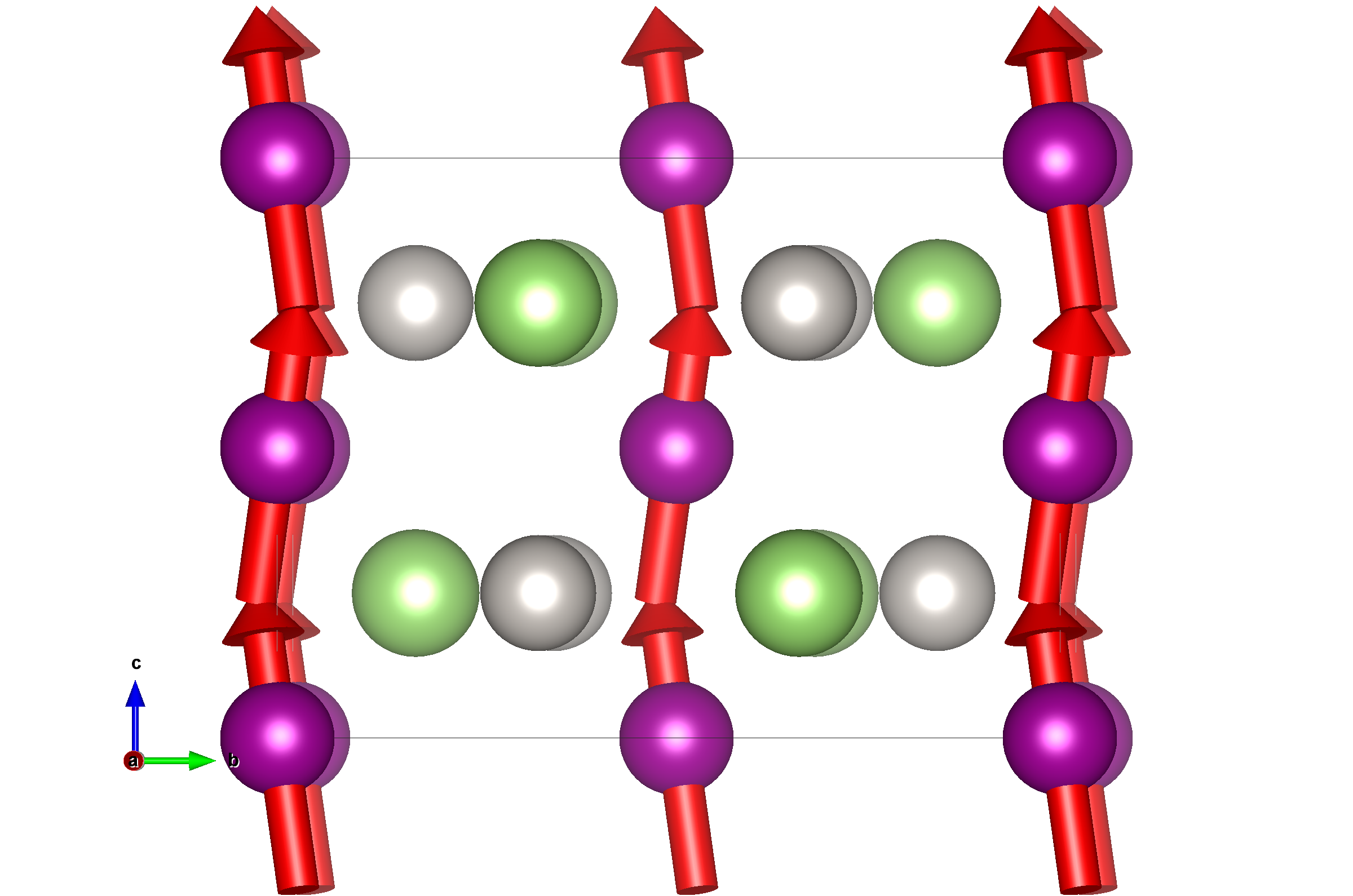}}}
    \end{mdframed}
    \end{minipage}
    \caption{Computed ground state magnetic configurations for two input structures (trials A and B) randomly perturbed from the experimentally measured magnetic structure for \ce{MnPtGa}; Comparisons are provided for the computed structures for \Fsf{PBE} versus PBE}
    \label{fig:MnGaPt_comparison}
\end{figure}

\begin{figure}[h]
    \centering
    \centering
    {\large Trial I} \\
    \vspace{2ex}
    \begin{minipage}{0.19\linewidth}
        \flushleft
        \textit{Random perturbation from experimental magnetic structure:}
        \subfloat[\centering Initial perturbed structure, trial I]{{
        \includegraphics[width=0.13\paperwidth]{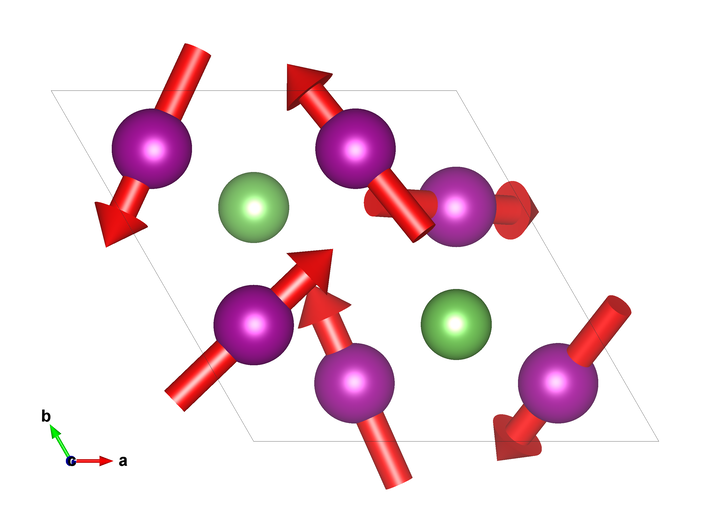}}}
    \end{minipage}
    \begin{minipage}{0.39\linewidth}
    \begin{mdframed}[roundcorner=10pt, linewidth=1.5pt]
        \centering
        \textit {Converged structure:} \\ PBE
        \subfloat[\centering PBE, trial I \newline \textit{viewed along [001]}]{{
        \includegraphics[width=0.13\paperwidth]{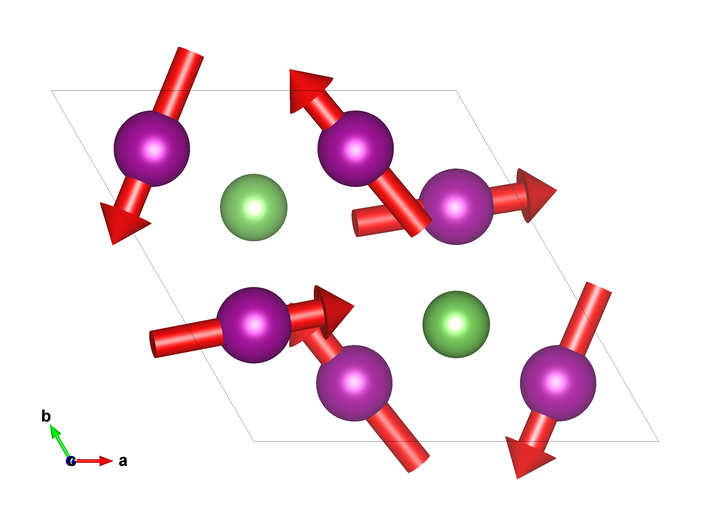}}}
        \subfloat[\centering PBE, trial I \newline \textit{viewed along [100]}]{{
        \includegraphics[width=0.13\paperwidth]{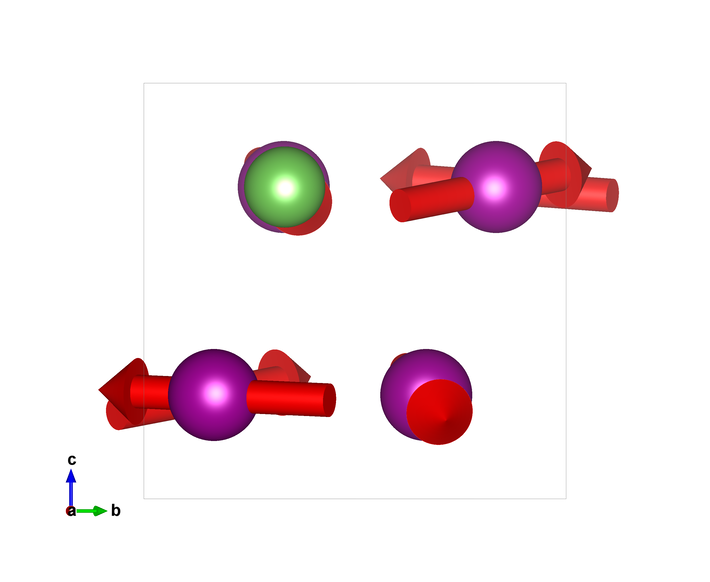}}}
    \end{mdframed}
    \end{minipage}
    \begin{minipage}{0.39\linewidth}
    \begin{mdframed}[roundcorner=10pt, linewidth=1.5pt]
        \centering
        \textit {Converged structure:} \\ Source-free PBE
        \subfloat[\centering \Fsf{PBE}, trial I \newline \textit{viewed along [001]}]{{
        \includegraphics[width=0.13\paperwidth]{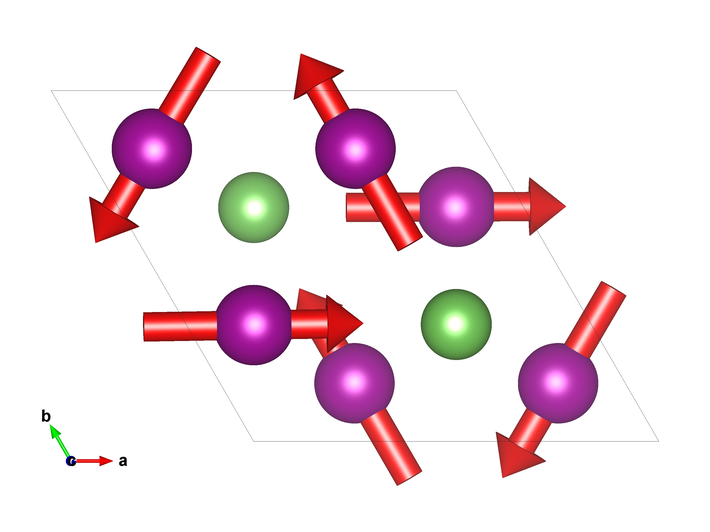}}}
        \subfloat[\centering \Fsf{PBE}, trial I \newline \textit{viewed along [100]}]{{
        \includegraphics[width=0.13\paperwidth]{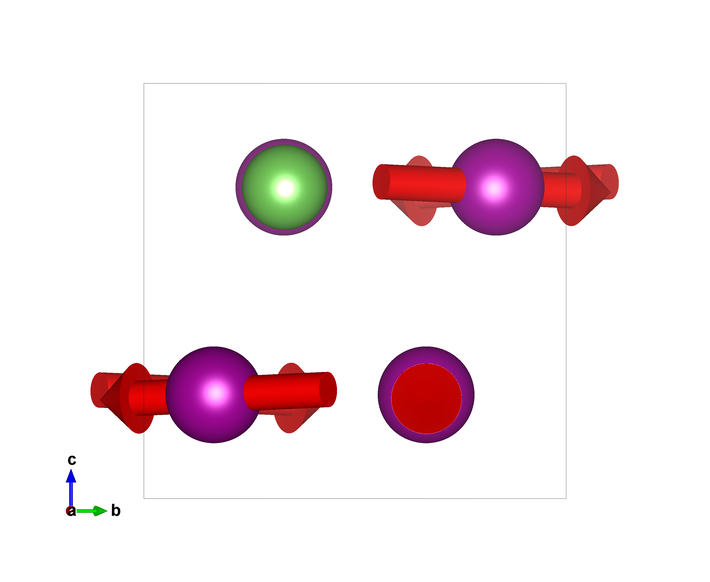}}}
    \end{mdframed}
    \end{minipage}
    \centering
    \vspace{2ex} \\
    {\large Trial II} \\
    \vspace{2ex}
    \begin{minipage}{0.19\linewidth}
        \flushleft
        \textit{Random perturbation from experimental magnetic structure:}
        \subfloat[\centering Initial perturbed structure, trial II]{{
        \includegraphics[width=0.13\paperwidth]{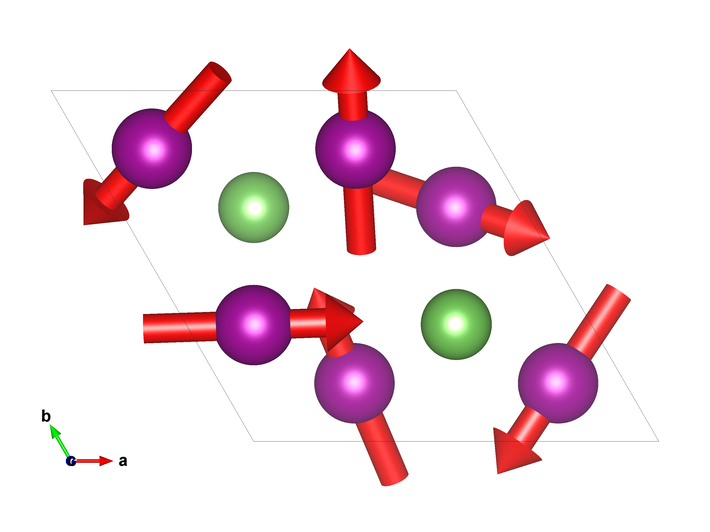}}}
    \end{minipage}
    \begin{minipage}{0.39\linewidth}
    \begin{mdframed}[roundcorner=10pt, linewidth=1.5pt]
        \centering
        \textit {Converged structure:} \\ PBE
        \subfloat[\centering PBE, trial II \newline \textit{viewed along [001]}]{{
        \includegraphics[width=0.13\paperwidth]{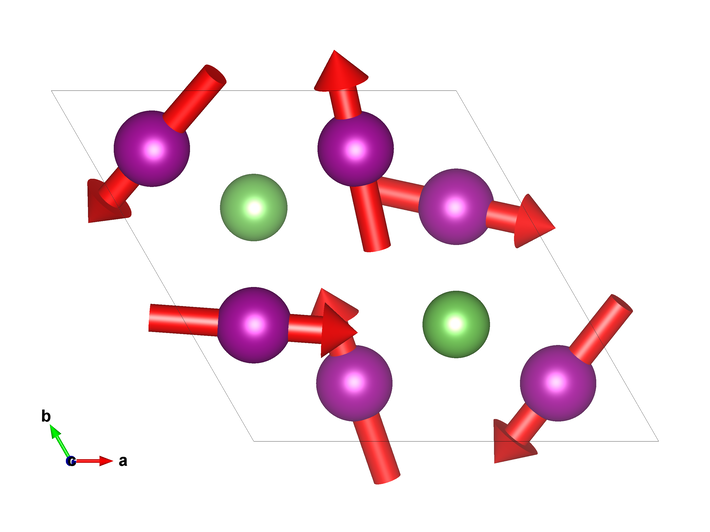}}}
        \subfloat[\centering PBE, trial II \newline \textit{viewed along [100]}]{{
        \includegraphics[width=0.13\paperwidth]{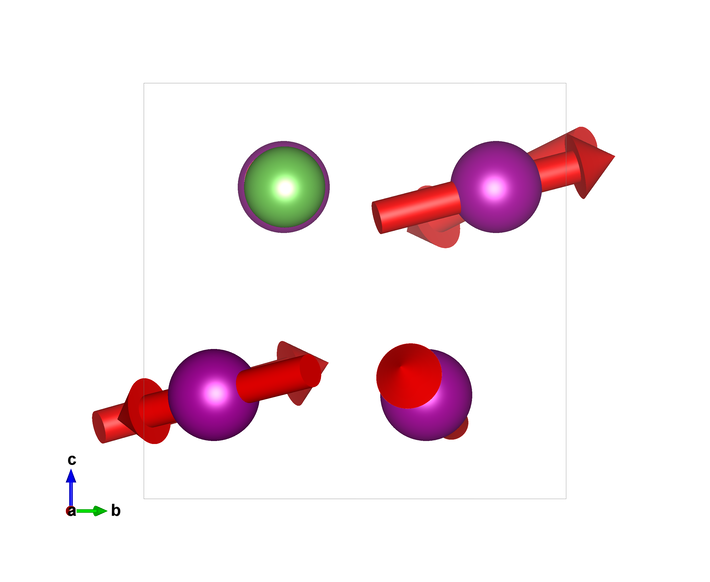}}}
    \end{mdframed}
    \end{minipage}
    \begin{minipage}{0.39\linewidth}
    \begin{mdframed}[roundcorner=10pt, linewidth=1.5pt]
        \centering
        \textit {Converged structure:} \\ Source-free PBE
        \subfloat[\centering \Fsf{PBE}, trial II \newline \textit{viewed along [001]}]{{
        \includegraphics[width=0.13\paperwidth]{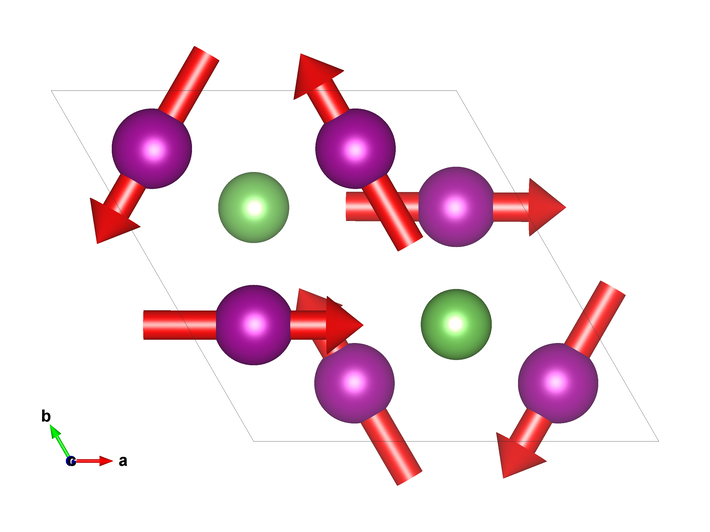}}}
        \subfloat[\centering \Fsf{PBE}, trial II \newline \textit{viewed along [100]}]{{
        \includegraphics[width=0.13\paperwidth]{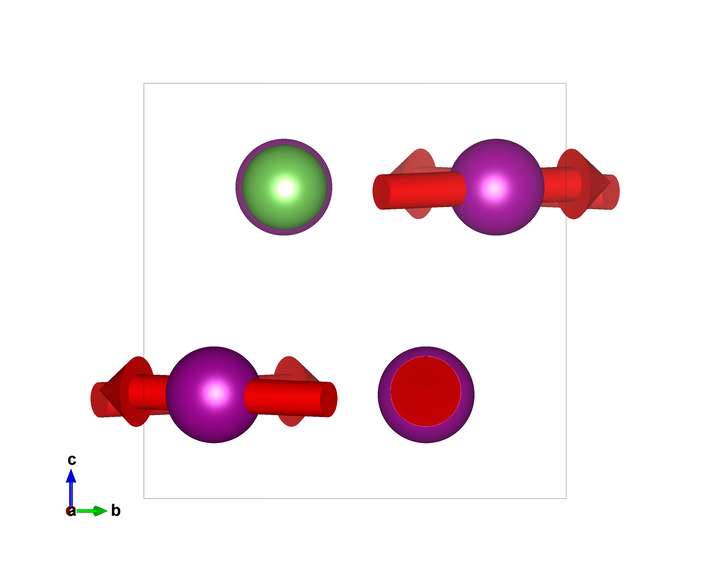}}}
    \end{mdframed}
    \end{minipage}
    \caption{Computed ground state magnetic configurations for two input structures (trials A and B) randomly perturbed from the experimentally measured magnetic structure for \ce{Mn3As}; Comparisons are provided for the computed structures for \Fsf{PBE} versus PBE; For \Fsf{PBE} and PBE runs, input atomic positions and cell shape were first determined by performing structural relaxations, with moments initialized in the symmetric orientation}
    \label{fig:Mn3As_comparison}
\end{figure}

\begin{figure}[h]
    \centering
    \centering
    {\large Trial I} \\
    \vspace{2ex}
    \begin{minipage}{0.49\linewidth}
    \begin{mdframed}[roundcorner=10pt, linewidth=1.5pt]
        \centering
        \textit {Converged structure:} \\ PBE \\
        \subfloat[\centering PBE, trial I \newline \textit{viewed along [001]}]{{
        \includegraphics[width=0.6\linewidth]{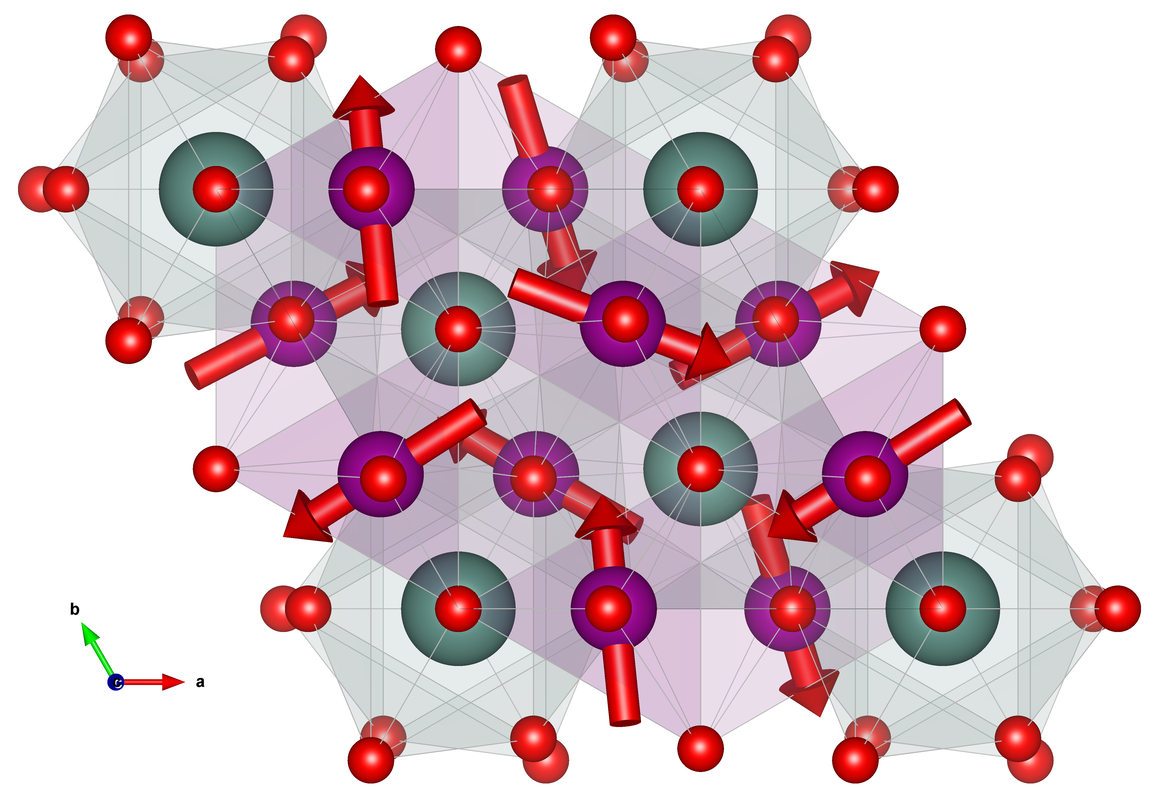}}}
        \subfloat[\centering PBE, trial I \newline \textit{viewed along [010]}]{{
        \includegraphics[width=0.6\linewidth,angle=90,origin=c]{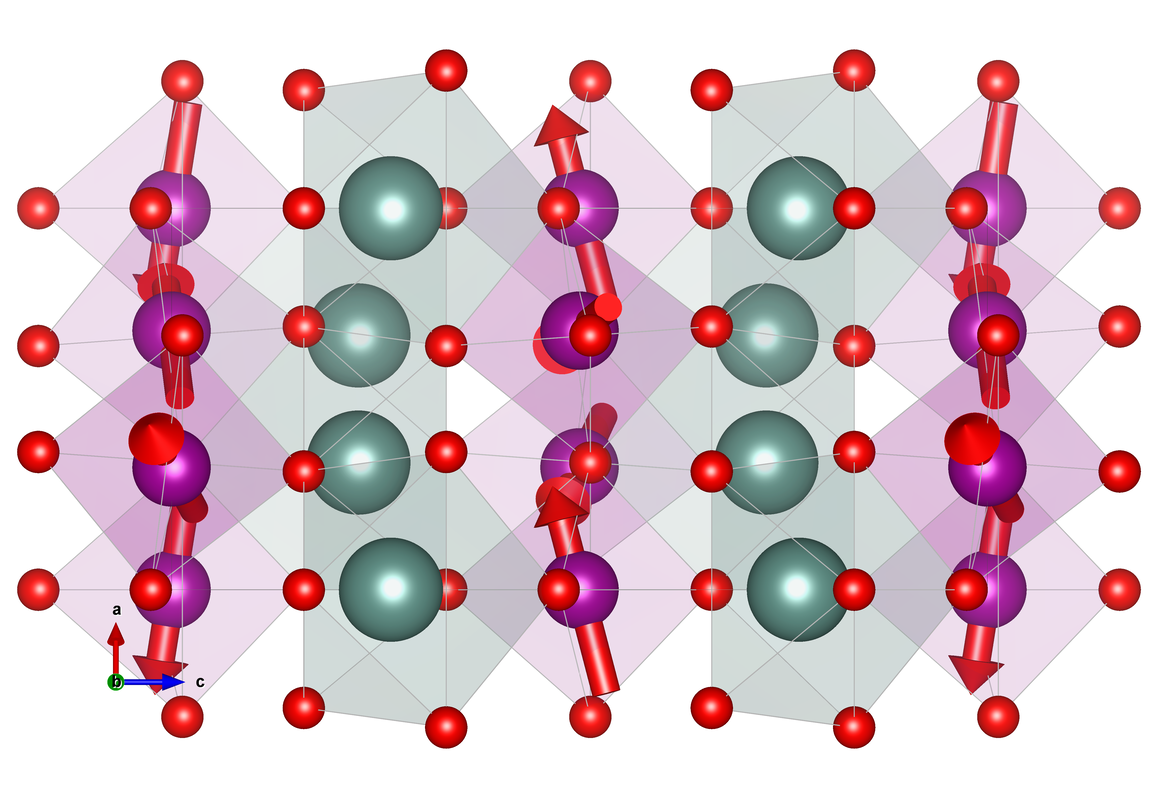}}}
    \end{mdframed}
    \end{minipage}
    \begin{minipage}{0.49\linewidth}
    \begin{mdframed}[roundcorner=10pt, linewidth=1.5pt]
        \centering
        \textit {Converged structure:} \\ Source-free PBE \\
        \subfloat[\centering \Fsf{PBE}, trial I \newline \textit{viewed along [001]}]{{
        \includegraphics[width=0.6\linewidth]{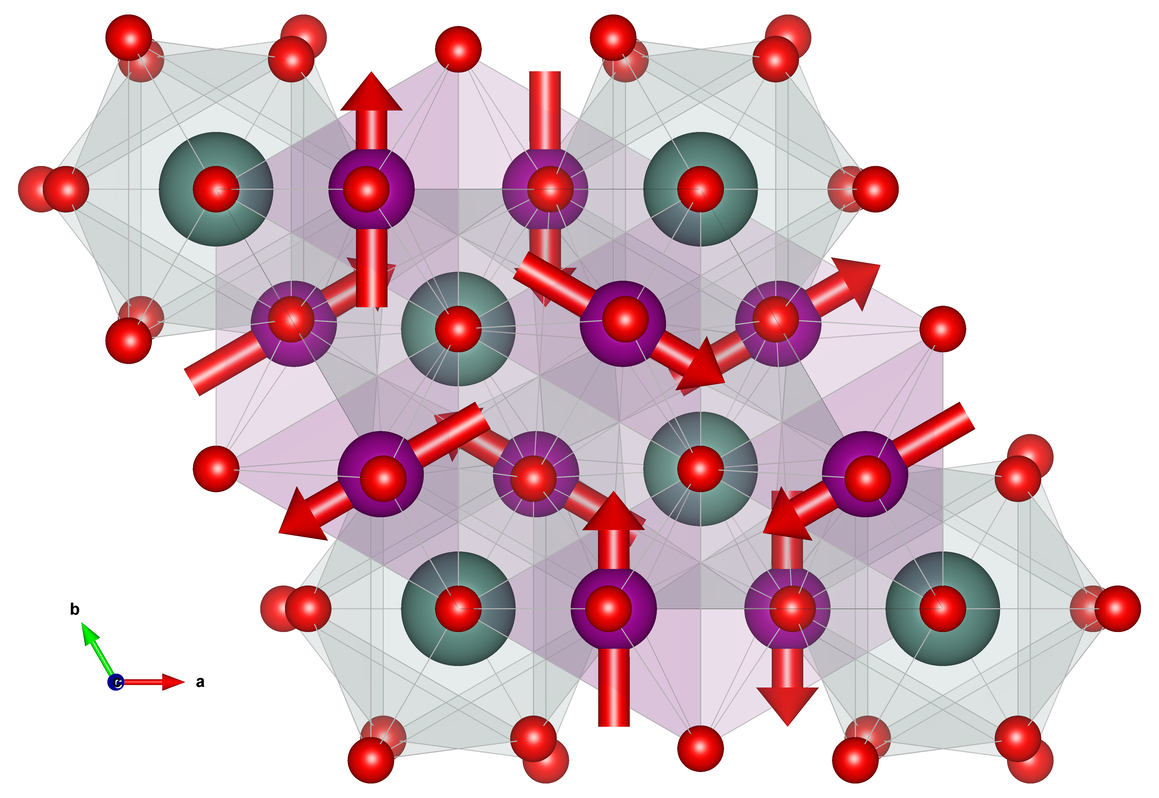}}}
        \subfloat[\centering \Fsf{PBE}, trial I \newline \textit{viewed along [010]}]{{
        \includegraphics[width=0.6\linewidth,angle=90,origin=c]{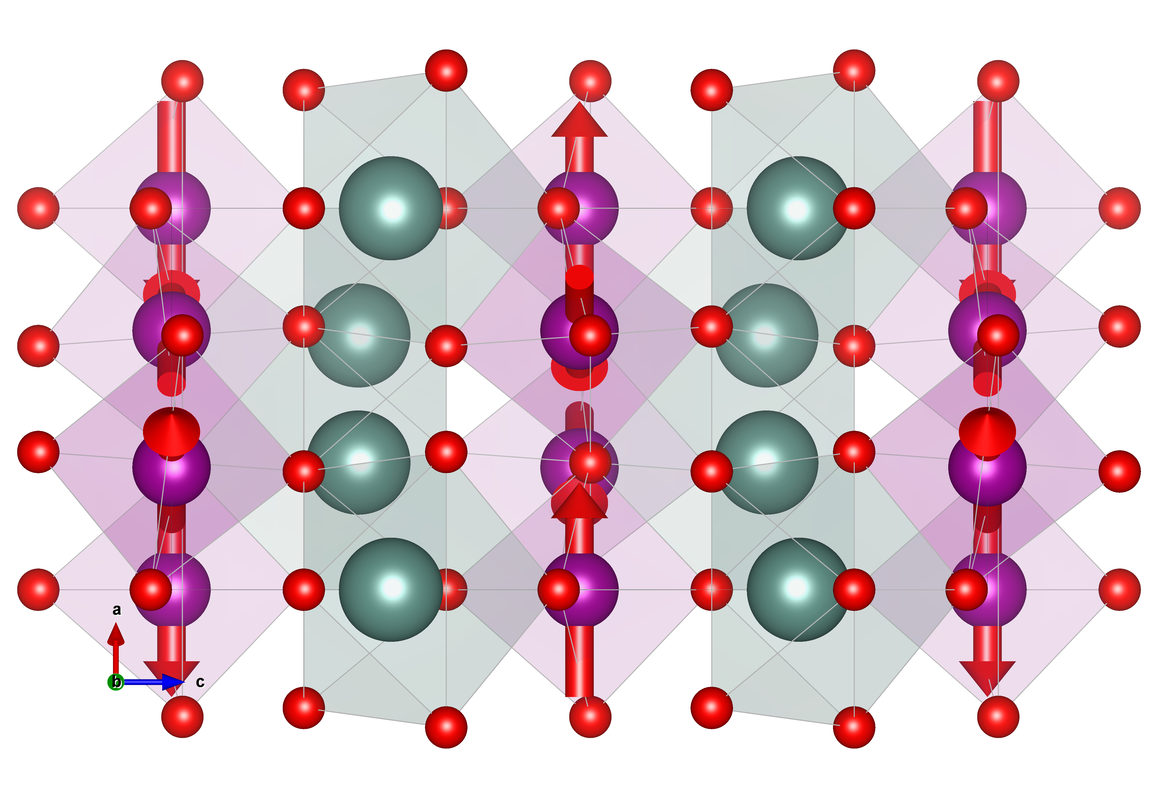}}}
    \end{mdframed}
    \end{minipage}
    \centering
    \vspace{2ex} \\
    {\large Trial II} \\
    \vspace{2ex}
    \begin{minipage}{0.49\linewidth}
    \begin{mdframed}[roundcorner=10pt, linewidth=1.5pt]
        \centering
        \textit {Converged structure:} \\ PBE \\
        \subfloat[\centering PBE, trial I \newline \textit{viewed along [001]}]{{
        \includegraphics[width=0.6\linewidth]{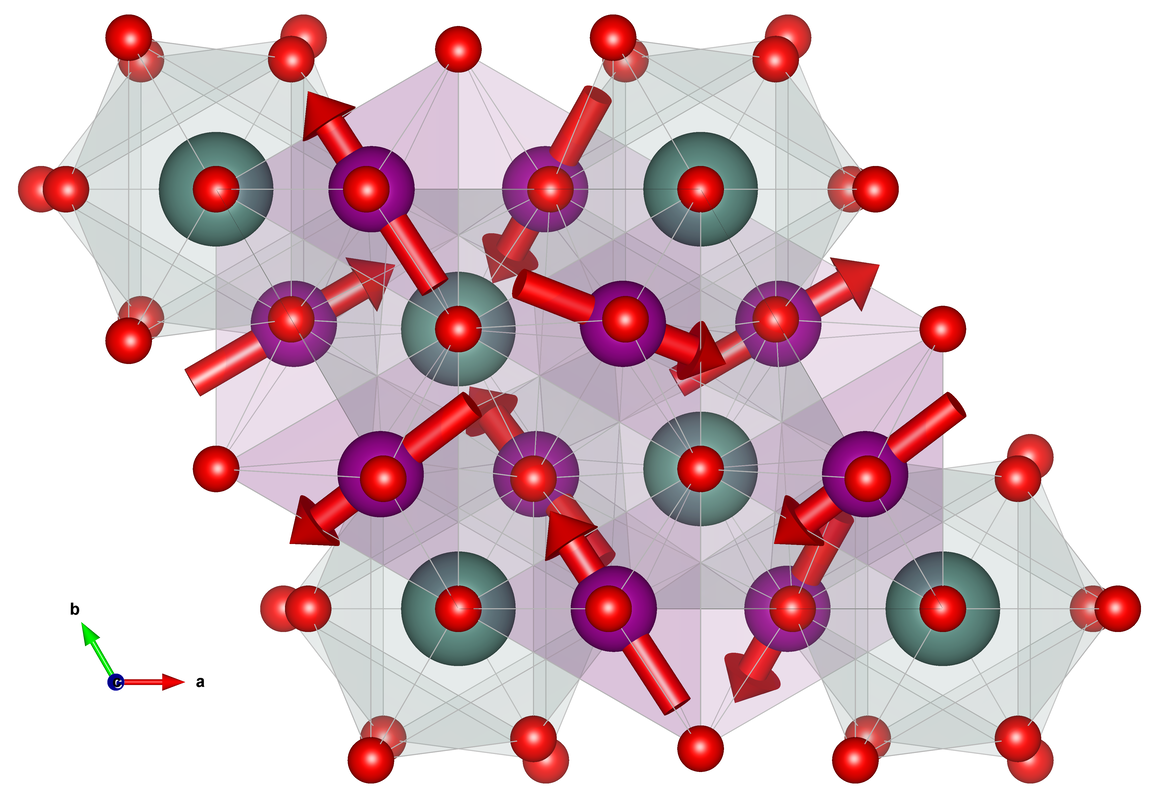}}}
        \subfloat[\centering PBE, trial I \newline \textit{viewed along [010]}]{{
        \includegraphics[width=0.6\linewidth,angle=90,origin=c]{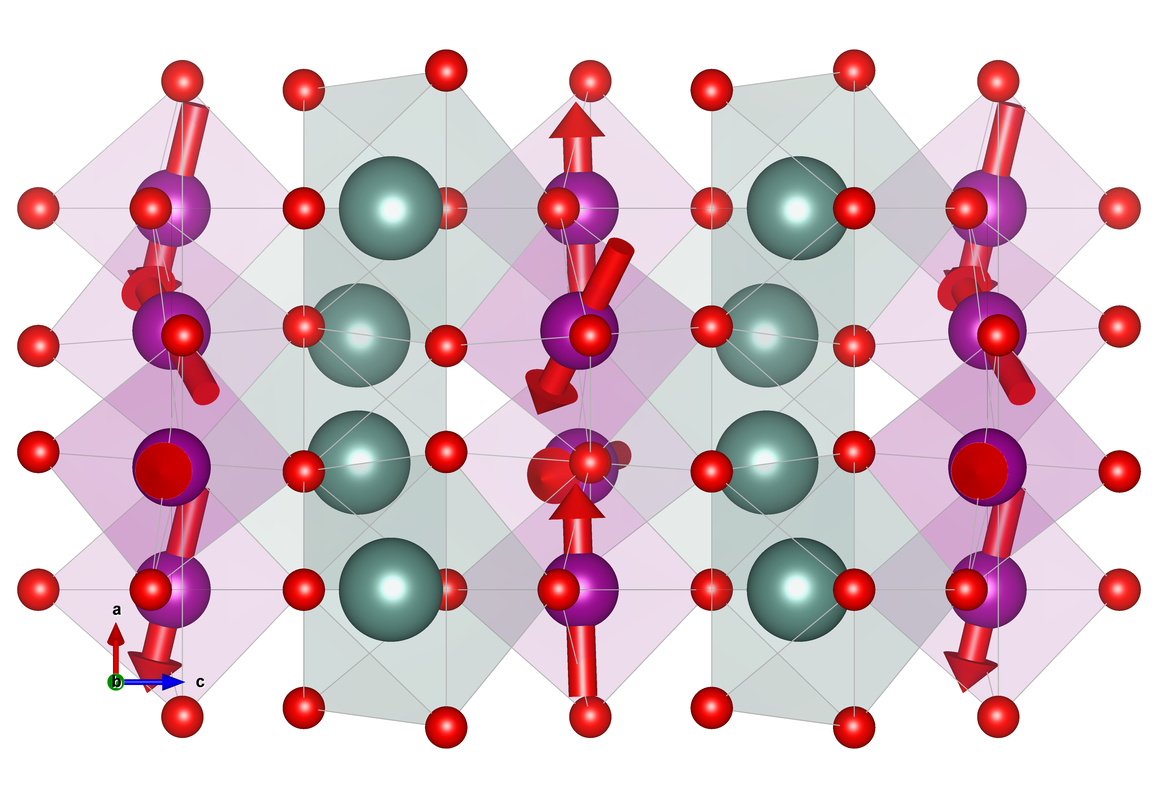}}}
    \end{mdframed}
    \end{minipage}
    \begin{minipage}{0.49\linewidth}
    \begin{mdframed}[roundcorner=10pt, linewidth=1.5pt]
        \centering
        \textit {Converged structure:} \\ Source-free PBE \\
        \subfloat[\centering \Fsf{PBE}, trial I \newline \textit{viewed along [001]}]{{
        \includegraphics[width=0.6\linewidth]{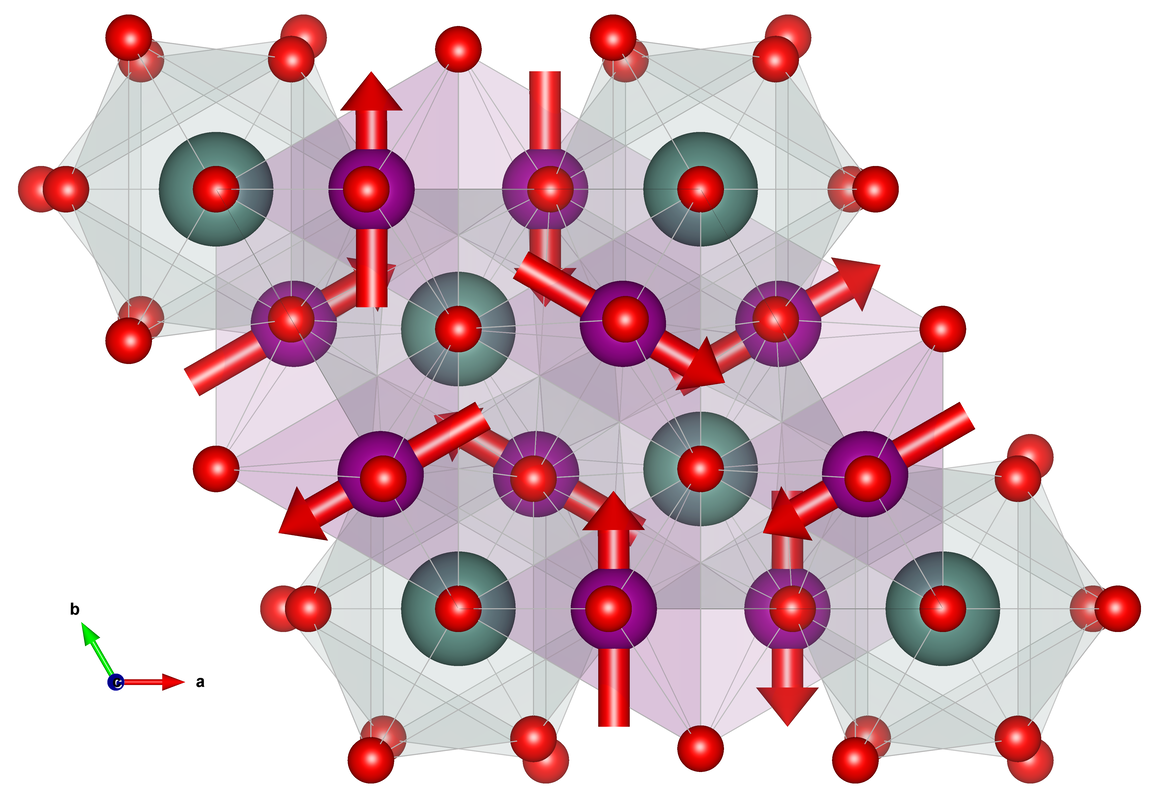}}}
        \subfloat[\centering \Fsf{PBE}, trial I \newline \textit{viewed along [010]}]{{
        \includegraphics[width=0.6\linewidth,angle=90,origin=c]{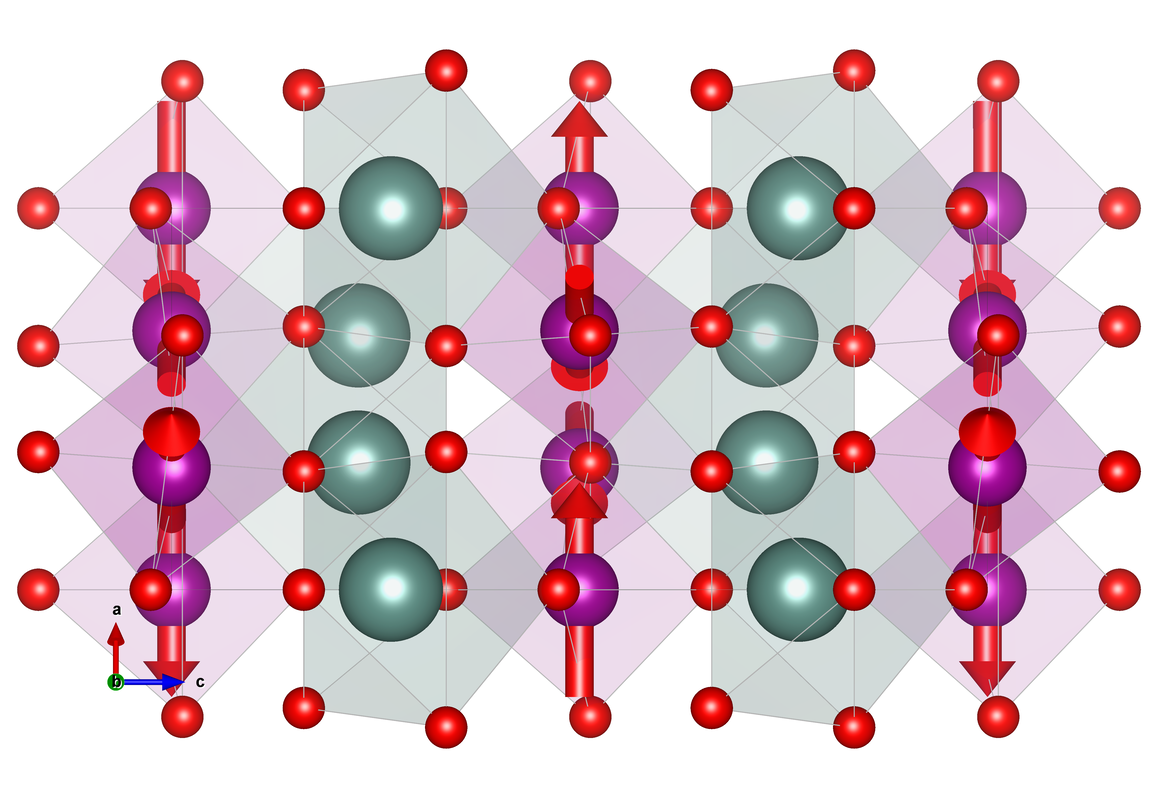}}}
    \end{mdframed}
    \end{minipage}
    \caption{Computed ground state magnetic configurations for two input structures (trials A and B) randomly perturbed from the experimentally measured magnetic structure for \ce{YMnO3}; Comparisons are provided for the computed structures for \Fsf{PBE} versus PBE; For \Fsf{PBE} and PBE runs, input atomic positions and cell shape were first determined by performing structural relaxations, with moments initialized in the symmetric orientation}
    \label{fig:YMnO3_comparison}
\end{figure}

\begin{figure}[h]
    \centering
    \includegraphics[width=0.98\linewidth]{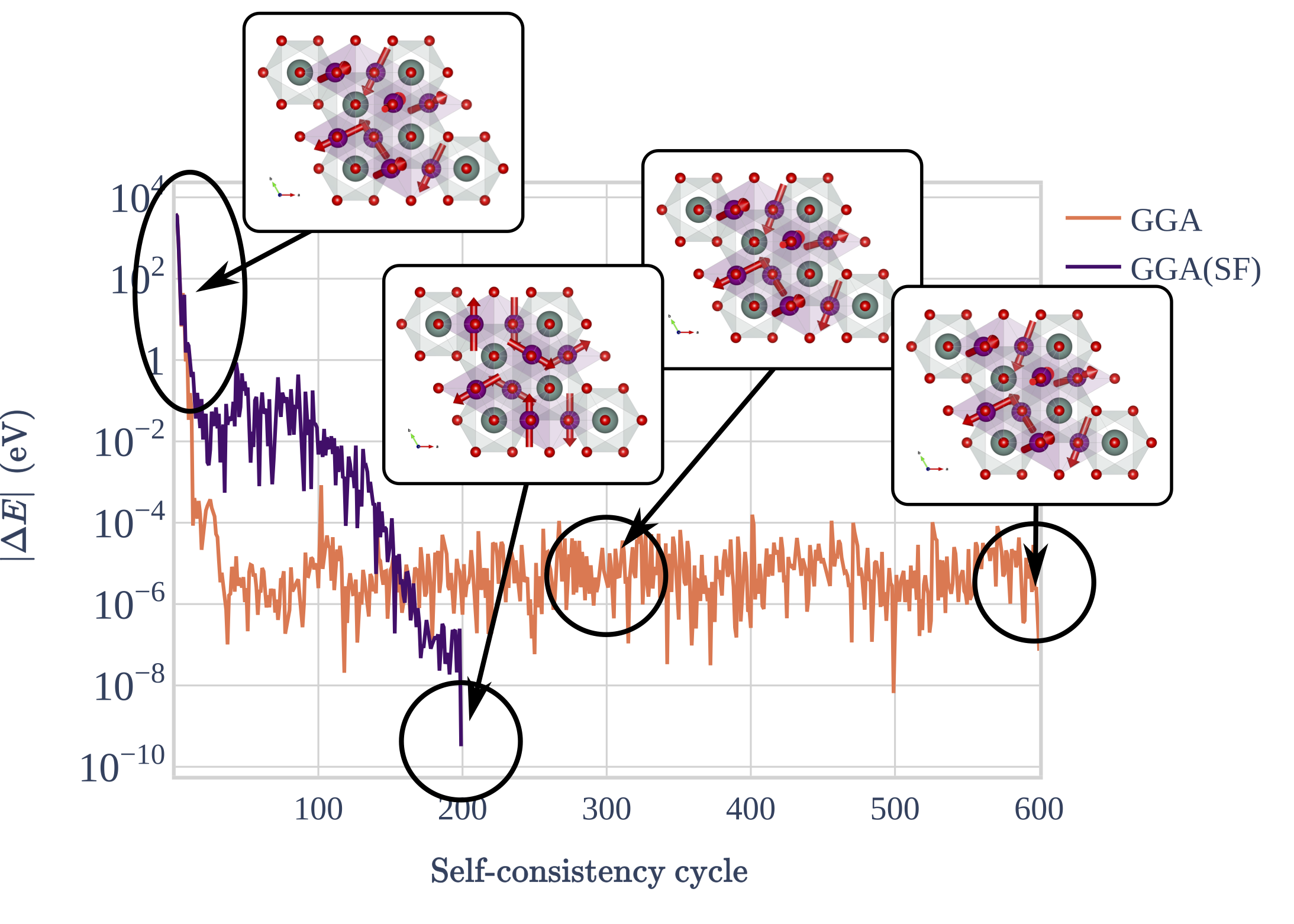} 
    \caption{Convergence of \Fsf{GGA} compared against GGA for \ce{YMnO3}, with moments randomly perturbed from the symmetric ground state structure. Snapshots of the magnetic configuration at each self-consistency step are shown to convey the improved convergence characteristics of the source-free functional. The $y$-axis is the absolute energy difference between subsequent self-consistency iterations, plotted on a logarithmic scale.}
    \label{fig:convergence_sf_YMnO3}
\end{figure}

\begin{figure}[h]
    \centering
    \includegraphics[width=0.68\linewidth]{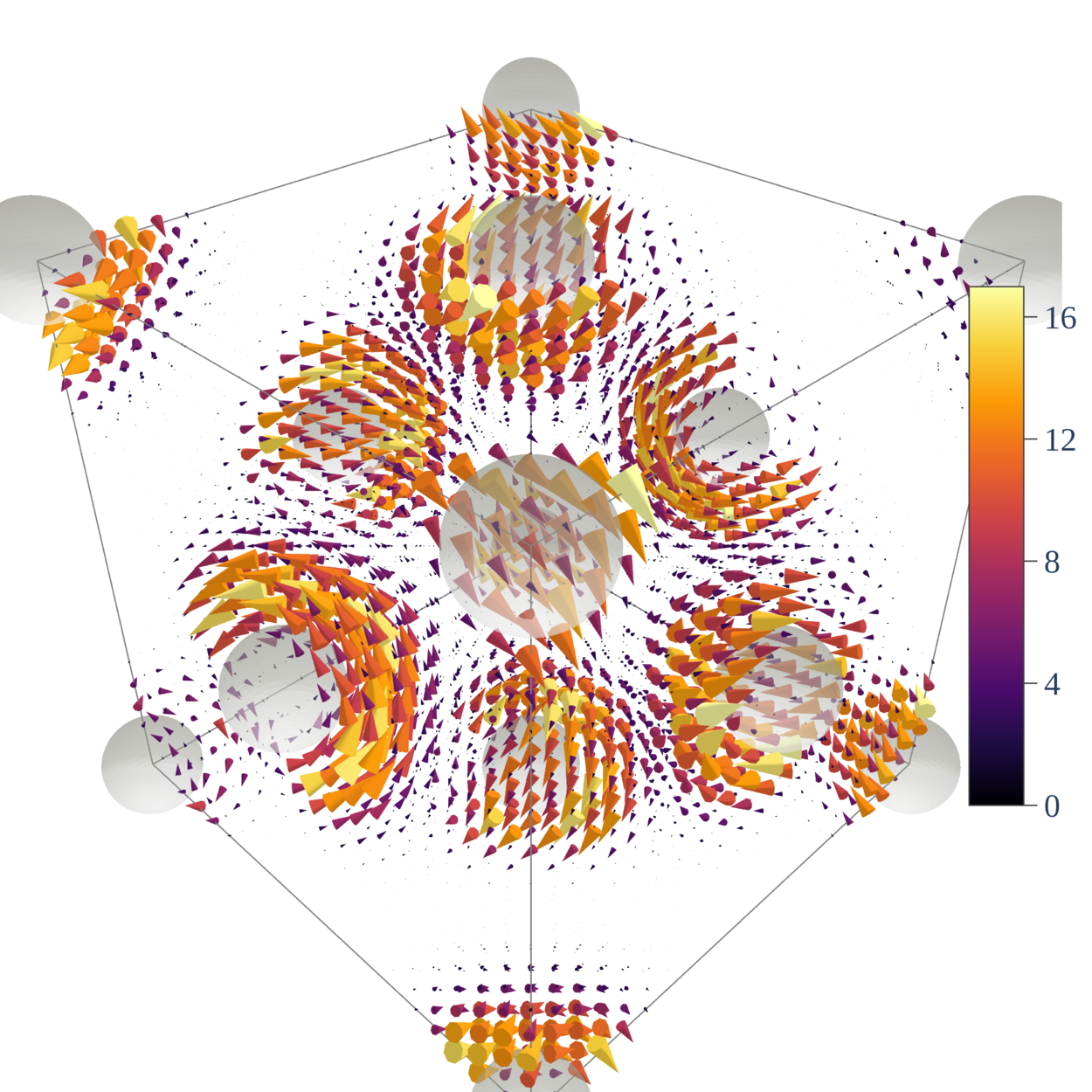} 
    \caption{Vector field visualization of the probability current density, $\bm j_p$, within \ce{UO2}. Strong circulation of $\bm j_p$ is observed surrounding the uranium atoms, in grey. }
    \label{fig:UO2_jparamag}
\end{figure}

\begin{table}[h]
\caption{Tabulated magnetic moment components of the four uranium atoms in the magnetic unit cell of \ce{UO2}. The spin and orbital contributions to the total computed moments by VASP are included below, computed using both \FsfUJ{PBE} and \FpUJ{PBE}. The on-site corrections of $U, J = 3.46, 0.3$ eV were applied to U-$f$ states to maintain consistency with Ref.~\onlinecite{dudarevParametrizationMathrmLSDANoncollinear2019}. Additionally, we compare to $U, J = 4.5, 1.0$ eV, which is the largest $U$ value reported in Ref.~\onlinecite{dudarevParametrizationMathrmLSDANoncollinear2019}, and a ``rounded-up" $J$ value of 1.0 eV to examine the effect of a larger Hund $J$ parameter on the spin and orbital uranium magnetic moments.}
\label{tab:magmomUO2}
\begin{ruledtabular}
\begin{tabular}{lc|rrrr|rrrr|rrrr}
Functional	&	Atom	&	\multicolumn{4}{c}{Spin}							&	\multicolumn{4}{c}{Orbital}							&	\multicolumn{4}{c}{Total}							\\				
	&		&	$m_x$	&	$m_y$	&	$m_z$	&	$| \bm m |$	&	$m_x$	&	$m_y$	&	$m_z$	&	$| \bm m |$	&	$m_x$	&	$m_y$	&	$m_z$	&	$| \bm m |$	\\				
\hline
\multirow{4}{*}{\shortstack{\FpUJ{PBE} \\ $U, J = 3.46, 0.30$ eV}} 	&	\rom{1}	&	-0.577	&	-0.583	&	-0.581	&	1.01	&	1.764	&	1.783	&	1.778	&	3.07	&	1.19	&	1.20	&	1.20	&	2.07	\\				
	&	\rom{2}	&	-0.584	&	0.576	&	0.581	&	1.01	&	1.787	&	-1.763	&	-1.776	&	3.08	&	1.20	&	-1.19	&	-1.20	&	2.07	\\				
	&	\rom{3}	&	0.578	&	-0.568	&	0.594	&	1.00	&	-1.769	&	1.739	&	-1.817	&	3.07	&	-1.19	&	1.17	&	-1.22	&	2.07	\\				
	&	\rom{4}	&	0.599	&	0.571	&	-0.57	&	1.00	&	-1.831	&	-1.747	&	1.744	&	3.07	&	-1.23	&	-1.18	&	1.17	&	2.07	\\				
\hline
\multirow{4}{*}{\shortstack{\FpUJ{PBE} \\ $U, J = 4.5, 1.0$ eV}} 	&	\rom{1}	&	-0.19	&	-0.19	&	-0.20	&	0.34	&	1.41	&	1.42	&	1.43	&	2.46	&	1.22	&	1.23	&	1.23	&	2.12	\\				
	&	\rom{2}	&	-0.20	&	0.20	&	0.19	&	0.34	&	1.43	&	-1.43	&	-1.40	&	2.46	&	1.24	&	-1.23	&	-1.21	&	2.12	\\				
	&	\rom{3}	&	0.19	&	-0.19	&	0.20	&	0.34	&	-1.40	&	1.39	&	-1.47	&	2.46	&	-1.21	&	1.20	&	-1.27	&	2.12	\\				
	&	\rom{4}	&	0.21	&	0.19	&	-0.18	&	0.34	&	-1.51	&	-1.40	&	1.35	&	2.46	&	-1.30	&	-1.21	&	1.16	&	2.12	\\				
\hline
\multirow{4}{*}{\shortstack{\FsfUJ{PBE}  \\ $U, J = 3.46, 0.30$ eV}} 	&	\rom{1}	&	0.21	&	0.17	&	0.24	&	0.36	&	0.73	&	0.73	&	0.81	&	1.31	&	0.93	&	0.90	&	1.04	&	1.66	\\				
	&	\rom{2}	&	0.21	&	-0.23	&	-0.19	&	0.36	&	0.73	&	0.73	&	-0.67	&	1.23	&	0.94	&	0.50	&	-0.86	&	1.37	\\				
	&	\rom{3}	&	-0.18	&	0.19	&	-0.24	&	0.36	&	-0.66	&	-0.66	&	-0.83	&	1.25	&	-0.84	&	-0.47	&	-1.07	&	1.44	\\				
	&	\rom{4}	&	-0.21	&	-0.21	&	0.21	&	0.36	&	-0.73	&	-0.73	&	0.74	&	1.26	&	-0.93	&	-0.93	&	0.94	&	1.62	\\				
\hline
\multirow{4}{*}{\shortstack{\FsfUJ{PBE}  \\ $U, J = 4.5, 1.0$ eV}} 	&	\rom{1}	&	0.08	&	0.08	&	0.09	&	0.14	&	0.95	&	0.91	&	1.04	&	1.68	&	1.03	&	0.98	&	1.13	&	1.82	\\				
	&	\rom{2}	&	0.08	&	-0.08	&	-0.08	&	0.14	&	0.99	&	-0.99	&	-0.92	&	1.68	&	1.07	&	-1.07	&	-1.00	&	1.82	\\				
	&	\rom{3}	&	-0.08	&	0.08	&	-0.09	&	0.14	&	-0.96	&	0.91	&	-1.03	&	1.68	&	-1.04	&	0.99	&	-1.12	&	1.82	\\				
	&	\rom{4}	&	-0.08	&	-0.08	&	0.08	&	0.14	&	-0.94	&	-1.01	&	0.96	&	1.68	&	-1.01	&	-1.09	&	1.04	&	1.82	\\				
\end{tabular}
\end{ruledtabular}					
\end{table}

\end{widetext}


\bibliography{main}

@article{bencheikhSpinOrbitCoupling2003,
  title = {Spin\textendash Orbit Coupling in the Spin-Current-Density-Functional Theory},
  author = {Bencheikh, K},
  year = {2003},
  month = dec,
  journal = {Journal of Physics A: Mathematical and General},
  volume = {36},
  number = {48},
  pages = {11929--11936},
  issn = {0305-4470, 1361-6447},
  doi = {10.1088/0305-4470/36/48/002},
}

@article{bravermanFastSpectralSubtractional2004,
  title = {A {{Fast Spectral Subtractional Solver}} for {{Elliptic Equations}}},
  author = {Braverman, Elena and Epstein, Boris and Israeli, Moshe and Averbuch, Amir},
  year = {2004},
  month = aug,
  journal = {Journal of Scientific Computing},
  volume = {21},
  number = {1},
  pages = {91--128},
  issn = {0885-7474},
  doi = {10.1023/B:JOMP.0000027957.39059.6b},
}

@article{capelleSpinCurrentsSpin2001,
  title = {Spin Currents and Spin Dynamics in Time-Dependent Density-Functional Theory},
  author = {Capelle, K. and Vignale, G. and Gyorffy, B. L.},
  year = {2001},
  month = oct,
  journal = {Physical Review Letters},
  volume = {87},
  number = {20},
  eprint = {cond-mat/0106021},
  eprinttype = {arxiv},
  pages = {206403},
  issn = {0031-9007, 1079-7114},
  doi = {10.1103/PhysRevLett.87.206403},
  archiveprefix = {arXiv},
}

@article{capelleSpinDensityFunctionalsCurrentDensity1997a,
  title = {Spin-{{Density Functionals}} from {{Current-Density Functional Theory}} and {{Vice Versa}}:{{A Road}} towards {{New Approximations}}},
  shorttitle = {Spin-{{Density Functionals}} from {{Current-Density Functional Theory}} and {{Vice Versa}}},
  author = {Capelle, K. and Gross, E. K. U.},
  year = {1997},
  month = mar,
  journal = {Physical Review Letters},
  volume = {78},
  number = {10},
  pages = {1872--1875},
  issn = {0031-9007, 1079-7114},
  doi = {10.1103/PhysRevLett.78.1872},
}

@article{dudarevParametrizationMathrmLSDANoncollinear2019,
  title = {Parametrization of {LSDA+$U$} for Noncollinear Magnetic Configurations: {Multipolar} Magnetism in \ce{UO2}},
  shorttitle = {Parametrization of \$\textbackslash mathrm\{\vphantom\}{{LSDA}}\vphantom\{\}+{{U}}\$ for Noncollinear Magnetic Configurations},
  author = {Dudarev, S. L. and Liu, P. and Andersson, D. A. and Stanek, C. R. and Ozaki, T. and Franchini, C.},
  year = {2019},
  month = aug,
  journal = {Physical Review Materials},
  volume = {3},
  number = {8},
  pages = {083802},
  publisher = {{American Physical Society}},
  doi = {10.1103/PhysRevMaterials.3.083802},
}

@article{eschrigCurrentDensityFunctional1985,
  title = {Current Density Functional Theory of Quantum Electrodynamics},
  author = {Eschrig, H. and Seifert, G. and Ziesche, P.},
  year = {1985},
  month = dec,
  journal = {Solid State Communications},
  volume = {56},
  number = {9},
  pages = {777--780},
  issn = {0038-1098},
  doi = {10.1016/0038-1098(85)90307-2},
}

@article{furnessCurrentDensityFunctional2015a,
  title = {Current {{Density Functional Theory Using Meta-Generalized Gradient Exchange-Correlation Functionals}}},
  author = {Furness, James W. and Verbeke, Joachim and Tellgren, Erik I. and Stopkowicz, Stella and Ekstr{\"o}m, Ulf and Helgaker, Trygve and Teale, Andrew M.},
  year = {2015},
  month = sep,
  journal = {Journal of Chemical Theory and Computation},
  volume = {11},
  number = {9},
  pages = {4169--4181},
  issn = {1549-9618, 1549-9626},
  doi = {10.1021/acs.jctc.5b00535},
}

@article{krishnaCompleteDescriptionMagnetic2019,
  title = {Complete Description of the Magnetic Ground State in Spinel Vanadates},
  author = {Krishna, Jyoti and Singh, N. and Shallcross, S. and Dewhurst, J. K. and Gross, E. K. U. and Maitra, T. and Sharma, S.},
  year = {2019},
  month = aug,
  journal = {Physical Review B},
  volume = {100},
  number = {8},
  pages = {081102},
  issn = {2469-9950, 2469-9969},
  doi = {10.1103/PhysRevB.100.081102},
}

@article{peraltaNoncollinearMagnetismDensity2007,
  title = {Noncollinear Magnetism in Density Functional Calculations},
  author = {Peralta, Juan E. and Scuseria, Gustavo E. and Frisch, Michael J.},
  year = {2007},
  month = mar,
  journal = {Physical Review B},
  volume = {75},
  number = {12},
  pages = {125119},
  issn = {1098-0121, 1550-235X},
  doi = {10.1103/PhysRevB.75.125119},
}

@article{pluharExchangecorrelationMagneticFields2019,
  title = {Exchange-Correlation Magnetic Fields in Spin-Density-Functional Theory},
  author = {Pluhar, Edward A. and Ullrich, Carsten A.},
  year = {2019},
  month = sep,
  journal = {Physical Review B},
  volume = {100},
  number = {12},
  pages = {125135},
  issn = {2469-9950, 2469-9969},
  doi = {10.1103/PhysRevB.100.125135},
}

@article{sharmaComparisonExactexchangeCalculations2007,
  title = {Comparison of Exact-Exchange Calculations for Solids in Current-Spin-Density- and Spin-Density-Functional Theory},
  author = {Sharma, S. and Pittalis, S. and Kurth, S. and Shallcross, S. and Dewhurst, J. K. and Gross, E. K. U.},
  year = {2007},
  month = sep,
  journal = {Physical Review B},
  volume = {76},
  number = {10},
  pages = {100401},
  issn = {1098-0121, 1550-235X},
  doi = {10.1103/PhysRevB.76.100401},
}

@article{sharmaSourceFreeExchangeCorrelationMagnetic2018a,
  title = {Source-{{Free Exchange-Correlation Magnetic Fields}} in {{Density Functional Theory}}},
  author = {Sharma, S. and Gross, E. K. U. and Sanna, A. and Dewhurst, J. K.},
  year = {2018},
  month = mar,
  journal = {Journal of Chemical Theory and Computation},
  volume = {14},
  number = {3},
  pages = {1247--1253},
  publisher = {{American Chemical Society}},
  issn = {1549-9618},
  doi = {10.1021/acs.jctc.7b01049},
}

@article{taoExplicitInclusionParamagnetic2005,
  title = {Explicit Inclusion of Paramagnetic Current Density in the Exchange-Correlation Functionals of Current-Density Functional Theory},
  author = {Tao, Jianmin},
  year = {2005},
  month = may,
  journal = {Physical Review B},
  volume = {71},
  number = {20},
  pages = {205107},
  issn = {1098-0121, 1550-235X},
  doi = {10.1103/PhysRevB.71.205107},
}

@article{tchenkoueForceBalanceApproach2019a,
  title = {Force {{Balance Approach}} for {{Advanced Approximations}} in {{Density Functional Theories}}},
  author = {Tchenkoue, Mary-Leena M. and Penz, Markus and Theophilou, Iris and Ruggenthaler, Michael and Rubio, Angel},
  year = {2019},
  month = oct,
  journal = {The Journal of Chemical Physics},
  volume = {151},
  number = {15},
  eprint = {1908.02733},
  eprinttype = {arxiv},
  pages = {154107},
  issn = {0021-9606, 1089-7690},
  doi = {10.1063/1.5123608},
  archiveprefix = {arXiv},
}

@article{vignaleCurrentSpindensityfunctionalTheory1988,
  title = {Current- and Spin-Density-Functional Theory for Inhomogeneous Electronic Systems in Strong Magnetic Fields},
  author = {Vignale, G. and Rasolt, Mark},
  year = {1988},
  month = jun,
  journal = {Physical Review B},
  volume = {37},
  number = {18},
  pages = {10685--10696},
  issn = {0163-1829},
  doi = {10.1103/PhysRevB.37.10685},
}

@article{vignaleDensityfunctionalTheoryStrong1987,
  title = {Density-Functional Theory in Strong Magnetic Fields},
  author = {Vignale, G. and Rasolt, Mark},
  year = {1987},
  month = nov,
  journal = {Physical Review Letters},
  volume = {59},
  number = {20},
  pages = {2360--2363},
  issn = {0031-9007},
  doi = {10.1103/PhysRevLett.59.2360},
}

@incollection{vignaleMagneticFieldsDensity1990,
  title = {Magnetic {{Fields}} and {{Density Functional Theory}}},
  booktitle = {Advances in {{Quantum Chemistry}}},
  author = {Vignale, G. and Rasolt, Mark and Geldart, D. J. W.},
  editor = {L{\"o}wdin, Per-Olov},
  year = {1990},
  month = jan,
  series = {Density {{Functional Theory}} of {{Many-Fermion Systems}}},
  volume = {21},
  pages = {235--253},
  publisher = {{Academic Press}},
  doi = {10.1016/S0065-3276(08)60599-7},
}

@article{BilbaoServer,
    author = {Perez-Mato, J. and Orobengoa, D. and Tasci, E. and De la Flor Martin, G. and Kirov, A.},
    year = {2011},
    month = {01},
    pages = {183-197},
    title = {Crystallography Online: Bilbao Crystallographic Server},
    volume = {43},
    journal = {Bulgarian Chemical Communications}
}

@book{SakuraiNapolitano2017, 
    place={Cambridge}, 
    edition={2}, 
    title={Modern Quantum Mechanics}, 
    DOI={10.1017/9781108499996}, 
    publisher={Cambridge University Press}, 
    author={Sakurai, J. J. and Napolitano, Jim}, 
    year={2017}
}

@article{socVASP,
  title = {Calculation of the magnetic anisotropy with projected-augmented-wave methodology and the case study of disordered {${\mathrm{Fe}}_{1\ensuremath{-}x}{\mathrm{Co}}_{x}$} alloys},
  author = {Steiner, Soner and Khmelevskyi, Sergii and Marsmann, Martijn and Kresse, Georg},
  journal = {Phys. Rev. B},
  volume = {93},
  issue = {22},
  pages = {224425},
  numpages = {6},
  year = {2016},
  month = {Jun},
  publisher = {American Physical Society},
  doi = {10.1103/PhysRevB.93.224425},
  url = {https://link.aps.org/doi/10.1103/PhysRevB.93.224425}
}

@incollection{hafner_vienna_1997,
	address = {Boston, MA},
	title = {The {Vienna} {Ab}-{Initio} {Simulation} {Program} {VASP}: {An} {Efficient} and {Versatile} {Tool} for {Studying} the {Structural}, {Dynamic}, and {Electronic} {Properties} of {Materials}},
	isbn = {978-1-4615-5943-6},
	shorttitle = {The {Vienna} {Ab}-{Initio} {Simulation} {Program} {VASP}},
	url = {https://doi.org/10.1007/978-1-4615-5943-6_10},
	urldate = {2022-03-10},
	booktitle = {Properties of {Complex} {Inorganic} {Solids}},
	publisher = {Springer US},
	author = {Hafner, J. and Kresse, G.},
	editor = {Gonis, Antonios and Meike, Annemarie and Turchi, Patrice E. A.},
	year = {1997},
	doi = {10.1007/978-1-4615-5943-6_10},
	pages = {69--82},
}

@article{PhysRevLett.69.2307,
  title = {Enhanced orbital magnetic moment on Co atoms in {Co/Pd} multilayers: A magnetic circular x-ray dichroism study},
  author = {Wu, Y. and St\"ohr, J. and Hermsmeier, B. D. and Samant, M. G. and Weller, D.},
  journal = {Phys. Rev. Lett.},
  volume = {69},
  issue = {15},
  pages = {2307--2310},
  numpages = {0},
  year = {1992},
  month = {Oct},
  publisher = {American Physical Society},
  doi = {10.1103/PhysRevLett.69.2307},
  url = {https://link.aps.org/doi/10.1103/PhysRevLett.69.2307}
}

@article{PhysRevLett.123.207201,
  title = {Noncollinear Ordering of the Orbital Magnetic Moments in Magnetite},
  author = {Elnaggar, H. and Sainctavit, Ph. and Juhin, A. and Lafuerza, S. and Wilhelm, F. and Rogalev, A. and Arrio, M.-A. and Brouder, Ch. and van der Linden, M. and Kakol, Z. and Sikora, M. and Haverkort, M. W. and Glatzel, P. and de Groot, F. M. F.},
  journal = {Phys. Rev. Lett.},
  volume = {123},
  issue = {20},
  pages = {207201},
  numpages = {6},
  year = {2019},
  month = {Nov},
  publisher = {American Physical Society},
  doi = {10.1103/PhysRevLett.123.207201},
  url = {https://link.aps.org/doi/10.1103/PhysRevLett.123.207201}
}

@article{restaElectricalPolarizationOrbital2010,
	title = {Electrical polarization and orbital magnetization: the modern theories},
	volume = {22},
	issn = {0953-8984},
	shorttitle = {Electrical polarization and orbital magnetization},
	url = {https://dx.doi.org/10.1088/0953-8984/22/12/123201},
	doi = {10.1088/0953-8984/22/12/123201},
	number = {12},
	urldate = {2023-04-07},
	journal = {Journal of Physics: Condensed Matter},
	author = {Resta, Raffaele},
	month = mar,
	year = {2010},
	pages = {123201},
}

@article{changBerryCurvatureOrbital2008,
	title = {Berry curvature, orbital moment, and effective quantum theory of electrons in electromagnetic fields},
	volume = {20},
	issn = {0953-8984},
	url = {https://dx.doi.org/10.1088/0953-8984/20/19/193202},
	doi = {10.1088/0953-8984/20/19/193202},
	number = {19},
	urldate = {2023-04-07},
	journal = {Journal of Physics: Condensed Matter},
	author = {Chang, Ming-Che and Niu, Qian},
	month = apr,
	year = {2008},
	pages = {193202},
}

@article{streltsovOrbitalPhysicsTransition2017,
	title = {Orbital physics in transition metal compounds: new trends},
	volume = {60},
	issn = {1063-7869},
	shorttitle = {Orbital physics in transition metal compounds},
	url = {https://iopscience.iop.org/article/10.3367/UFNe.2017.08.038196/meta},
	doi = {10.3367/UFNe.2017.08.038196},
	number = {11},
	urldate = {2023-09-24},
	journal = {Physics-Uspekhi},
	author = {Streltsov, S. V. and Khomskii, D. I.},
	month = nov,
	year = {2017},
	note = {Publisher: IOP Publishing},
	pages = {1121},
}

@book{AshcroftMermin1976,
    author = {Ashcroft, Neil W. and 
    Mermin, David N.},
    address = {Australia},
    booktitle = {Solid state physics},
    isbn = {0030839939},
    publisher = {Brooks/Cole},
    title = {Solid state physics },
    year = {1976 - 1976},
}

@article{hamadaPhaseInstabilityMagnetic2012b,
	title = {Phase instability of magnetic ground state in antiperovskite \ce{Mn3ZnN}: {Giant} magnetovolume effects related to magnetic structure},
	volume = {111},
	issn = {0021-8979},
	shorttitle = {Phase instability of magnetic ground state in antiperovskite {Mn3ZnN}},
	url = {https://aip.scitation.org/doi/full/10.1063/1.3670052},
	doi = {10.1063/1.3670052},
	number = {7},
	urldate = {2023-04-16},
	journal = {Journal of Applied Physics},
	author = {Hamada, T. and Takenaka, K.},
	month = apr,
	year = {2012},
	note = {Publisher: American Institute of Physics},
	pages = {07A904},
}

@article{lawQuantitativeCriterionDetermining2018,
	title = {A quantitative criterion for determining the order of magnetic phase transitions using the magnetocaloric effect},
	volume = {9},
	copyright = {2018 The Author(s)},
	issn = {2041-1723},
	url = {https://www.nature.com/articles/s41467-018-05111-w},
	doi = {10.1038/s41467-018-05111-w},
	number = {1},
	urldate = {2023-04-16},
	journal = {Nature Communications},
	author = {Law, Jia Yan and Franco, Victorino and Moreno-Ramírez, Luis Miguel and Conde, Alejandro and Karpenkov, Dmitriy Y. and Radulov, Iliya and Skokov, Konstantin P. and Gutfleisch, Oliver},
	month = jul,
	year = {2018},
	note = {Number: 1
Publisher: Nature Publishing Group},
	pages = {2680},
}

@misc{mooreHighthroughputDeterminationHubbard2022a,
	title = {High-throughput determination of {Hubbard} {U} and {Hund} {J} values for transition metal oxides via linear response formalism},
	url = {http://arxiv.org/abs/2201.04213},
	urldate = {2023-04-16},
	publisher = {arXiv},
	author = {Moore, Guy C. and Horton, Matthew K. and Ganose, Alexander M. and Siron, Martin and Linscott, Edward and O'Regan, David D. and Persson, Kristin A.},
	month = oct,
	year = {2022},
	note = {arXiv:2201.04213 [cond-mat]},
}

@article{PhysRevLett.95.057205,
  title = {Spin Current and Magnetoelectric Effect in Noncollinear Magnets},
  author = {Katsura, Hosho and Nagaosa, Naoto and Balatsky, Alexander V.},
  journal = {Phys. Rev. Lett.},
  volume = {95},
  issue = {5},
  pages = {057205},
  numpages = {4},
  year = {2005},
  month = {Jul},
  publisher = {American Physical Society},
  doi = {10.1103/PhysRevLett.95.057205},
  url = {https://link.aps.org/doi/10.1103/PhysRevLett.95.057205}
}

@misc{hawkhead2023firstprinciples,
  title={First-principles calculations of magnetic states in pyrochlores using a source-corrected exchange and correlation functional}, 
  author={Z. Hawkhead and N. Gidopoulos and S. J. Blundell and S. J. Clark and T. Lancaster},
  year={2023},
  eprint={2302.08564},
  archivePrefix={arXiv},
  primaryClass={cond-mat.mtrl-sci}
}

@article{dewhurstDevelopmentElkLAPW,
	title = {Development of the {Elk} {LAPW} {Code}},
	url = {https://www2.mpi-halle.mpg.de/fileadmin/templates/images/articles/elk/article.pdf},
	journal = {Max Planck Institute of Microstructure Physics Theory Department},
	author = {Dewhurst, J K and Sharma, S},
}

@misc{elk-code,
  title = {\texttt{Elk} codebase},
  year = {2023},
  publisher = {SourceForge},
  journal = {SourceForge package},
  howpublished = {\url{https://elk.sourceforge.io}},
}

@misc{stokesISOTROPYSoftwareSuite,
    title = {{ISOTROPY} {Software} {Suite}},
    url = {https://iso.byu.edu/iso/isotropy.php},
    urldate = {2023-09-23},
    author = {Stokes, Harold T. and Hatch, Dorian M. and Campbell, Branton J.},
}

@misc{sf-patch-github,
  title = {Source-free {$\bm B_{xc}$} code patch repository},
  year = {2023},
  publisher = {GitHub},
  journal = {GitHub repository},
  howpublished = {\url{https://github.com/guymoore13/source_free_Bxc_VASP}},
}

@incollection{UO2-exp,
	title = {Chapter 1.5. {Magnetic} properties},
	volume = {D},
	copyright = {Copyright (c) 2013 International Union of Crystallography},
	url = {https://onlinelibrary.wiley.com/iucr/itc/Db/ch1o5v0001/fn/},
	urldate = {2023-09-22},
	booktitle = {International {Tables} for {Crystallography}},
	author = {Borovik-Romanov, A. S. and Grimmer, H. and Kenzelmann, M.},
	year = {2013},
	doi = {10.1107/97809553602060000904},
	note = {Publisher: International Union of Crystallography},
}

@article{YMnO3-exp,
	title = {Magnetic structure of hexagonal \ce{RMnO3} ({R = Y, Sc}): {Thermal} evolution from neutron powder diffraction data},
	volume = {62},
	shorttitle = {Magnetic structure of hexagonal \${R}\{{\textbackslash}mathrm\{{MnO}\}\}\_\{3\}\$ \$({R}={\textbackslash}mathrm\{{Y}\},{\textbackslash}mathrm\{ \}{\textbackslash}mathrm\{{Sc}\})},
	url = {https://link.aps.org/doi/10.1103/PhysRevB.62.9498},
	doi = {10.1103/PhysRevB.62.9498},
	number = {14},
	urldate = {2023-09-22},
	journal = {Physical Review B},
	author = {Muñoz, A. and Alonso, J. A. and Martínez-Lope, M. J. and Casáis, M. T. and Martínez, J. L. and Fernández-Díaz, M. T.},
	month = oct,
	year = {2000},
	note = {Publisher: American Physical Society},
	pages = {9498--9510},
}

@article{MnF2-exp,
	title = {Neutron scattering study of the classical antiferromagnet \ce{MnF2}: a perfect hands-on neutron scattering teaching issue on {Neutron} {Scattering} in {Canada}.},
	volume = {88},
	issn = {0008-4204},
	shorttitle = {Neutron scattering study of the classical antiferromagnet {MnF2}},
	url = {https://cdnsciencepub.com/doi/full/10.1139/P10-081},
	doi = {10.1139/P10-081},
	number = {10},
	urldate = {2023-09-22},
	journal = {Canadian Journal of Physics},
	author = {Yamani, Z. and Tun, Z. and Ryan, D. H.},
	month = oct,
	year = {2010},
	note = {Publisher: NRC Research Press},
	pages = {771--797},
}

@article{MnPtGa-exp,
	title = {Evolution of noncollinear magnetism in magnetocaloric {MnPtGa}},
	volume = {4},
	url = {https://link.aps.org/doi/10.1103/PhysRevMaterials.4.044405},
	doi = {10.1103/PhysRevMaterials.4.044405},
	number = {4},
	urldate = {2023-09-22},
	journal = {Physical Review Materials},
	author = {Cooley, Joya A. and Bocarsly, Joshua D. and Schueller, Emily C. and Levin, Emily E. and Rodriguez, Efrain E. and Huq, Ashfia and Lapidus, Saul H. and Wilson, Stephen D. and Seshadri, Ram},
	month = apr,
	year = {2020},
	note = {Publisher: American Physical Society},
	pages = {044405},
}

@article{Mn3Pt-exp,
	title = {Investigation of the {First}‐{Order} {Magnetic} {Transformation} in \ce{Mn3Pt}},
	volume = {38},
	issn = {0021-8979},
	url = {https://doi.org/10.1063/1.1709571},
	doi = {10.1063/1.1709571},
	number = {3},
	urldate = {2023-09-22},
	journal = {Journal of Applied Physics},
	author = {Krén, E. and Kádár, G. and Pál, L. and Szabó, P.},
	month = jun,
	year = {2004},
	pages = {1265--1266},
}

@article{Mn3ZnN-exp,
	title = {DIFFRACTION NEUTRONIQUE DE \ce{Mn3ZnN}},
	volume = {32},
	issn = {0449-1947, 2777-3418},
	url = {http://dx.doi.org/10.1051/jphyscol:19711309},
	doi = {10.1051/jphyscol:19711309},
	number = {C1},
	urldate = {2023-09-22},
	journal = {Le Journal de Physique Colloques},
	author = {Fruchart, D. and Bertaut, E. F. and Madar, R. and Fruchart, R.},
	month = feb,
	year = {1971},
	note = {Publisher: EDP Sciences},
	pages = {C1--877},
}



\renewcommand{\thesection}{S\arabic{section}}
\renewcommand{\thesubsection}{\Alph{subsection}}
\renewcommand{\theequation}{S\arabic{equation}}
\renewcommand{\thetable}{S\arabic{table}}
\renewcommand{\thefigure}{S\arabic{figure}}

\setcounter{section}{0}
\setcounter{equation}{0}
\setcounter{table}{0}
\setcounter{figure}{0}

\begin{widetext}

\clearpage


\newcommand{\refM}[2]{\ref{#1}}



\begin{center}

\vspace{4ex}

{
\LARGE
Supplementary information: \\ 
\textit{
Realistic non-collinear ground states of solids \\ 
with source-free exchange correlation functional \\
}
}

\vspace{4ex}

{
\large
Guy C. Moore, 
Matthew K. Horton, \\
Aaron D. Kaplan, 
Sin\'ead M. Griffin, \\
Kristin A. Persson
}

\vspace{4ex}

\today

\vspace{4ex}

\end{center}


\section{Numerical \& computational details}

\subsection{DFT specifications}

In all magnetic systems and VASP calculation settings that we tested, we found improved convergence of the SF XC functional using the blocked Davidson (BD) algorithm (\texttt{ALGO = Normal}, \texttt{IALGO = 38}) compared with the preconditioned conjugated gradient (PCG) algorithm (\texttt{ALGO = All}, \texttt{IALGO = 58}). 
Using the PCG algorithm, we found in most instances that the energy decreased up to a threshold, and then increased up to a self-consistency step, after which the absolute energy difference between self-consistency steps decreased very slowly. By comparison, the BD algorithm often required an order of magnitude fewer number of self-consistency steps to converge. Therefore, we strongly recommend that for the SF XC implementation in VASP, users use the BD algorithm (\texttt{ALGO = Normal}, \texttt{IALGO = 38}).
In many instances the source-free constrained PBE calculations took significantly longer to converge than the conventional PBE counterparts. However, this is not entirely surprising because one would expect the additional constraint on $\bm B_{xc}$ to require a longer convergence trajectory. In a significant number of calculations, on the other hand, we found that the SF calculations converged within a few iterations, similar to their non-SF equivalent. For small enough perturbation from the ground-state, the \Fsf{PBE} calculations converged quicker than conventional noncollinear PBE, such as in the case of \ce{Mn3As}.

\section{Additional explanations \& observations}

\subsection{Noncollinear nature of $\bm B_{xc}$}
\label{sec:NclBxc}

The source-free constraint forces the resultant $\bm B_{xc}$ to be noncollinear, and therefore this constrained functional is not applicable to collinear magnetic functionals. To illustrate this, consider the example where we define the magnetic field vector field $\bm B_{xc} (\bm r)$ over $\mathbb R^3$.
If we only allow one component of this three-dimensional vector field to be non-zero (in the $z$-direction for instance, where $\bm B_{xc} (\bm r) = B^z_{xc} \bm{\hat z}$) we can rewrite the divergence-free constraint of $\bm B_{xc} (\bm r)$ as 
\begin{align}
    \nabla \cdot \bm B_{xc} (\bm r) = \frac{\partial}{\partial z} B^z_{xc} = 0
\end{align}
Therefore, $B^z_{xc}$ must be constant with respect to $z$, where $\bm B_{xc} (\bm r) = B^z_{xc}(x,y) \bm{\hat z}$. In general, we would expect the ground-state magnetization field $\bm m (\bm r)$ to align parallel to the corresponding $\bm B_{xc} (\bm r)$, and therefore this constraint on $\bm B_{xc} (\bm r)$ severely limits the possible $\bm m (\bm r)$ in ways that are unphysical.

\subsection{Expressing kinetic energy contributions in terms of projected spin currents}
\label{sec:KinEprojSpinCurrents}

It has been established in Ref.~\cite{vignaleCurrentSpindensityfunctionalTheory1988} that the density and spin current matrices within SCDFT are related via the following relations
\begin{align}
\rho_{\alpha \beta}(\bm{r}) &= \rho(\bm{r}) \delta_{\alpha \beta}+\sum_{\lambda=1}^3 s_\lambda(\bm{r}) \sigma_{\alpha \beta}^\lambda \label{eq:rho_spinor} \\
\bm{j}_{p, \alpha \beta}(\bm{r}) &= \bm{j}_p(\bm{r}) \delta_{\alpha \beta}+\sum_{\lambda=1}^3 \bm{j}_{p, \lambda}(\bm{r}) \sigma_{\alpha \beta}^\lambda.
\end{align} \label{eq:jp_spinor}
From Ref.~\cite{vignaleMagneticFieldsDensity1990}, $\hat{s}$ denotes the ``direction of the local spin polarization,'' from which it is possible to define a set of variables projected along spin orientation
\begin{align}
|\bm{s}(\bm{r})| &= \left(\sum_\lambda s_\lambda(\bm{r})^2\right)^{1 / 2} \label{eq:s_mag} \\
\rho_{\pm \hat{\bm{s}}}(\bm{r}) &= \frac{1}{2} \left( \rho(\bm{r}) \pm|\bm{s}(\bm{r})| \right) \label{eq:rho_up_down} \\
\bm{j}_{p \pm \hat{\bm{s}}}(\bm{r}) &= \frac{1}{2} \left( \bm{j}_p(\bm{r}) \pm \bm{j}_{p \|}(\bm{r}) \right) \label{eq:jp_up_down} \\
\bm{j}_{p \|}(\bm{r}) &= \sum_{\lambda=1}^3 \bm{j}_{p \lambda}(\bm{r}) s_\lambda(\bm{r}) /|\bm{s}(\bm{r})| \label{eq:jp_parallel}
\end{align}
where $\bm{j}_{p \|}(\bm{r})$ is the ``longitudinal (in spin-space) current.'' Armed with these definitions, it is straightforward to show that
\begin{align}
    \left| \bm j_{p,\uparrow} \right|^2 + \left| \bm j_{p,\downarrow} \right|^2
    &= \frac{1}{2} \left| \bm j_p \right|^2 + \frac{1}{2} \sum_\lambda \left| \bm j_{p,\lambda} \right|^2
\end{align}
Therefore, we can express the relationship between current contributions to kinetic energy densities (Equation~\refM{eq:kin_density_spin}{19}) as 
\begin{align}
    \frac{\rho_\uparrow}{\rho}\overline{\tau}_{\uparrow} + \frac{\rho_\downarrow}{\rho} \overline{\tau}_{\downarrow}
    &= \frac{1}{2} \left\{ \overline{\tau} + \sum_\lambda \frac{\rho_\lambda}{\rho} \overline{\tau}_{\lambda} \right\},
\end{align}
%
%
where $\overline{\tau}_{\mu} = - m \left| \bm j_{p,\mu} \right|^2 / \rho_\mu $.


\subsection{Vanishing integrals under periodic boundary conditions}
\label{sec:VanishIntPBC}

We can see by the fundamental theorem of calculus, that the following integral is zero under periodic boundary conditions (PBCs)
\begin{align}
  \int_\Omega \pderiv{\phi}{x_i} \ d \bm r
  &= \int_0^{L_j} \prod_{j \ne i} d x_j \ \int_0^{L_i} d x_i \ \pderiv{\phi}{x_i} \nonumber \\
  &= \int_0^{L_j} \prod_{j \ne i} d x_j \ \left[ \phi(x_i = L_i, ..., x_{j \ne i}) - \phi(x_i = 0, ..., x_{j \ne i}) \right] \nonumber \\
  &= 0,
  \label{eq:pderiv_int_periodic}
\end{align}
because $\phi(x_i = L_i, ..., x_{j \ne i}) = \phi(x_i = 0, ..., x_{j \ne i})$ by the definition of PBCs. From Equation~\ref{eq:pderiv_int_periodic}, it is straightforward to show that the following integrals are also zero under PBCs:
\begin{align}
  \int_\Omega \nabla \cdot \bm L \ d \bm r &= 0 \label{eq:div_thm_periodic} \\
  \int_\Omega \nabla \phi \ d \bm r &= \bm 0 \label{eq:grad_int_periodic} \\
  \int_\Omega \nabla \times \bm L \ d \bm r &= \bm 0. \label{eq:stokes_periodic}
\end{align}


\end{widetext}


\end{document}